\documentclass[12pt, letter]{article}
\usepackage{natbib}
\usepackage{float}
\usepackage{eurosym}
\usepackage{amsfonts}
\usepackage{enumitem}
\usepackage{amssymb}
\usepackage{subcaption,graphicx}
\usepackage{lscape}
\usepackage{amsmath}
\usepackage{setspace}
\usepackage{rotating}
\usepackage{longtable}
\usepackage{multirow}
\usepackage{booktabs}
\usepackage{appendix}
\usepackage{tabularx}
\usepackage{changepage}
\newcolumntype{Y}{>{\centering\arraybackslash}X}
\usepackage[labelfont=normalfont, margin = -1.5cm]{caption}
\DeclareCaptionLabelSeparator{dot}{.\space}
\newlength{\paperlength}\newlength{\paperwidthx}
\def \sym#1{\ifmmode^{#1}\else \(^{#1}\) \fi}

\newlength{\leftmarginodd}\newlength{\tmargin}
\newlength{\bmargin}\newlength{\xvar}\newlength{\rightmarginodd}
\newlength{\hcomp}\newlength{\vcomp}
\usepackage{titling}
\renewcommand{\thepage}{-- \arabic{page} --}
\renewcommand{\thesection}{\arabic{section}}

\usepackage{dcolumn}
\newcolumntype{R}[1]{>{\raggedleft\arraybackslash}p{#1}}
\newcolumntype{L}[1]{>{\raggedright\arraybackslash}p{#1}}
\newcolumntype{C}[1]{>{\centering\arraybackslash}p{#1}}
\newcolumntype {d}[1]{D{.}{.}{#1}} 
\newcolumntype {.}{D{.}{.}{-1}}
\usepackage[hyperfootnotes=true]{hyperref}
\hypersetup {colorlinks=true,raiselinks=true,breaklinks=false}
\hypersetup{colorlinks,
            linkcolor=darkred,
            anchorcolor=darkred,
            citecolor=darkred}
            \usepackage{color}
    \definecolor{darkred}{rgb}{0.6,0.0,0.0}
\newcommand{\SVS}{\operatorname{SVIX}}
\newcommand{\calloption}{\operatorname{call}}
\newcommand{\putoption}{\operatorname{put}}

\newcommand{\E}{\operatorname{\mathbb{E}}}

\begin{document}
\title{\textbf{Disaster Resilience and Asset Prices}\thanks{\noindent{\footnotesize \noindent Marco Pagano gratefully acknowledges financial support from the Italian 
Ministry for Education, University and Research (MIUR) and the Einaudi Institute for 
Economics and Finance (EIEF). Christian Wagner is an associate member of the Center for 
Financial Frictions (FRIC) and acknowledges support from grant no. DNRF102. Pagano:  pagano56@gmail.com. Wagner: christian.wagner@wu.ac.at. Zechner: josef.zechner@wu.ac.at}}
}

\author{ Marco Pagano\\
\normalsize{University of Naples Federico II, CSEF, EIEF and CEPR} \and 
Christian Wagner \\
\normalsize{WU Vienna University of Economics and Business}\\ \normalsize{and Vienna Graduate School of Finance (VGSF})
\and Josef Zechner\\
\normalsize{WU Vienna University of Economics and Business,} \\ \normalsize{Vienna Graduate School of Finance (VGSF) and CEPR}}

\date{May 2020}

\maketitle
\begin{abstract}
{\small \noindent \noindent 
This paper investigates whether security markets price the effect of social 
distancing on firms’ operations. We document that firms that are more resilient 
to social distancing significantly outperformed those with lower resilience during 
the COVID-19 outbreak, even after controlling for the standard risk factors.
Similar cross-sectional return differentials already emerged before the COVID-19 crisis: 
the 2014-19 cumulative return differential between more and less resilient firms is of 
similar size as during the outbreak, suggesting growing 
awareness of pandemic risk well in advance of its materialization. Finally, we use 
stock option prices to infer the market’s return expectations after the onset of the 
pandemic: even at a two-year horizon, stocks of more pandemic-resilient firms are 
expected to yield significantly lower returns than less resilient ones, reflecting 
their lower exposure to disaster risk. Hence, going forward, markets appear to price 
exposure to a new risk factor, namely, pandemic risk.}
\end{abstract}

\smallskip


\thispagestyle{empty}\onehalfspacing

\setcounter{page}{0}%
\pagebreak
\section{Introduction}
\label{intro}

The COVID-19 pandemic and the resulting lockdown are currently inflicting 
massive harm on the economy, in the form of unprecedented output losses, massive 
redundancies and countless bankruptcies. The effects of the pandemic are 
widely perceived not to be purely transient, as witnessed by current changes 
in expectations \citep{coibion/gorodnichenko/weber:20labormarketcovid19, 
hanspal/weber/wohlfart:20wp,coibion/gorodnichenko/weber:20costcovid19} 
and in asset prices \citep{gormsen/koijen:20bfi}. But its effects are also 
proving to be quite heterogeneous: some firms, especially in high-tech industries,
appear to have adapted quite well to social distancing requirements, 
for instance by resorting extensively to teleworking, while others, such as food 
catering, travel and tourism, could not do so, as the nature of their 
business requires close contact with customers and/or among employees. This 
heterogeneity is also visible in the diverging stock price performance of companies: 
in the first quarter of 2020, stocks such as Apple, Microsoft and Google outperformed 
the market, yielding market-adjusted returns of 19\%, 12\% and 33\% respectively, 
while others such as Marriott, United Airlines and Royal Caribbean massively underperformed, with market-adjusted returns of $-38\%$, $-53\%$ and $-66\%$.

\smallskip

Hence, the COVID-19 shock has unearthed a hitherto hidden economic watershed, namely, 
that between disaster-resilient activities and non-resilient ones. Insofar as the 
COVID-19 pandemic persists or revives in the near future, or similar disasters may  
occur further in the future, resilience may become a key firm attribute, one which 
will be relevant to investors' portfolio choices, banks' lending policies, and 
managers' investment decisions. 

\smallskip

In this paper, we investigate whether asset prices reveal growing investors' awareness 
that pandemic resilience, defined as reliance on technologies and/or organizational 
structures that are robust to social distancing, is priced by security markets. To 
measure firms' resilience to pandemics, we rely on measures recently introduced 
in labor economics by \cite{dingel/neiman:20nber}, \cite{hensvik/lebarbanchon/rathelot:20cepr} 
and  \cite{koren/peto:20covid}, intended to capture the extent to which firms'  
operations are compatible with social distancing. These measures quantify the degree 
to which jobs can be done from home and do not rely on human interaction in physical
proximity. 

\smallskip

We then test whether the stocks of more resilient companies have generated excess 
returns, after controlling for market risk and other established risk factors, 
\textit{before} and/or \textit{during} the COVID-19 shock. We find this to be 
the case: pandemic-resilient stocks outperformed less resilient ones 
not only between late February and March 2020, i.e. during the outbreak of the 
pandemic, but also in the previous six years, in which their cumulative excess 
return was of similar magnitude as during the crisis.

\smallskip

Furthermore, we investigate the returns that investors expect the two stock classes 
to generate  \textit{after} the COVID-19 shock, by extracting the expected stock 
returns implied by option prices. We find that, going forward, pandemic-resilient 
stocks are expected to generate lower excess returns, indicating that, since  
the COVID-19 pandemic materialized, the market has priced a disaster premium. 

\smallskip

Thus, our research question is not only \textit{whether}, but also 
\textit{when} financial markets started pricing disaster risk and resilience to 
it. Interestingly, our evidence indicates that investors became gradually aware of such 
risk even before the COVID-19 shock, and they consider such risk to be still 
price-relevant, even after it has materialized. 

\smallskip

Given the largely unanticipated nature of the current pandemic, it may appear 
unrealistic that investors became cognizant to its threat in advance. But it should 
be recalled that as early as five years before the COVID-19 outbreak, high-profile 
business and political leaders already issued public warnings of the risk of 
devastating epidemics. For instance, in a speech delivered on 2 December 2014 at 
the NIH about the response to the Ebola epidemic, U.S. President Barack Obama stated:

\begin{quote}
``There may and likely will come a time in which we have both an airborne disease that is 
deadly. And in order for us to deal with that effectively, we have to put in place an 
infrastructure -- not just here at home, but globally -- that allows us to see it quickly, isolate it quickly, respond to it quickly. [...] So that if and when a new strain of flu, 
like the Spanish flu, crops up five years from now or a decade from now, we've made the 
investment and we're further along to be able to catch it. It is a smart investment for 
us to make. It's not just insurance; it is knowing that down the road we're going to 
continue to have problems like this -- particularly in a globalized world where you 
move from one side of the world to the other in a day.''
\end{quote}

Similarly, in 2015, Microsoft co-founder Bill Gates gave a  TED Talk about pandemics that attracted widespread attention. In this talk he warned that in 2014 the world had barely 
avoided a global outbreak of Ebola, largely because of pure luck, and, just like 
Obama, alerted the audience to the need to prepare for future pandemics, from scenario 
planning to vaccine research and health worker training. Hence, it cannot be ruled out 
that the most alert investors may have started taking into account such concerns in 
their portfolio choices well in advance of the current pandemic, shying away from the 
stocks of companies that would be less resilient to it and starting to overweight those 
likely to be more resilient. 

\smallskip

One may also wonder why even after the occurrence of COVID-19 the market still prices 
pandemic risk to some extent, being willing to accept lower expected returns on more 
resilient stocks, as revealed by option prices according to our estimates. However, 
currently there is still high uncertainty regarding the duration of the COVID-19 
pandemic. Medical experts have repeatedly warned about the risk of a second wave of 
contagion as the lockdown measures are gradually relaxed \citep[see][among others]{xu_li:20lancet}. Indeed, top health officials do not rule out that the disease may 
become endemic. On 13 May 2020, Dr. Mike Ryan, executive director of emergencies at 
the World Health Organization (WHO) stated:

\begin{quote}
``I think it's important to put this on the table. This virus may become just another 
endemic virus in our communities. And this virus may never go away. HIV has not gone 
away, we've come to terms with the virus [...] I think it is important that we're 
realistic and I don't think anyone can predict when or if this disease will disappear.''
\end{quote}

Such uncertainty, possibly coupled with heightened awareness that similar disasters may 
occur again in the future, could explain why, since COVID-19, pandemic risk is priced 
in the cross-section of returns, as shown by our evidence based on option prices. Hence,
going forward, asset pricing models will have to include exposures to this additional 
risk factor among those used to explain the cross-section of returns, and 
asset managers will have to take such exposures into account in portfolio selection. 

\smallskip

Our analysis is related to the asset pricing literature on rare disasters, starting with \cite{rietz:88jme}, who extends the \citet{lucas:78ecta} model to include a rare disaster state and shows that this leads to high equity risk premia and low risk-free returns, even with reasonable time discounting and risk preferences. \cite{Barro_2006:QJE} and \cite{Barro_2009:AER} extend this model and show that empirically calibrated disaster 
probabilities may suffice to explain the observed high equity premium, low risk-free 
rate and stock return volatility. The theoretical literature on disaster risk has 
also been extended to allow for learning (see, e.g.,\cite{veronesi:04jedc}, 
\cite{Wachter_Zhu_2019}, \cite{Gillman_et_al_RF:2014}, \cite{Lu_Siemer_2016}), and/or 
for stochastic disaster risk (see, e.g., \cite{wachter:13jf} and \cite{Gabaix_2012:QJE}).

One common feature of these models is that risk premia that would appear abnormally 
high conditioning on no disaster occurring, are in fact justified, being merely an 
equilibrium compensation for the expected loss in a disaster plus a risk premium 
as this loss occurs when the marginal utility of consumption is high. In particular, 
the model by \cite{Gabaix_2012:QJE} implies that, in the cross-section of stocks, 
those that are expected to be less resilient to disasters, should carry a higher 
risk premium, conditioning on no disaster occurring, but also unconditionally. 
Our empirical analysis speaks to these predictions.   

\smallskip

Our work is also related to a fast growing literature on the stock market response 
to the COVID-19 pandemic in the first quarter of 2020. The early comprehensive 
study by \citet{ramelli/wagner:20cepr} documents that during the `outbreak' period 
(which they define as 20 January to 21 February 2020), U.S. firms with high exposure 
to China and, more generally, to international trade, as well as firms with high 
leverage and low cash holdings experienced the sharpest stock price declines. The 
leverage- and cash holding effects also persist through the `fever' period (from 24 
February to 20 March 2020). \citet{bretscher/hsu/tamoni:20ssrn} provide evidence 
for supply chain effects in the cross-section of stocks during COVID-19. 
Moreover, \citet{albuquerque/koskinen/yang/zhang:20ssrn} find that firms with 
high environmental and social (ES) ratings offered comparably higher returns and  
lower return volatility in the first quarter of 2020. 

\smallskip

Some studies relate the price response of different stocks to the pandemic to 
the corresponding firms' exposure to the disease. \citet{ramelli/wagner:20cepr} and 
\citet{hassan/hollander/vanlent/tahoun:20nber} analyze conference call data, which 
the latter use to construct text-based, firm-level measures for exposures to epidemic 
diseases, and find that stock returns are significantly and negatively related to 
disease exposures, with demand- and supply-chain related concerns being primary drivers. 
\citet{alfaro/chari/greenland/schott:20ssrn} analyse the effect of unanticipated changes 
in infections during the SARS and the COVID-19 epidemics on stock returns, and show 
that stock prices drop in response to high unexpected infections. In the cross-section, 
exposure to pandemic risk turns out to be greater for larger and more capital 
intensive firms, and, consistently with \cite{ramelli/wagner:20cepr}, more levered and 
less profitable ones. 

\smallskip

Some of the results obtained for the response of U.S. stock returns to the pandemic 
appear to extend to non-U.S. stock returns. \citet{ding/levine/lin/xie:20nber} 
show for a sample of over 50 countries that firms with better financials, less 
supply chain exposures and more corporate social responsibility (CSR) activities 
experienced milder stock price reactions in the first quarter of 2020. Other studies 
focus on the response of country-level stock market indices to COVID-19: 
\citet{ru/yang/zou:20ssrn} find that stock markets in countries with 2003 SARS 
experience reacted more quickly to the outbreak than countries without prior 
experience, while \citet{gerding/martin/nagler:20ssrn} document that market declines 
were more severe in countries with lower fiscal capacity, defined as higher debt/GDP 
ratio. Finally, \citet{croce/farroni/wolfskeil:20ssrn} use Twitter news to study 
the (real-time) COVID-19 caused contagion in global equity markets. 

\smallskip

Our analysis differs from that in all these papers since it focuses on the asset pricing 
implications of companies' exposure to social distancing, which is the main economic 
consequence of the epidemic, and studies such implications not only for the period 
of the COVID-19 outbreak, but also prior and after its occurrence. This enables us 
to investigate whether learning about pandemic risk occurred in advance of 
the outbreak, and whether investors kept pricing it since the outbreak. 

\smallskip

The rest of the paper is structured as follows. Section \ref{theory} provides a framework to interpret the relationship between disaster risk and the stock returns of firms 
featuring different disaster resilience. Section \ref{measuringresilience} provides a brief discussion of alternative measures of firms' disaster resilience. The data and results 
are presented in Section \ref{results}. Section \ref{conclusions} concludes.

\section{Disaster Awareness and Risk Premia}
\label{theory}

In this section we sketch three distinct and mutually exclusive models of how financial markets may respond to disaster risk and to its materialization. As we shall see, 
each model has different predictions about the stock return differential of 
resilient \textit{vs.} non-resilient firms  prior, during and after the occurrence 
of a disaster, as shown in Table \ref{tab:predictions}. Our empirical analysis in 
Section \ref{results} will investigate which of the three sets of predictions 
is most consistent with the data. 

\medskip

\begin{table}[h!]
\centering
\caption{\textbf{Predicted return differential of resilient vs. non-resilient firms}}
\begin{tabular}{c|c|c|c}
\hline \hline
Theory  & Before COVID-19 & During COVID-19 & After COVID-19 \\
\hline
Unpriced disaster risk  & Zero     & Positive & Zero \\
Priced disaster risk    & Negative & Positive & Negative \\
Pre-disaster learning   & Positive & Positive & Negative \\  
\hline \hline   
\end{tabular}
\label{tab:predictions}
\end{table}

\medskip

\noindent {\it Unpriced disaster risk.} We start with a model where disaster is 
completely unexpected. In the case of COVID-19, this amounts to assuming that 
before the first quarter of 2020 financial market participants were unaware of the 
danger posed by the virus and of its consequences in terms of social distancing.  
Going forward, a new disaster is again regarded as a zero-probability event, or anyway
as being price irrelevant (for instance, in the case of COVID-19, due to development 
of vaccines and/or effective drugs). In this model, disaster risk would not be priced 
before COVID-19 nor after it. Such market expectations could result either 
from bounded rationality, i.e. investors assigning a zero probability weight to a 
positive probability event, or from a disaster truly having a negligible probability 
both before and after its occurrence. In the latter case, unpriced disaster risk would 
be consistent with rational expectations. 

If this is the correct model in the case of the COVID-19 pandemic, one should observe 
(i) no return pattern related to firm pandemic resilience prior to 2020; (ii) a 
sharper price drop for less resilient firms than for more resilient ones at the time of 
the outbreak, as investors take into account that the cash flows of the former will be 
hurt more than those of the latter; (iii) no differential response of the expected 
returns of the two classes of firms after the crisis, since COVID-19 does not lead to 
any updating of the return-generating process, i.e. disaster risk remains unpriced. 
As disaster is considered as a one-time event, its occurrence leaves the stochastic 
discount factors of all stocks unchanged.

\medskip

\noindent {\it 	Priced disaster risk.} The second model is one where disasters, 
however rare, are rationally anticipated, so that more resilient firms are priced 
at a premium relative to less resilient ones, i.e. offer a lower expected return 
in no-disaster periods, as predicted by \cite{Barro_2006:QJE}, \cite{Barro_2009:AER} 
and \cite{Gabaix_2012:QJE}. Such models predict that, in equilibrium, securities 
more exposed to a disaster, i.e. those issued by less resilient firms, pay a risk 
premium to compensate investors for this risk. Hence, they predict that (i) prior to 
the disaster, stocks' excess returns are related to firms' disaster resilience; 
(ii) during the disaster, investors take into account that the cash flows of less 
resilient firms drop by more than those of the more resilient ones, so that 
their disaster-time stock performance is worse; (iii) after the disaster, the 
pre-disaster excess return pattern re-emerges, i.e. disaster risk remains priced. 
In principle, this hypothesis does not preclude that investors may update the 
probability of disasters upon the occurrence of one: if they have increased 
this probability as a result of COVID-19, the expected return differential 
between non-resilient and resilient firms should become more pronounced after the 
pandemics than it was before it.  

\medskip

\noindent {\it	Prior learning about disaster risk.} Finally, we consider a model 
where investors learn about the probability of a disaster occurring or about 
its implications before its occurrence. In the case of COVID-19, as mentioned in 
the introduction, they may have revised upwards the probability of a pandemic 
occurring or become more keenly aware of its social distancing implications, for 
instance as a result of SARS, Ebola, and/or the statements by Bill Gates, Barack 
Obama and other opinion leaders. By the same token, investors may have become more 
aware of the characteristics that make firms more resilient to pandemics. 
Any of these forms of learning implies a demand shift by investors from less 
to more resilient stocks before the pandemic, leading the latter to outperform the 
former, once their respective exposures to ``traditional'' risk factors are 
controlled for. Once disaster strikes, one would again observe the stocks of 
more resilient firms outperforming less resilient ones. But this pattern should 
reverse in the post-disaster phase: as at that stage learning would be complete, 
the stocks of more resilient firms will be priced at a premium relative to 
less resilient ones, i.e. should offer a lower expected return. This scenario 
has parallels  with learning about the ``importance'' of  Environmental, Social and 
Governance (ESG) scores by investors. In a rational expectations equilibrium 
without learning, portfolios with strict ESG rules should underperform, but in the 
presence of gradual learning about the importance of ESG ratings for investors, the 
stocks of firms with high ESG ratings appreciate, and ESG mutual funds outperform \citep[see][]{pastor/stambaugh/taylor:20ssrn}. 

Hence, this model predicts that (i) excess returns of resilient relative to non-resilient 
firms should be observed prior to the disaster; (ii) when disaster strikes, investors take 
into account that the cash flows of less resilient firms drop by more than those 
of more resilient ones; (iii) after the disaster, resilient firms offer lower expected 
returns that non-resilient ones, as learning about disaster risk has taken place.

\section{Measuring Firm Resilience to Pandemics}
\label{measuringresilience}

This section describes social distancing measures that may be relevant for the pricing of stocks, as firms with operations requiring less direct physical interaction and more easily performed from home should be more resilient to social distancing rules than other firms. To gauge the effect of social distancing on firms, recent research in labor economics has developed measures of the extent to which jobs can be done from home and rely on human interaction in physical proximity. Some studies have developed such measures starting from  worker-level survey responses, while others have done so by characterizing the tasks required by each occupation based on the Occupational Information Network (O*Net) and on the authors' own judgement.

\smallskip

\citet[][HLR]{hensvik/lebarbanchon/rathelot:20cepr} use the first approach: they rely on the `American Time Use Survey'  (2011-2018) to estimate the prevalence of working at `home' and at the `workplace' and, starting from worker-level survey responses, estimate the fraction of employees that work at home and at the workplace, as well as the hours worked at home and at the workplace at the industry-level.\footnote{As the authors mention, their measures for working at home should provide a lower bound for the current situation, as the COVID-19 crisis is likely to have prompted additional substitution of work at the workplace with work at home.} Instead, both \citet[][DN]{dingel/neiman:20nber} and \citet[][KP]{koren/peto:20covid} use data from O*Net surveys. DN use this data, and their own judgement, to assess the teleworkability of occupations and provide industry-level estimates for the percentage of jobs that can be done at home as well as for the percentage of wages associated with teleworkable occupations. KP construct three types of industry-level measures of face-to-face interactions, depending on whether these are due to internal communication (`teamwork'), external communication (`customers'), or physical proximity to others (`presence'). They also aggregate `teamwork' and `customers' to a measure of `communication' intensity and construct an industry-level measure of the percentage of employees `affected'  by social distancing regulations due to their occupations being communication-intensive and/or requiring close physical proximity to others.\footnote{The estimates of these studies are available for NAICS industry classifications, at the 2- and 3-digit level for DN, at the 4-digit level for HLR, and at the 3-digit-level for KP. For details on the data, see Table \ref{tab:empirical.measures} in the Appendix.} 

\smallskip

In our estimates, we rely primarily on the measures proposed by KP, but also check 
whether our results are robust to the use of those produced by HLR and DN. The results 
presented in the next section focus mostly on KP's `affected\_share' variable. 
We choose this as our main variable because, beside teleworkability, it also 
explicitly accounts for physical proximity to others, i.e. exactly what social 
distancing rules aim to avoid.  Additionally, we consider DN's measure of the 
fraction of wages accounted for by jobs that can be performed at home 
(`teleworkable\_manual\_wage') and HLR's measure of daily work hours at the 
workplace (`dur\_workplace'). However, we also discuss results obtained using 
all other variables suggested by KP, DN, and HLR. Table \ref{tabA1} in the 
Appendix lists all the measures and presents their definitions.

\section{Empirical Results}
\label{results}

We use the NAICS industry classification of stocks to assign the DN, HLR, and KP metrics 
to firms, and code each firm as a `High' or `Low' resilience one, depending on how 
its industry scores relative to the median value of the relevant metric. Then, we  
analyze whether and how the resulting variation in U.S. firms' pandemic resilience 
affects the cross-section of their stock returns at the time of the COVID-19 shock 
(Subsection \ref{during}), before the shock (Subsection \ref{before}) and after 
its occurrence (Subsection \ref{after}).

\subsection{Data and Methodology}
\label{data}

We obtain prices for all common stocks listed at the NYSE, AMEX and NASDAQ from the 
Compustat Capital IQ North America Daily database and compute daily returns, accounting 
for price-adjustments and dividends. We also retrieve data on daily risk-free, market 
and standard factor returns from Kenneth French’s website. Our estimates in Subsection 
\ref{during} require daily data from January 2019 to March 2020: after estimating 
firms' exposures to common factors from 2019 data, we use these exposures to compute 
factor model-adjusted stock returns in the first quarter of 2020.\footnote{This 
approach follows \citet{ramelli/wagner:20cepr} and other related papers.} 

\smallskip

For the regressions of daily stock returns on factor returns, we require a minimum of 
127 daily observations, and estimate the following specifications: we alternatively 
regress stock returns on market excess returns (CAPM), on the returns of the three 
\citet{fama:1993:jfe} factors (FF3, i.e. market, size, and value) and the five 
\citet{fama/french:2015jfe} factors (FF5, i.e. market, size, value, investment, 
and profitability). In addition, we augment the FF3 and FF5 specifications by the 
momentum factor \citep[][]{carhart:97jf} so as to obtain FF4 and FF6 exposures. The 
2019 exposures are then used to measure factor model-adjusted stock returns in the 
first quarter of 2020 as the difference between a stock's daily excess return and 
its CAPM beta multiplied by the daily market excess return; we proceed analogously 
for the FF specifications. 

\smallskip

Next, stock return data are matched with the resilience proxies based on KP, DN, and HLR 
metrics by industry, based on firms' 2-, 3-, or 4-digit NAICS codes. Only industries 
for which resilience measures are available are retained in the data set. Moreover, 
firms with equity market capitalization below USD 10 million are dropped from the 
sample. For the first quarter of 2020, this results in a sample with a total of 227,812 
observations for 3,614 firms in 75 industries at the NAICS 3-digit level for KP and DN, 
and 222 industries at the NAICS 4-digit level for HLR.

\smallskip

For the analysis in Subsection \ref{after}, we additionally obtain S\&P 500 index 
option and individual stock options data from OptionMetrics for the first quarter of 
2020. We use the volatility surface data to compute $\SVS$-measures of the risk-neutral 
variance, keeping data for firms for which we can compute $\SVS$-measures for horizons 
of 30, 91, 182, 365, and 730 days, and match this with the stock data. This results in 
a sample of 160,951 observations for 2,721 firms in 74 industries at the NAICS 3-digit 
level for KP and DN and 212 industries at the NAICS 4-digit level for HLR.

\subsection{Returns and Firm Resilience during the Disaster}
\label{during}

\begin{figure}\caption{Attention to Covid19 and US equity returns \label{fig1}}\footnotesize
\begin{adjustwidth}{-1.75cm}{-1.75cm}
\begin{spacing}{1.0}
Panel A illustrates the attention to Covid19 in the United States, as measured by the Google trends index for the term ``Coronavirus'' in the US, and the cumulative returns of the US stock market, as measured by the Fama-French market factor, during the first quarter of 2020. Panel B plots the cumulative returns of portfolios sorted by firms' resilience to disaster risk. On any given day, we assign a firm to the `High' portfolio if its `affected\_share' \citep[as defined by][]{koren/peto:20covid} is below the median value and to the `Low' portfolio if it is above. We plot the cumulative value-weighted portfolio returns for the `High' portfolio (in green) and the Low portfolio (in red) as well as the High-Low differential return (in blue). The dashed vertical lines mark February 24, the day after Italy introduced its lockdown, and March 20, the last trading day before the Fed announced its intervention.
\end{spacing}
\end{adjustwidth}

\bigskip\bigskip\bigskip
\begin{adjustwidth}{-2cm}{-2cm}\centering
\vspace{-6mm}
	{Panel A. Google trends index and aggregate US stock market}\\ 
	{\includegraphics[scale = 0.85]{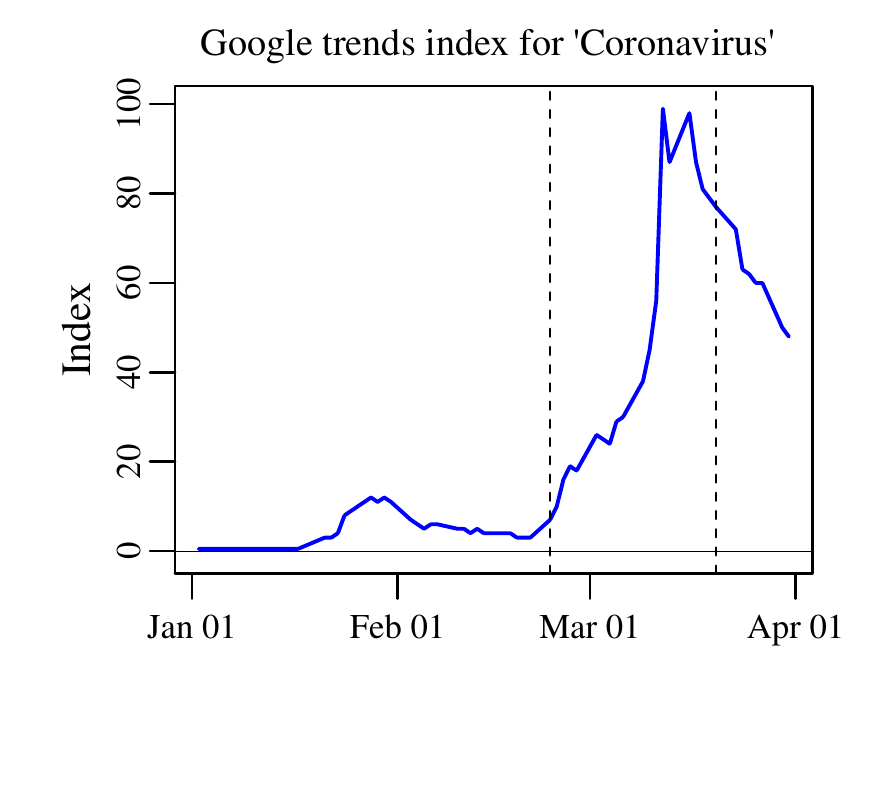}}
	{\includegraphics[scale = 0.85]{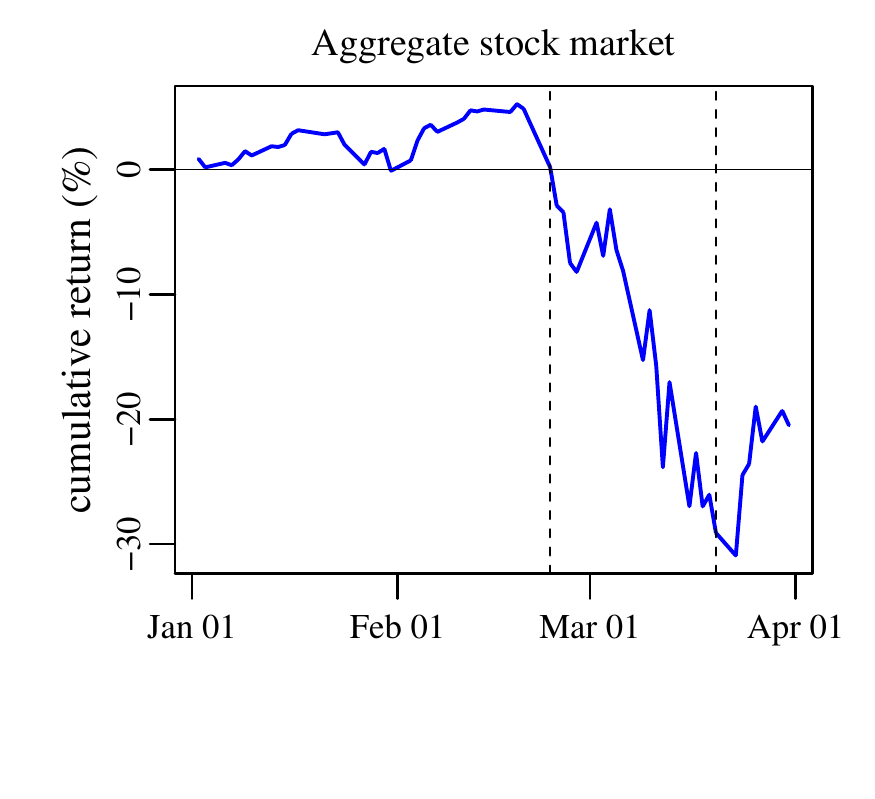}}

\vspace{-3mm}
	{Panel B. Excess returns of firms with high and low resilience to social distancing}  \\ 
	{\includegraphics[scale = 0.85]{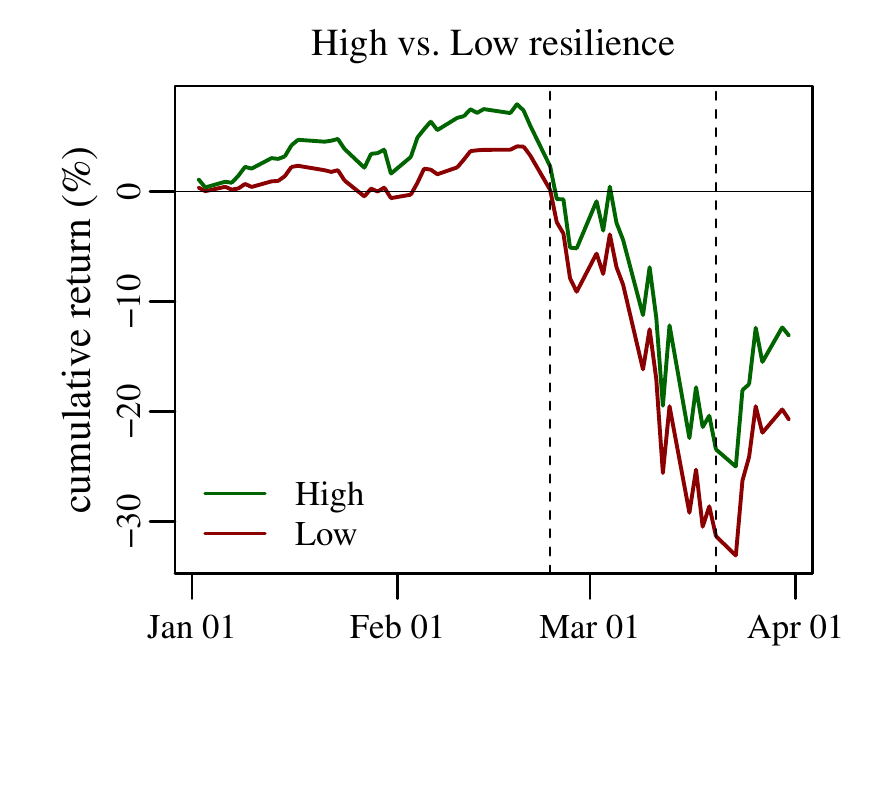}}
	{\includegraphics[scale = 0.85]{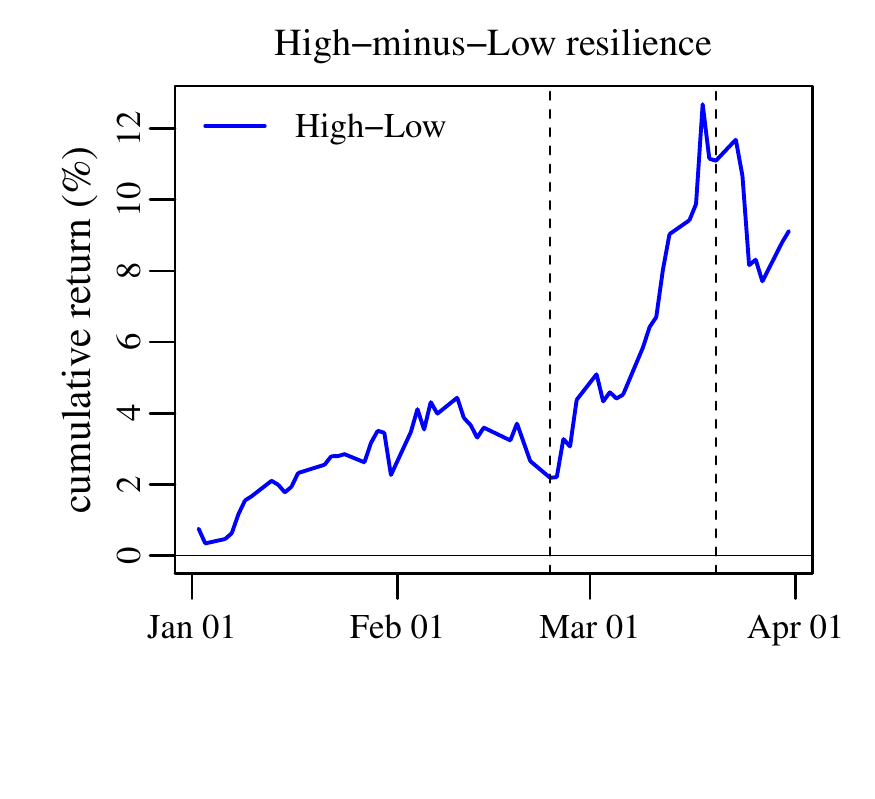}}
\end{adjustwidth}
\end{figure}

We study stock returns in the first quarter of 2020, and specifically from February 
24, the day after Italy introduced its lockdown, to March 20, the last trading 
day before the Fed announced aggressive action intended to soften the blow of the 
pandemic.\footnote{This period corresponds to the `fever'-period in 
\citet{ramelli/wagner:20cepr}; see their paper for a detailed account of the sequence 
of events. On Monday March 23, the Fed unveiled its plan to buy an unlimited amount 
of bonds with government guarantees, including some commercial mortgage debt. It 
also established the Secondary Market Corporate Credit Facility (SMCCF), in order 
to purchase  existing investment-grade corporate debt, including exchange-traded funds, 
as well as the Primary Market Corporate Credit Facility (PMCCF), to purchase newly 
issued corporate bonds, so as to prevent companies facing pandemic fallout from 
dismissing employees and terminating business relationships.}  As shown by Panel A 
of Figure \ref{fig1}, in this time window there was a surge in the public's attention 
to the COVID-19 epidemic (as measured by Google trends), while the prices of U.S. 
stocks (as measured by the Fama-French market factor) fell sharply.

\smallskip

A first look at the data suggests that the stocks of more pandemic-resilient firms, 
i.e. those included in the `High' resilience portfolio (based on the `affected\_share' 
metric by KP) performed better in this time window than those in the `Low' resilience 
portfolio. Panel B in Figure \ref{fig1}, which plots cumulative excess returns for 
the value-weighted portfolios of firms with high and low resilience, shows that 
both portfolios dropped sharply in price, but that of high-resilience firms 
depreciated far less: from February 24 to March 20, their shares outperformed those 
of the other group by approximately 10\%. 

\begin{center}
\textbf{[Insert Figure \ref{fig1}]}
\end{center}

\begin{figure}\caption{Risk-adjusted returns of stocks with high and low resilience to social distancing \label{fig2}}\footnotesize

\begin{adjustwidth}{-1.75cm}{-1.75cm}
\begin{spacing}{1.0}
This figure plots the cumulative risk-adjusted returns of portfolios sorted by firms' resilience to disaster risk during the first quarter of 2020. On any given day, we assign a firm to the `High' portfolio if its `affected\_share' \citep[as defined by][]{koren/peto:20covid} is below the median value and to the `Low' portfolio if it is above. In Panel A, we present CAPM-adjusted returns, i.e. controlling for exposure to market risk. Panels B and C present results controlling for the Fama-French three factor model exposures (i.e. market, size, value) and five factor model exposures (i.e. market, size, value, investments, profitability), respectively. We plot the cumulative value-weighted portfolio returns for the `High' portfolio (in green) and the Low portfolio (in red) as well as the High-Low differential return (in blue). The dashed vertical lines mark February 24, the day after Italy introduced its lockdown, and March 20, the last trading day before the Fed announced its intervention.
\end{spacing}
\end{adjustwidth}

\begin{adjustwidth}{-2cm}{-2cm}\centering
	
\vspace{-2mm}
	{Panel A. CAPM-adjusted returns}\\ 
	{\includegraphics[scale = 0.85]{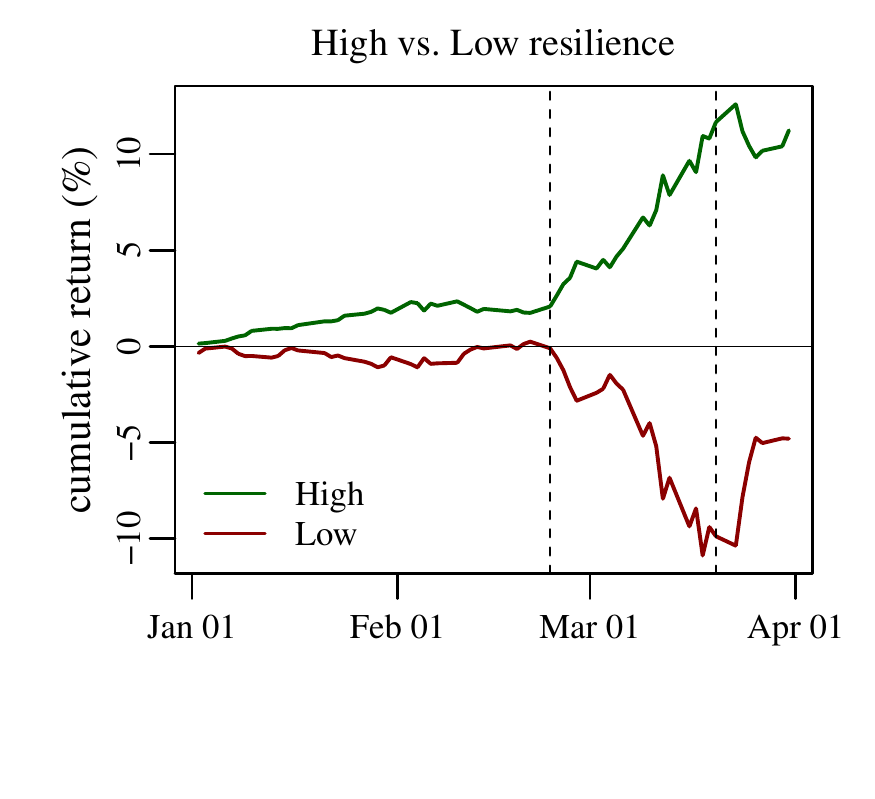}}
	{\includegraphics[scale = 0.85]{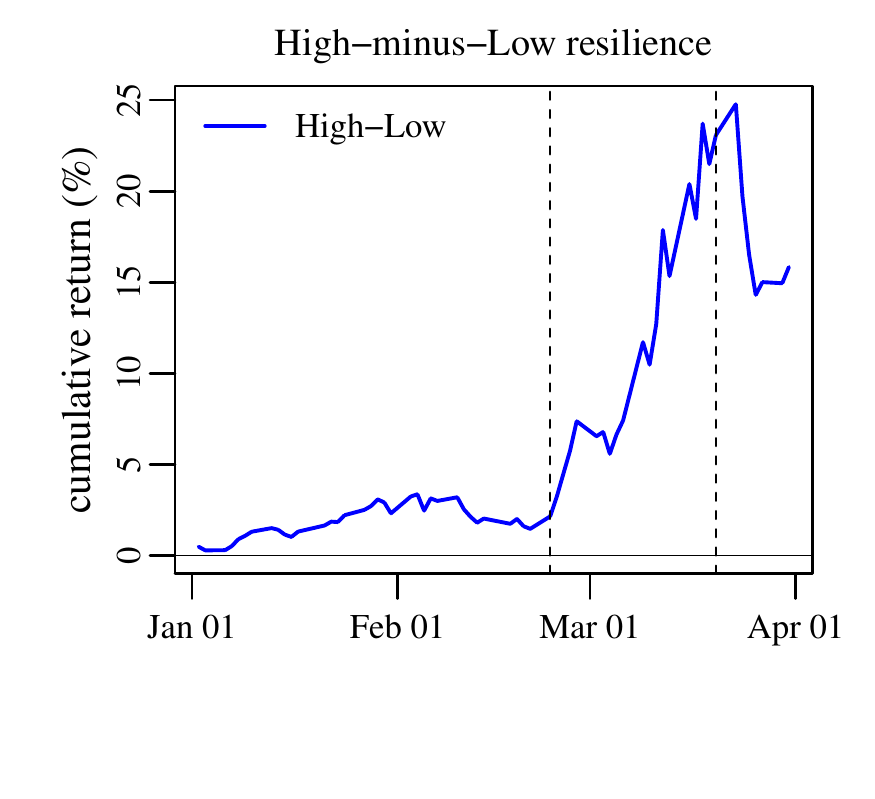}}
	
\vspace{-6mm}
	{Panel B. FF3-adjusted returns}\\ 
	{\includegraphics[scale = 0.85]{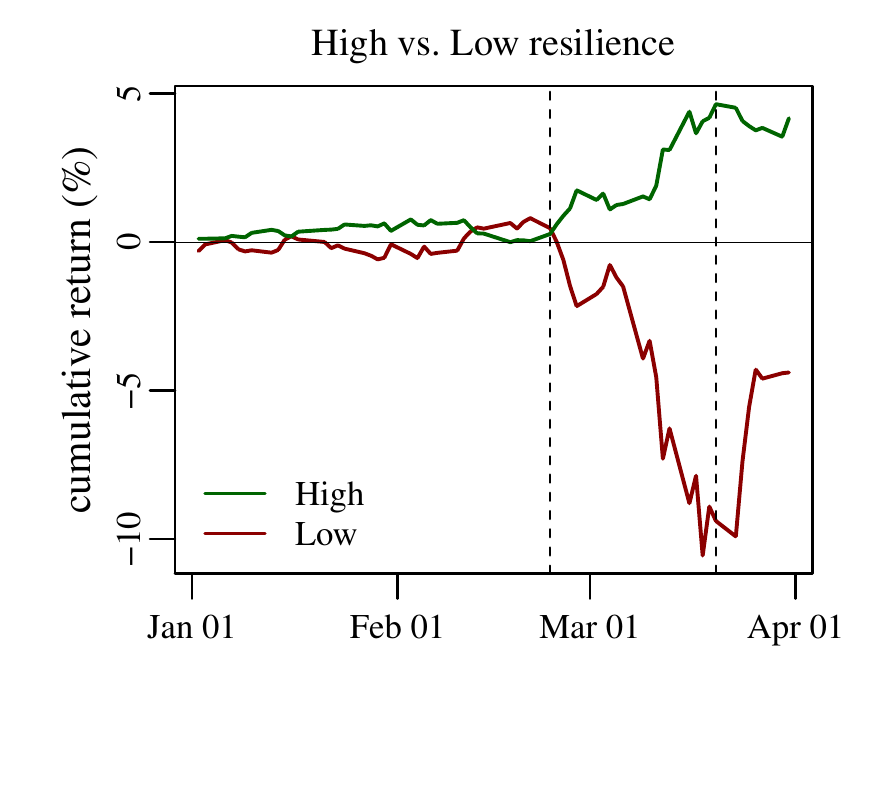}}
	{\includegraphics[scale = 0.85]{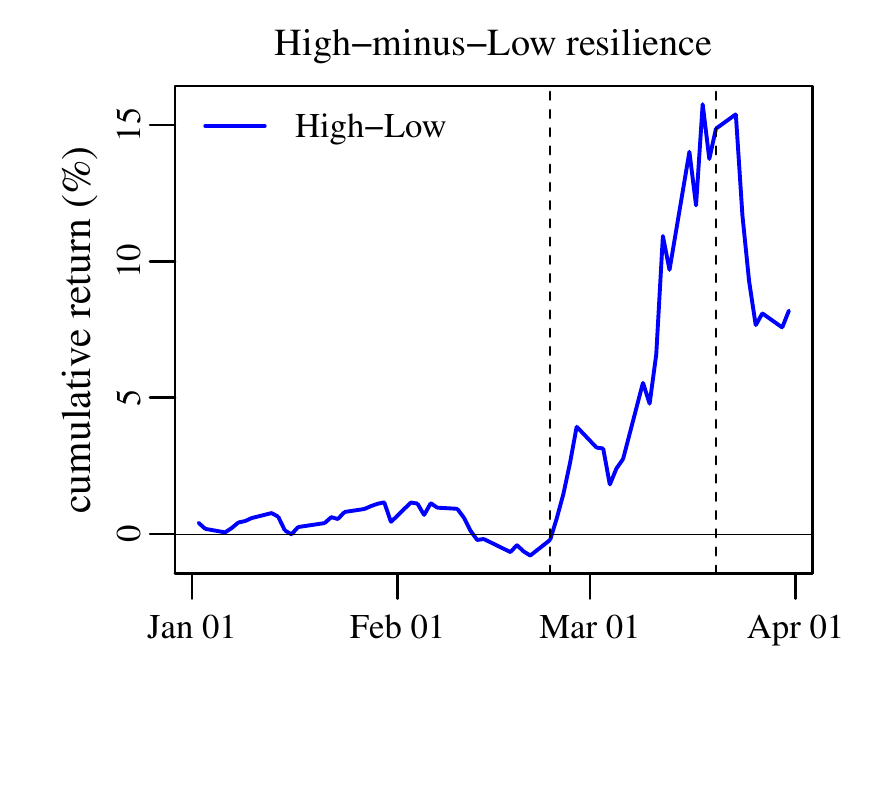}}
	
\vspace{-6mm}
	{Panel C. FF5-adjusted returns}\\ 
	{\includegraphics[scale = 0.85]{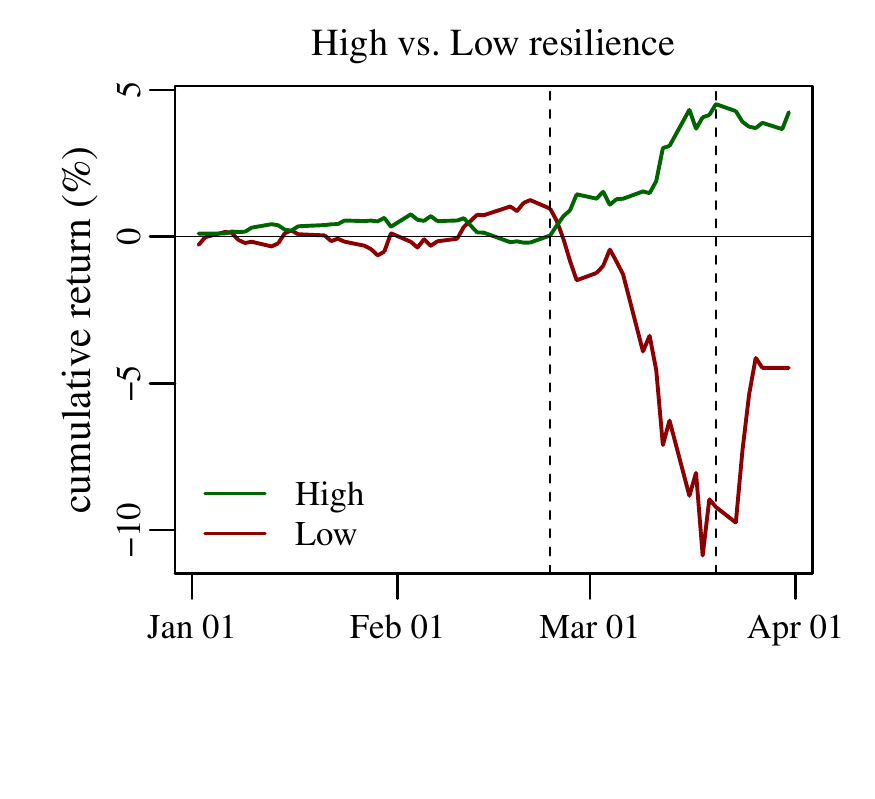}}
	{\includegraphics[scale = 0.85]{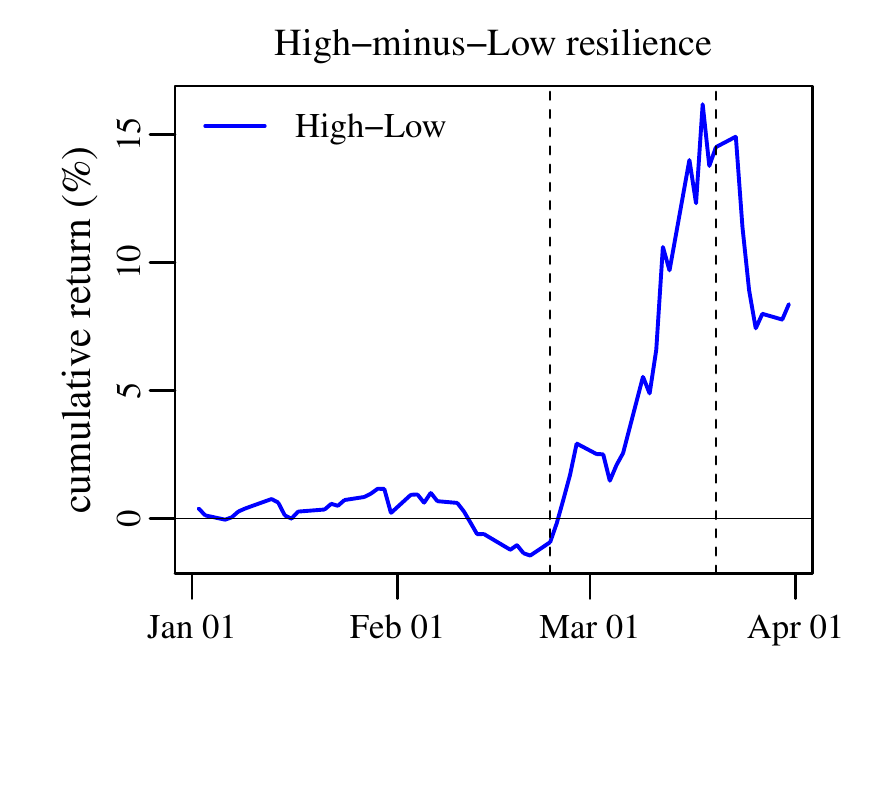}}
\end{adjustwidth}
\end{figure}

Since the different performances of the two portfolios shown in Figure \ref{fig1} may 
stem from their different exposure to standard risk factors, we estimate CAPM- 
and FF-exposures from daily data in 2019. We then use these exposures to compute 
daily CAPM-adjusted and FF-adjusted returns, i.e., excess returns net of such 
exposures multiplied by the respective factor returns (see Subsection \ref{data} 
for details). Figure \ref{fig2} presents the results: controlling for market factor 
risk (CAPM-adjusted returns in Panel A), the cumulative return of the High- and 
Low-resilience portfolio are about +10\% and  $-$15\%, respectively, from February 24 
to March 20, i.e. the cumulative CAPM-adjusted High-minus-Low return is approximately 
25\%. Panels B and C show that accounting for the FF-factors does not change the results  
qualitatively: the resulting risk-adjusted return differentials are in the range between 
15\% and 20\%. The plots suggest that the differential return between the two portfolios 
has been negligible from early January until late February, and that the results in 
the COVID-19 time window are mostly driven by the sharp decline of the Low-resilience 
portfolio. Once the Fed announced its intervention (on March 23) the High-resilience 
portfolio recovered slightly and, as a result, the Low-minus-High differential dropped. 
Interestingly, the time-series of the cumulative Low-minus-High returns resembles that 
of the Google trends index. Indeed, regressing daily returns on changes in the Google 
index confirms a statistically significant relation, with $R^2$ between 0.19 and 0.22, 
depending on which factor-model adjustment is used.

\begin{center}
\textbf{[Insert Figure \ref{fig2}]}
\end{center}

Figures \ref{figA1} and \ref{figA2} in the Appendix provide evidence on the robustness 
of the results shown in Figure \ref{fig2} with respect to other resilience measures. 
They plot the risk-adjusted returns and the differentials of the High- and 
Low-resilience portfolios based on  DN's `teleworkable\_manual\_wage' and HLR's 
`dur\_workplace'. The results are qualitatively similar to those in Figure \ref{fig2}. 

\smallskip

To analyze the statistical significance of these findings, Table \ref{tab2} reports 
the averages of daily excess returns (in Column 1) and risk-adjusted returns (in Columns 
2 to 6) for High-resilience and Low-resilience stocks from February 24 to March 20, 
as well as the difference between the two, using the resilience measures based on KP, 
DN and HLR.  Panel A shows that the differential return based on SKP's `affected\_share' 
metric is statistically significant, whether it is based on raw excess returns or 
adjusted for market exposure (CAPM), for the classic Fama-French factors (FF3 and FF5) 
or for those that also control for momentum (FF4 and FF6). For resilience measures 
based on DN (Panel B) and HLR (Panel C), the average CAPM-adjusted differential returns 
are statistically significant as well, but there is some variation in the significance 
of the FF-adjusted returns. A common feature across all resilience measures is that 
Low-resilience stocks generate (at least marginally) significant negative excess returns 
in all return specifications. For KP, we additionally find that all risk-adjusted 
returns of the High-resilience portfolio are significantly positive in all specifications. 

\begin{center}
\textbf{[Insert Table \ref{tab2}]}
\end{center}

Tables \ref{tabA2} to \ref{tabA4} in the Appendix present results for the other metrics proposed by KP, DN, and HLR, respectively. For KP (Table \ref{tabA2}), most return differentials are significantly different from zero when controlling for CAPM- and FF-exposures. For `presence\_share', the measure that aims to capture the necessity of working in close proximity to others, the results are even stronger than those based on `affected\_share'. For DN and HLR, relying on the other proxies generally reduces the significance of the results, but CAPM-adjusted return differentials remain statistically significant (see Tables \ref{tabA3} and \ref{tabA4}, respectively).

\smallskip

To better understand the source of the High-minus-Low differential returns, we study the cumulative risk-adjusted returns in the cross-section of (value-weighted) industry portfolios and present results for the 25 industries with the highest number of firms, in total 2,974 firms. Table \ref{tab3} presents summary statistics.

\begin{center}
\textbf{[Insert Table \ref{tab3}]}\\
\end{center}

\begin{figure}\caption{Resilience to social distancing and industry portfolio returns  \label{fig3}}\footnotesize
\begin{adjustwidth}{-1.75cm}{-1.75cm}
\begin{spacing}{1.0}
This figure plots the cumulative risk-adjusted returns of value-weighted industry portfolios against the industries' resilience to disaster risk. The sample period is from February 24 to March 20, 2020, i.e. from the day after Italy introduced its lockdown to the last trading day before the Fed announced its intervention. We define resilience as 100 (\%) minus the `affected\_share' defined by \citet{koren/peto:20covid} and present results for the 25 industries with the highest number of firms (in total 2,974). In Panel A, we present CAPM-adjusted returns, i.e. controlling for exposure to market risk. Panels B and C present results controlling for the Fama-French three factor model exposures (i.e. market, size, value) and five factor model exposures (i.e. market, size, value, investments, profitability), respectively. The plot labels indicate the industries' 3-digit NAICS codes. The plot legends report results for cross-sectional regressions with $t$-statistics based on \citet{white:80ecta} standard errors in square brackets.
\end{spacing}
\end{adjustwidth}
\bigskip
\begin{adjustwidth}{-2cm}{-2cm}\centering

	{Panel A. CAPM-ajdusted returns}\\ \vspace{-5mm}
	{\includegraphics[scale = 0.75]{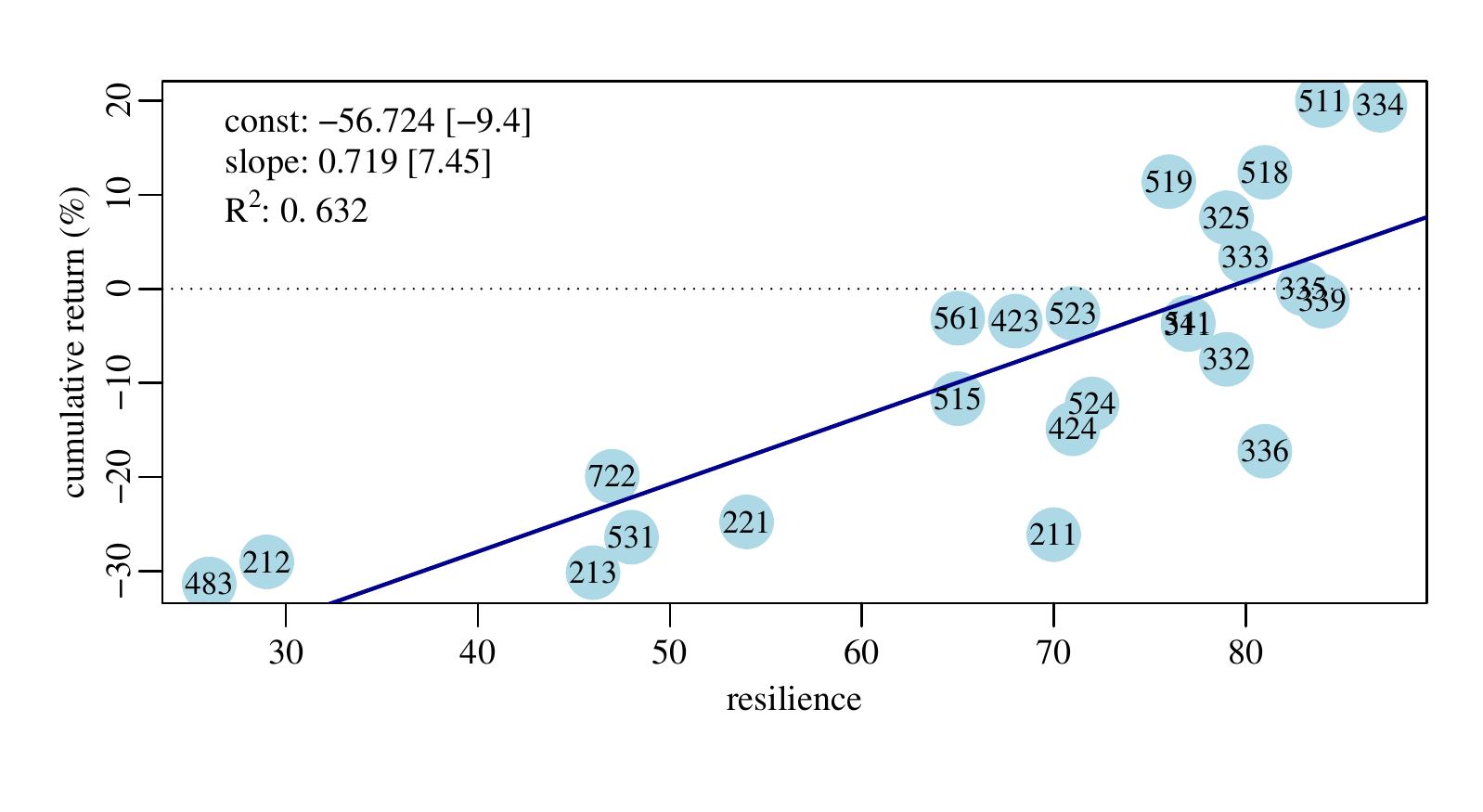}}

	{Panel B. FF3-ajdusted returns}\\ \vspace{-5mm}
	{\includegraphics[scale = 0.75]{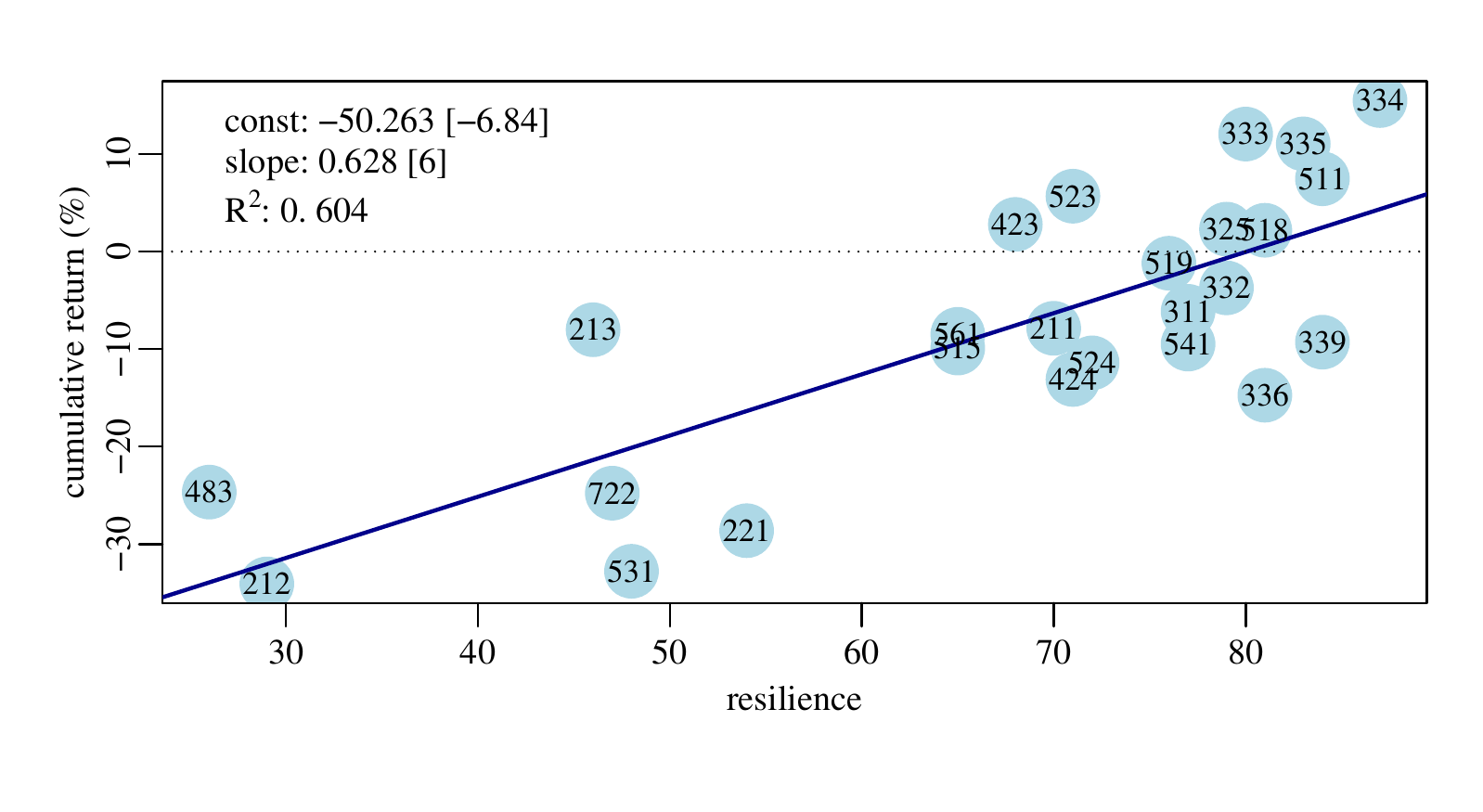}}

	{Panel C. FF5-ajdusted returns}\\ \vspace{-5mm}
	{\includegraphics[scale = 0.75]{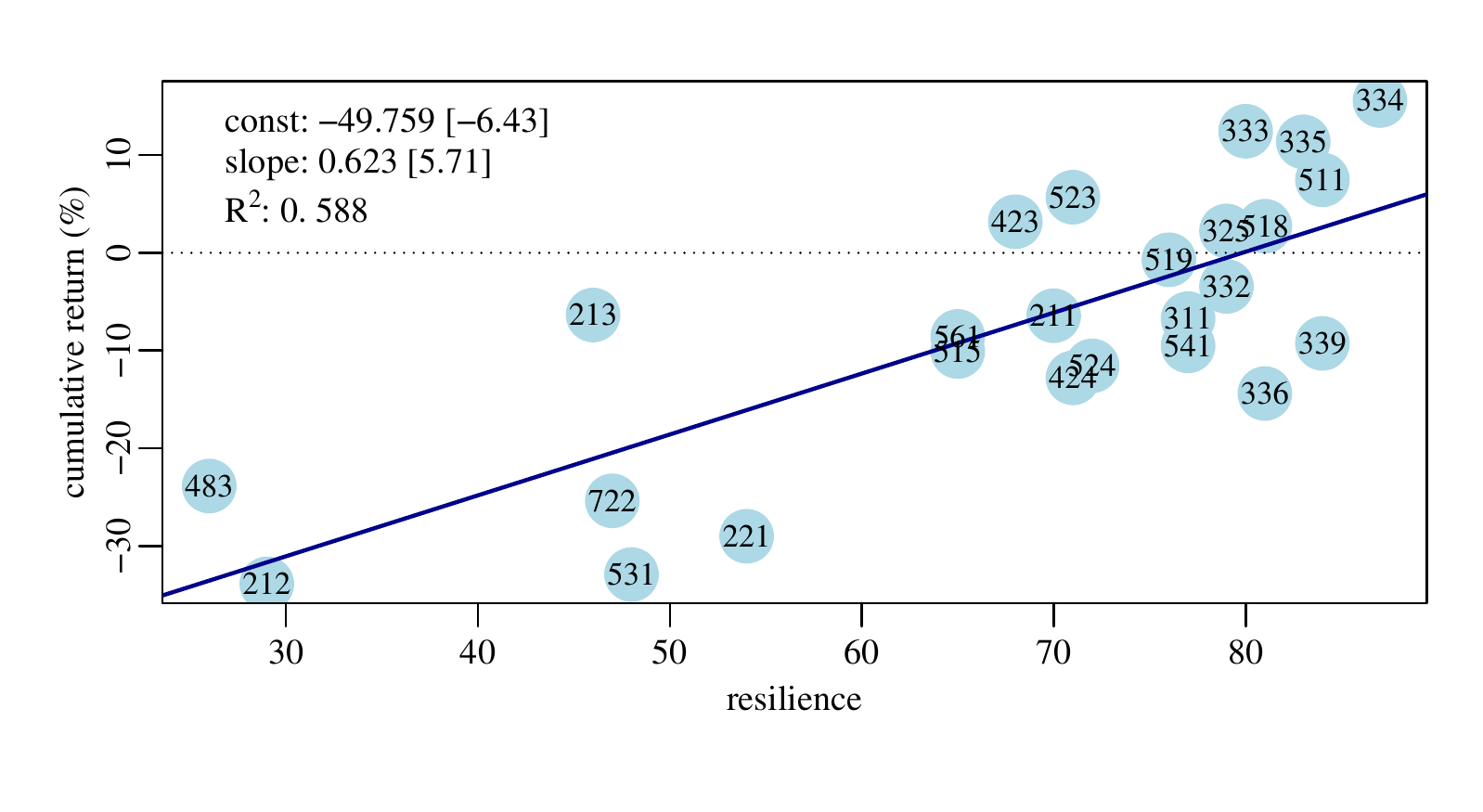}}

\end{adjustwidth}
\end{figure}

Figure \ref{fig3} plots the cumulative risk-adjusted returns of value-weighted industry 
portfolios, ranked by their resilience to pandemics, based on KP's  `affected\_share' 
variable. The figure shows that less resilient industries feature substantially lower 
cumulative risk-adjusted returns during the COVID-19 crisis: the stocks of the least 
resilient industries (such as those in NAICS-industry 212: Mining, except oil and gas; 
483: Water transportation) generated returns 40\% to 50\% lower than the most resilient 
ones (224: Computer and electronic products; 511: Publishing industries, except Internet), 
depending on the risk adjustment. The cross-sectional relationship between pandemic 
resilience and cumulative returns is not only highly statistically significant in all 
three panels of the figure, but also economically significant: for instance, a decrease 
of 10 percentage points in the resilience metric is associated with a drop of 7.2\% in 
the CAPM-adjusted cumulative return. In the Appendix, Figure \ref{tabA3} shows 
qualitatively similar results when resilience is measured on the basis of DN's 
`teleworkable\_manual\_wage' variable.

\begin{center}
\textbf{[Insert Figure \ref{fig3}]}\\
\end{center}

\subsection{Returns and Firm Resilience before the Disaster}
\label{before}

The evidence in the previous section indicates that \textit{when} the public became 
aware of the COVID-19 outbreak, stock returns reacted differently depending on 
companies' resilience to the social distancing rules triggered by the pandemic. However, 
in principle investors may have been aware of pandemic risk even before the COVID-19 
shock, so that this risk may have to some extent been priced by the stock market in 
advance. As explained in Section  \ref{theory}, if investors were fully aware of such 
risk in advance, they should have required lower expected returns on the stocks of 
pandemic-resilient companies than on those of non-resilient ones, controlling for their 
respective exposures to other risks. If instead investors had become gradually aware 
of such risk, one should observe the opposite pattern, namely, the stocks of 
pandemic-resilient companies outperforming non-resilient ones. Finally, if investors 
were completely unaware of such risk, one should not detect any difference in the 
stock market performance of the two types of companies, prior to the pandemic.

\smallskip

\begin{figure}\caption{Risk-adjusted returns of high and low resilience stocks prior to the Covid19-crisis}\footnotesize \label{fig4}

\begin{adjustwidth}{-1.75cm}{-1.75cm}

\begin{spacing}{1.0}
This figure plots the cumulative risk-adjusted returns of portfolios sorted by firms' resilience to disaster risk during from January 2014 through March 2020. On any given day, we assign a firm to the `High' portfolio if its `affected\_share' \citep[as defined by][]{koren/peto:20covid} is below the median value and to the `Low' portfolio if it is above. In Panel A, we present CAPM-adjusted returns, i.e. controlling for exposure to market risk. Panels B and C present results controlling for the Fama-French three factor model exposures (i.e. market, size, value) and five factor model exposures (i.e. market, size, value, investments, profitability), respectively. We plot the cumulative value-weighted portfolio returns for the `High' portfolio (in green) and the Low portfolio (in red) as well as the High-Low differential return (in blue). 
\end{spacing}

\end{adjustwidth}

\bigskip
\begin{adjustwidth}{-2cm}{-2cm}\centering

\vspace{-2mm}

	{Panel A. CAPM-adjusted returns}\\ 
	{\includegraphics[scale = 0.85]{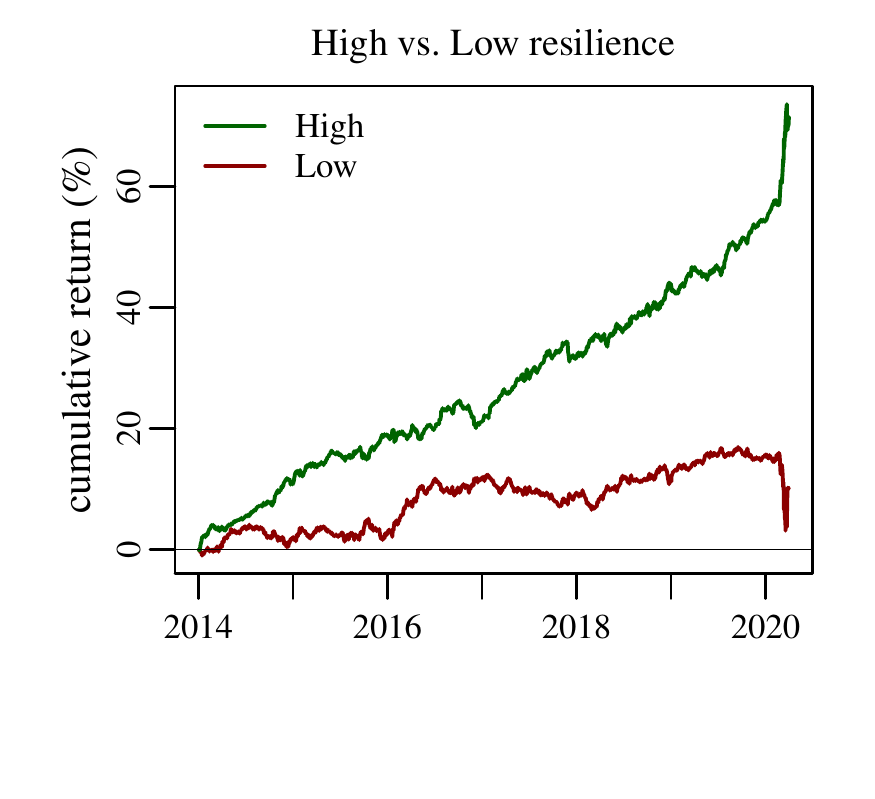}}
	{\includegraphics[scale = 0.85]{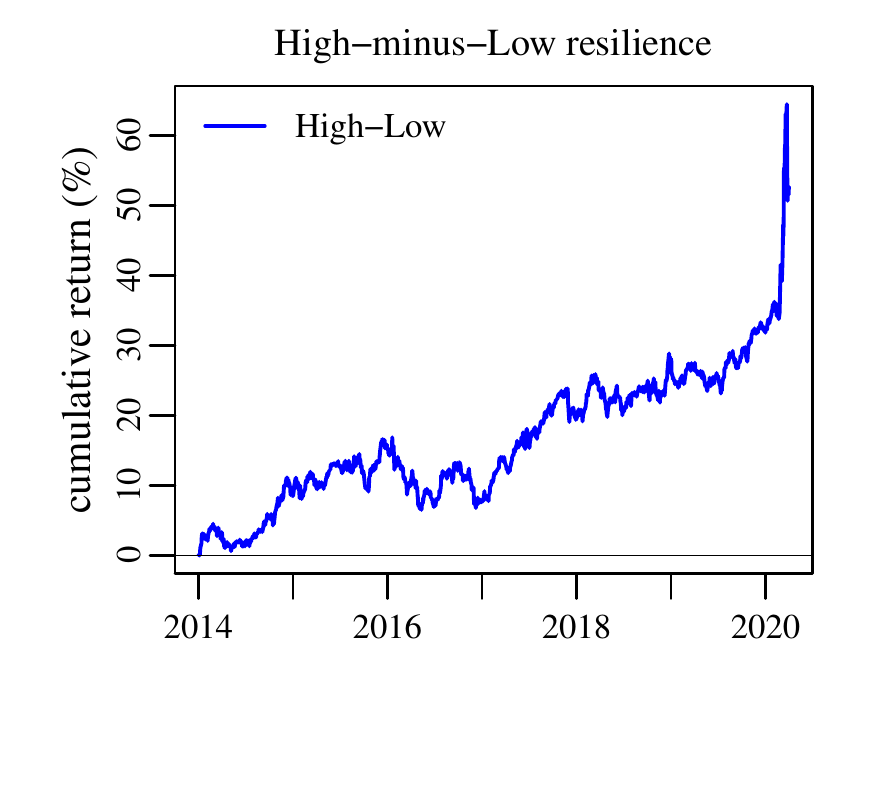}}

\vspace{-6mm}

	{Panel B. FF3-adjusted returns}\\ 
	{\includegraphics[scale = 0.85]{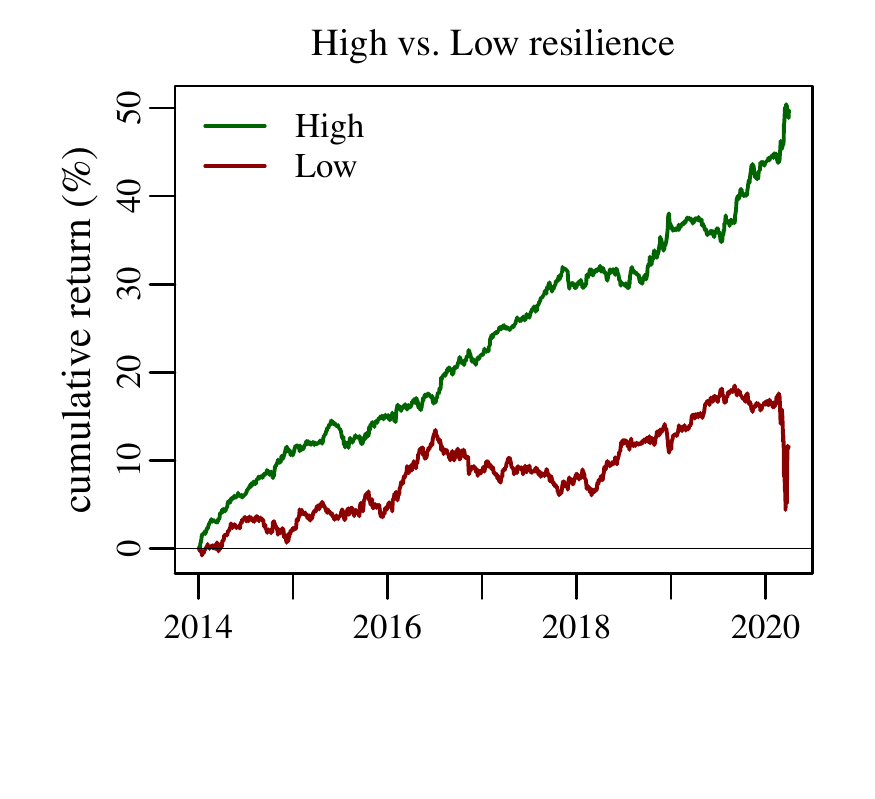}}
	{\includegraphics[scale = 0.85]{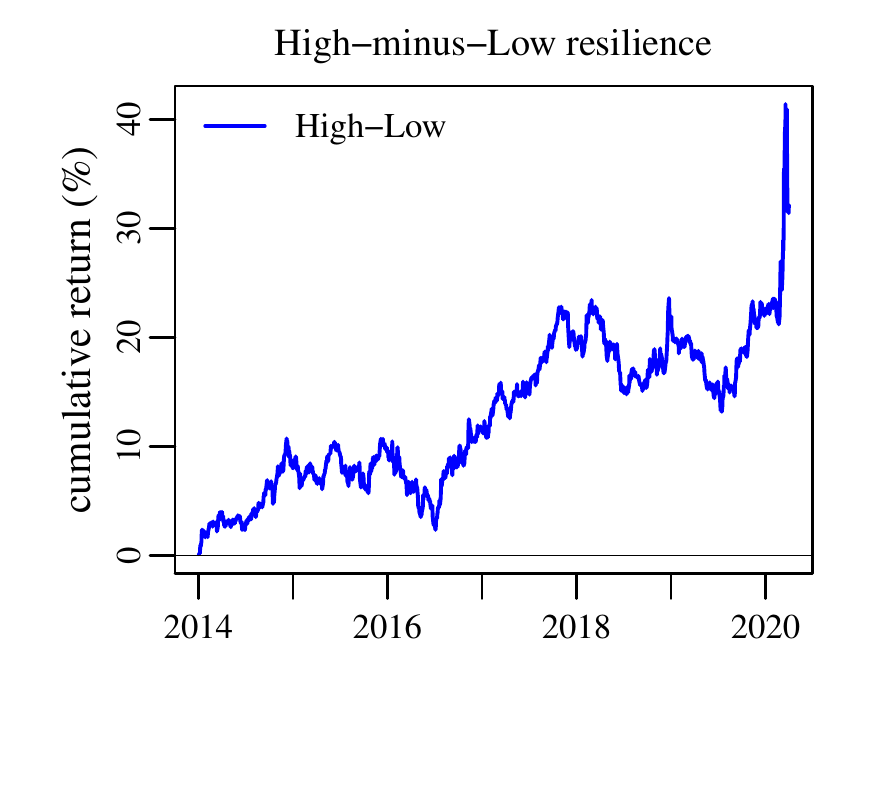}}

\vspace{-6mm}
	{Panel C. FF5-adjusted returns}\\ 
	{\includegraphics[scale = 0.85]{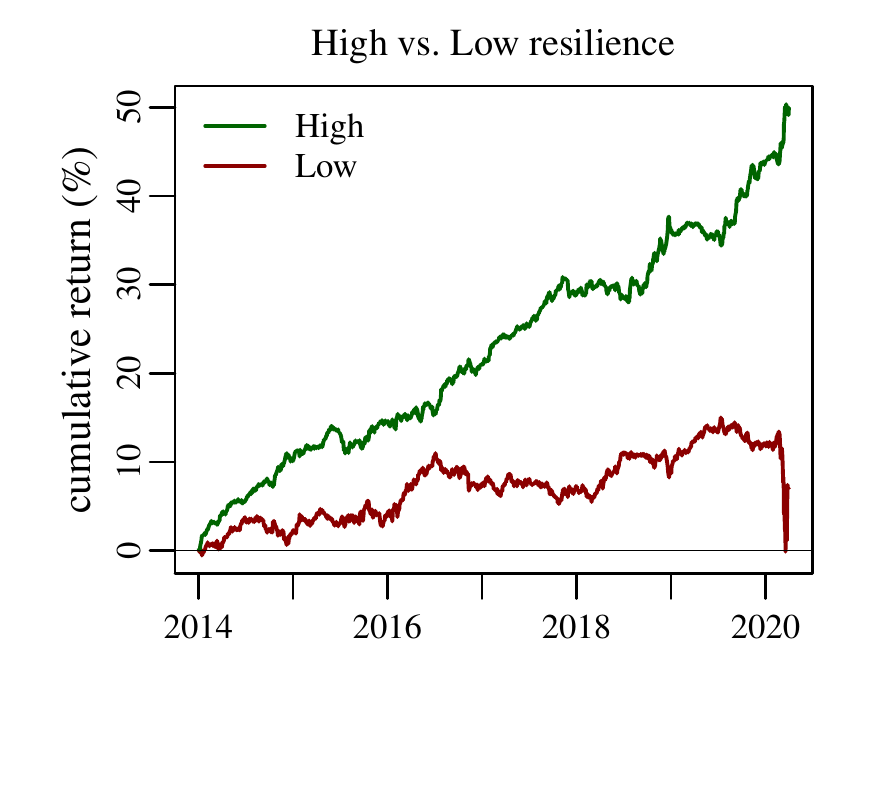}}
	{\includegraphics[scale = 0.85]{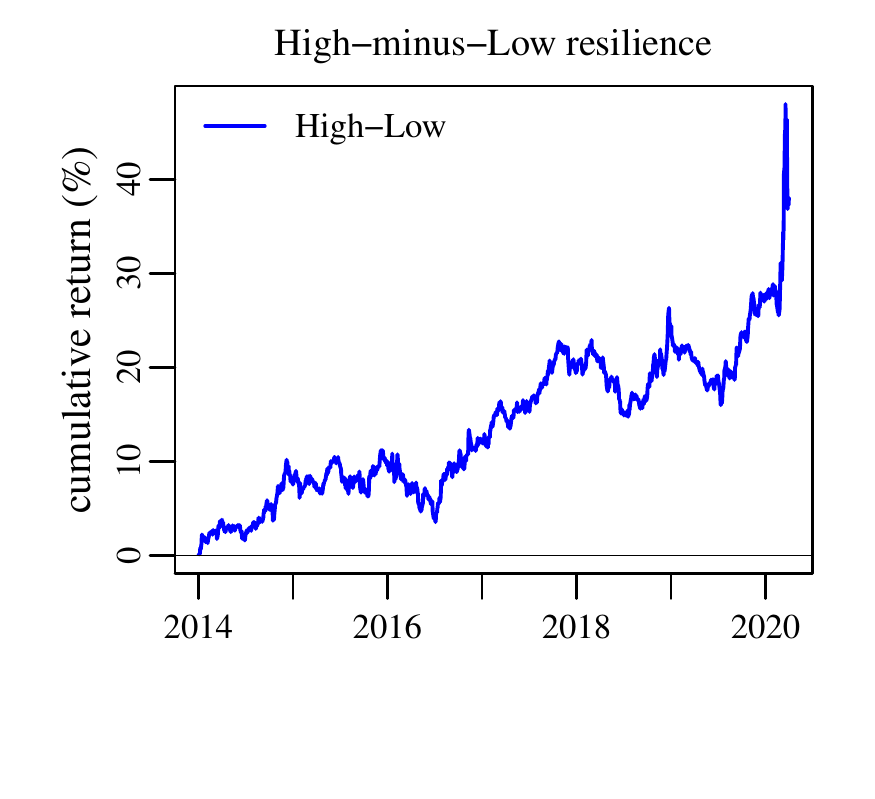}}

\end{adjustwidth}
\end{figure}

Figure \ref{fig4} provides evidence on this point, by displaying the time series 
pattern of risk-adjusted cumulative returns for High- and Low-resilience stocks for 
six years before the COVID-19 crisis, as well as for a High-minus-Low-resilience 
portfolio.\footnote{The empirical approach is the same as before, i.e. we estimate 
exposure from daily excess returns over a calendar year and use these exposures to 
compute risk-adjusted returns in the next year.} Irrespective of the risk 
adjustment considered (CAPM, FF3 or FF5), the figure shows that High-resilience 
stocks vastly outperformed Low-resilience ones, with most of the differential return 
stemming from the outperformance of the former rather than the underperformance of 
the latter. Moreover, about half of the cumulative risk-adjusted return 
differential between the two portfolios over the whole interval from 2014 
to early 2020 materialized \textit{before} the COVID-19 crisis, the spike 
occurring \textit{during} the crisis accounting for the other half. 

\begin{center}
\textbf{[Insert Figure \ref{fig4}]}
\end{center}

Hence, this evidence appears consistent with the third hypothesis outlined in 
Section \ref{theory}, namely, that investors have become gradually aware of disaster 
risk before the current pandemics, and therefore have sought to acquire the stocks 
of pandemic-resilient stocks at increasingly high prices, to insulate their 
portfolios against this previously unknown type of risk.    

\subsection{Option-Implied Expected Returns after the Disaster}
\label{after}

To study how the market prices resilience to disaster risk going forward, we rely on 
equity options data. Options prices are observable in real time and inherently 
forward-looking. These features are especially useful in the current crisis, 
in which -- due to its  unprecedented nature -- relying on historical data appears 
particularly questionable.
 
\smallskip

Recent research shows how prices of index options and stock options can be used to 
compute measures of expected market returns and expected stock returns. In our analysis, 
we follow the approaches suggested by \citet{martin:17qje} and \citet{martin/wagner:19jf}.
\citet{martin:17qje} argues that the risk-neutral variance of the market provides a 
lower bound on the equity premium. He also argues that, empirically, the lower bound 
is approximately tight, so that the risk-neutral variance of the market directly 
measures the equity premium. \citet{martin/wagner:19jf} derive a formula for the 
expected return on a stock in terms of the risk-neutral variance of the market and 
the stock's excess risk-neutral variance relative to that of the average stock.  

\smallskip

The three measures of risk-neutral variance -- for the market, a particular stock and 
the average stock -- can be computed from option prices using the approach of 
\citet{breeden/litzenberger:78jb}. The market risk-neutral variance, $\SVS_{t}^{2}$, 
is determined by the prices of index options:
\[
\SVS_{t}^{2} = \frac{2}{R_{f,t+1}S_{m,t}^{2}} \left[ \int_{0}^{F_{m,t}} \putoption_{m,t} (K) \, dK + \int_{F_{m,t}}^{\infty} \calloption_{m,t}(K) \,dK \right],
\]
where $R_{f,t+1}$ is the gross riskfree rate, $S_{m,t}$ and $F_{m,t}$ denote the spot 
and forward (to time $t+1$) prices of the market, and $\putoption_{m,t}(K)$ and 
$\calloption_{m,t}(K)$ denote the time $t$ prices of European puts and calls on the 
market that expire at time~$t+1$ with strike $K$. The length of the time interval from 
$t$ to  $t+1$ corresponds to the maturity of the options used in the computation. 

\smallskip

The risk-neutral variance at the individual stock level, $\SVS_{i,t}^{2}$, is defined 
in terms of individual stock option prices:
\[
\SVS_{i,t}^{2} = \frac{2}{R_{f,t+1}S_{i,t}^{2}} \left[ \int_{0}^{F_{i,t}} \putoption_{i,t} (K) \, dK + \int_{F_{i,t}}^{\infty} \calloption_{i,t}(K) \,dK \right], 
\]
where the subscripts $i$ indicate the underlying stock $i$.

\smallskip

Finally, using $\SVS_{i,t}^{2}$ for all firms available at time $t$, we calculate 
the risk-neutral average stock variance index as $\overline{\SVS}_{t}^{2}  =  \sum_{i} w_{i,t} \SVS_{i,t}^{2}$.

\smallskip

Using these three risk-neutral variances, one can compute the expected return on 
a stock using the formula derived by \citet{martin/wagner:19jf}: 
\begin{equation*}
	\frac{ \E_{t} R_{i,t+1} - R_{f,t+1}}{R_{f,t+1}} = \SVS_{t}^{2} + \frac{1}{2} \left( \SVS_{i,t}^{2} - \overline{\SVS}_{t}^{2} \right) . \label{eq:expected.return}
\end{equation*}
where $R_{i,t+1}$ denotes the one period gross return on stock $i$. Hence, in the 
cross-section, differences in expected returns reflect variation in $\SVS_{i,t}^{2}$.

\smallskip

In our analysis of expected returns, we use the measures proposed by KP, DN, and HLR 
to  construct risk-neutral variance indices for firms with high resilience to disaster 
risk, $\overline{\SVS}_{H}^2$, and for firms with low resilience $\overline{\SVS}_{L}^2$. 
For every day in our sample, we compute these indices for each resilience measure as 
the value-weighted sum of the underlying firms' $\SVS_{i,t}^{2}$, for maturities ranging 
from 30 to 730 days. We present results for all stocks for which options are available 
in OptionMetrics, but to face possible concerns about limited trading in options on 
small stocks we also present results for the subset of S\&P 500 firms in the Appendix.  

\smallskip

Figure \ref{fig5} plots the time series of $\overline{\SVS}_{L}^2$ and 
$\overline{\SVS}_{H}^2$, measuring resilience on the basis of KP's `affected\_share' 
measure, as well as their difference for maturities of 30, 91, 365 and 730 days. 
The results show that, until early March, the options-implied variance of High-resilience 
stocks slightly exceeded that for Low-resilience stocks. At the time of the Italian 
lockdown (February 24), the approximate 2\% (p.a.) difference $\overline{\SVS}_{L}^2 - 
\overline{\SVS}_{H}^2$ implies that High-resilience firms had higher expected returns 
than Low-resilience ones. During the Covid-19 crisis, the sign of this premium reverses 
and its magnitude surges to more than 30\% (p.a.) at the peak, as implied by 
one-month options decreasing with longer maturities to approximately 8\% (p.a.) for 
a forecast horizon of two years. Until the end of March these premia gradually declined 
to half (interestingly, starting shortly before the reversal in stock prices) and 
imply, as of March 31, that the stocks of Low-resilience firms are expected to carry 
a premium of about 5.5\% (p.a.) over the next year and about 4\% (p.a.) over the next 
two years. The results are qualitatively very similar when only large firms (i.e., 
constituents of the S\&P 500) are retained in the sample, although these firms 
feature  lower $\SVS$-levels, as shown in Figure \ref{figA4} of the Appendix.

\begin{center}
\textbf{[Insert Figure \ref{fig5}]}
\end{center}

To illustrate these results with reference to some well-known stocks, Figure 
\ref{fig6a} presents the expected returns of a selected group of S\&P firms, 
respectively featuring high and low resilience to the pandemic, plotting all of them 
on the same scale, for all stocks and maturities. As examples of very high-resilience 
firms, the figure plots the option-implied expected returns of Apple, Google, and 
Microsoft. At the opposite end of the resilience range, as examples of very 
low-resilience stocks, the figure plots the expected returns of Marriott, United 
Airlines and Royal Caribbean: travel and tourism have been among the industries 
hardest hit by the stay-at-home orders and social distancing rules. 

\begin{center}
\textbf{[Insert Figure \ref{fig6a}]}
\end{center}

Two results emerge strikingly from Figure \ref{fig6a}. First, at the outbreak of 
the COVID-19 crisis, all the option-implied expected returns rose, but those of 
low-resilience stocks (right-side charts) increased by an order of magnitude more 
than those of low-resilience ones (left-side charts). At the peak of the crisis, 
the expected return implied by short-term options became an enormous 300\% (p.a.) 
for United Airlines and Royal Caribbean, reflecting unprecedented uncertainty about 
the immediate future of their businesses. Second, this increase is much more 
persistent for low-resilience stocks: at the end of our sample, on March 31, their 
expected returns are still elevated, while for high-resilience stocks they revert 
back to pre-COVID-19 levels, especially for the two-year horizon. Third, for all 
stocks expected returns decrease in levels as maturities increase, indicating that 
there is a term structure to pandemic risk: it is perceived to decrease substantially 
as the horizon lengthens, though it far from vanishes for low-resilience stocks. 

\smallskip

Figure \ref{fig6b} provides a clearer view of the time-series patterns of the 
expected returns of the six stocks, as it adapts the vertical scale of the plot to 
their range of variation. The figure allows in particular to appreciate that, even at 
the end of our sample, almost two months after the outbreak, investors still require 
much higher expected returns from Marriott, United Airlines and Royal Caribbean than 
from Apple, Google and Microsoft, and that even at the two-year horizon the 
expected return is a multiple of what it was at the beginning of the year before 
the COVID-19 crisis. The most extreme case is Royal Caribbean, whose two-year 
options imply, as of March 31, an expected return of 60\% (p.a.) for a two-year horizon. 

\begin{center}
\textbf{[Insert Figure \ref{fig6b}]}
\end{center}

Taken together, these results indicate that disaster resilience is priced in equity 
options and that the COVID-19 crisis has greatly affected \emph{how} financial markets 
price resilience to disaster risk: While the two-year expected returns have reverted 
to their pre-crisis levels for the firms least affected by social distancing 
requirements, the expected returns for the firms most severely affected by the 
pandemic are much higher than before the crisis. Hence, it appears that, going 
forward, markets consider disaster risk, and specifically the resilience against a 
pandemic, to be much more important than they did before COVID-19.


\section{Conclusions}
\label{conclusions}

The contribution of this paper is threefold. First, it investigates whether 
the COVID-19 outbreak triggered a different stock return response depending 
on companies' resilience to social distancing, which is the most severe 
constraint that the pandemic has imposed on firms' operations. On this score, 
we find that more resilient companies greatly outperformed less resilient ones, 
even after controlling for all conventional measures of risk premia.

\smallskip

Second, the paper explores whether similar cross-sectional return differentials 
already emerged before the COVID-19 outbreak. Indeed this is the case: in the 
2014-19 interval, the cumulative return differential between more and less 
pandemic-resilient firms is of about the same magnitude as during the outbreak, 
i.e. between late February and early April 2020. We interpret this as evidence 
of learning by investors, i.e., of their growing awareness of the potential 
threat posed by pandemics well in advance of its materialization.

\smallskip

Finally, we exploit option price data to infer whether, after the COVID-19 outbreak,
investors price pandemic risk over different horizons, and find that they do: even 
on a 2-year horizon, stocks of more pandemic-resilient firms are expected to yield 
significantly lower returns than less resilient ones, reflecting lower exposure 
to disaster risk. Such differences are massive in the case of some stocks: for 
example, as late as early April 2020, the expected return on low-resilience 
stocks such as Royal Caribbean and United Airlines is around 60\% and 40\% 
respectively, while those of high-resilience stocks such as Apple and Microsoft 
are between 3\% and 4\%. 

\smallskip

Hence, going forward, markets appear to price exposure to a new risk factor, namely, 
pandemic risk. In future development of this work, we plan to investigate whether 
such risk is part of a wider sustainability risk factor, or at least whether the two 
types of risk are correlated. We also plan to investigate whether resilience to social 
distancing has not only direct effects on stock prices, but also indirect effects via 
demand and supply linkages, i.e. whether for instance the stocks of firms that 
depend heavily on low-resilience firms are themselves more exposed to pandemic risk, 
other things equal.


\pagebreak

\bibliography{socialdistancing}
\bibliographystyle{rfsnew}

\clearpage
\addtolength{\voffset}{-1.75cm}
\addtolength{\footskip}{1.75cm}

\clearpage
\newpage

\clearpage
\newpage

\clearpage
\newpage

\clearpage


\begin{figure}\caption{Stock options-implied risk-neutral variances}\label{fig5}\footnotesize

\vspace{-3mm}

\begin{adjustwidth}{-1.75cm}{-1.75cm}
This figure plots stock options-implied risk-neutral variance indices for firms with high and low resilience to social distancing during the first quarter of 2020. On any given day, we assign a firm to the high resilience index, $\overline{\SVS}^2_{H,t}$, if its `affected\_share' \citep[as defined by][]{koren/peto:20covid} is below the median value and to the low resilience index, $\overline{\SVS}^2_{L,t}$, if it is above. The indices are computed as the value-weighted sums of individual firms' risk-neutral variances, $\SVS^2_{i,t}$. The difference $\overline{\SVS}^2_{L,t} - \overline{\SVS}^2_{L,t}$ measures the expected return of low resilience in excess of high resilience stocks. Panels A to D present results using options maturities of 30, 91, 365 and 730 days, respectively. The dashed vertical lines mark February 24 and March 20. 

\begin{spacing}{1.0}
\end{spacing}

\end{adjustwidth}

\begin{adjustwidth}{-2cm}{-2cm}\centering

\vspace{-6mm}

	{Panel A. 30-day horizon}\\ \vspace{-1.5mm}
	{\includegraphics[scale = 0.775]{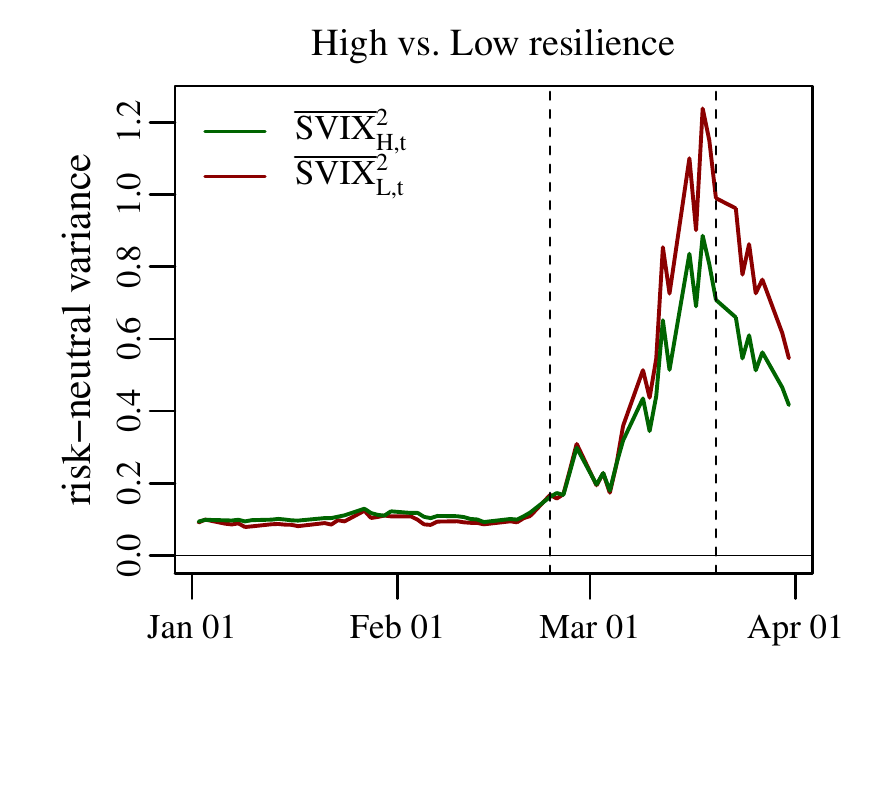}}
	{\includegraphics[scale = 0.775]{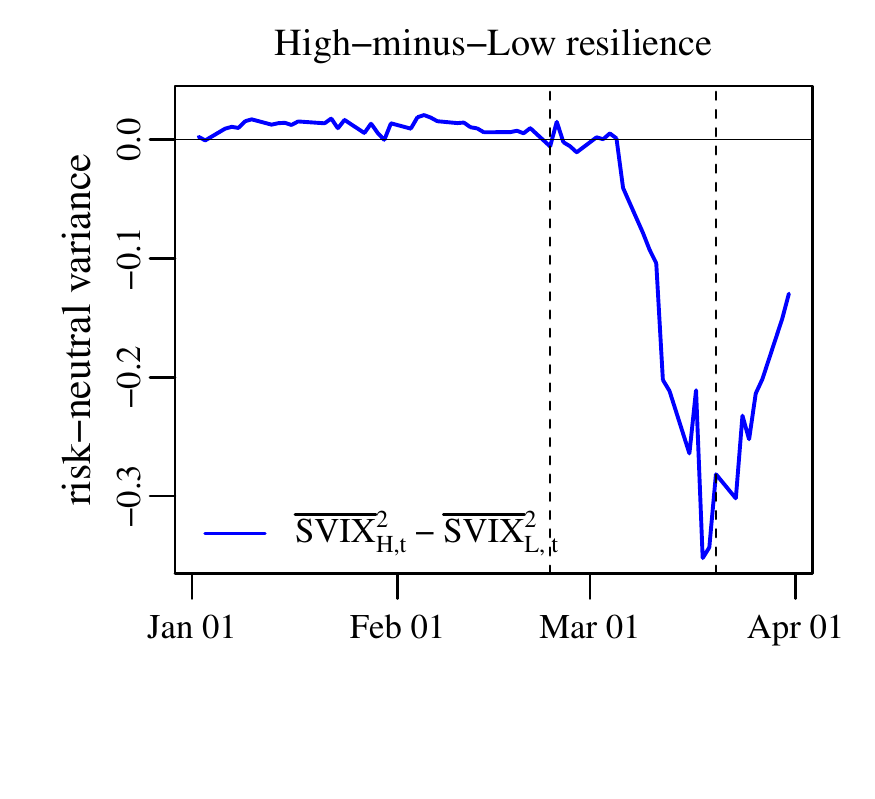}}

\vspace{-11mm}

	{Panel B. 91-day horizon}\\ \vspace{-1.5mm}
	{\includegraphics[scale = 0.775]{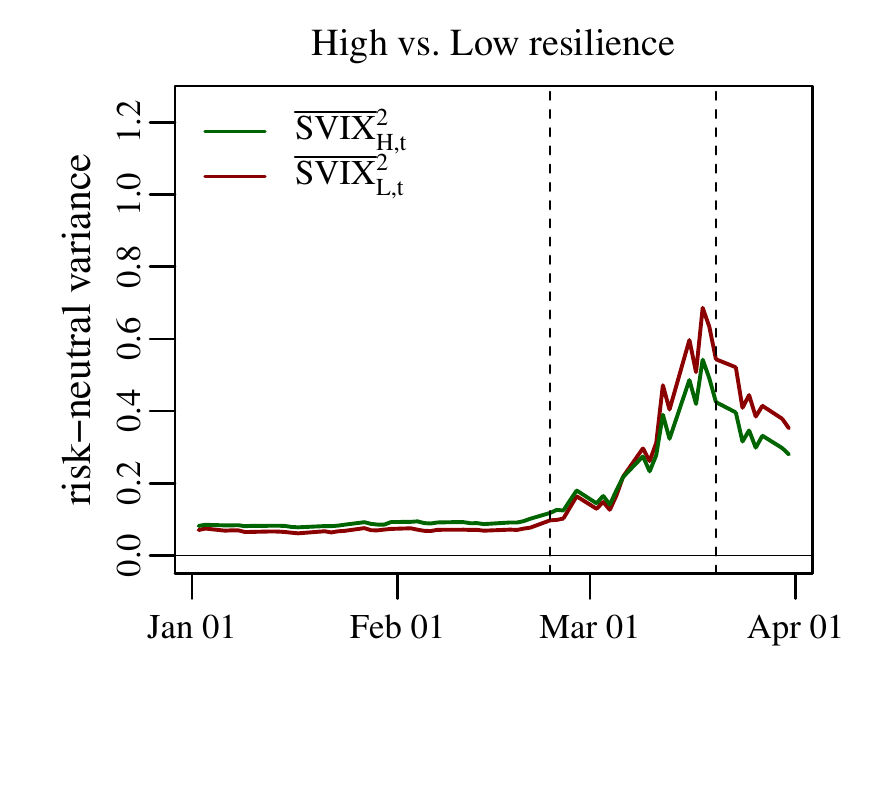}}
	{\includegraphics[scale = 0.775]{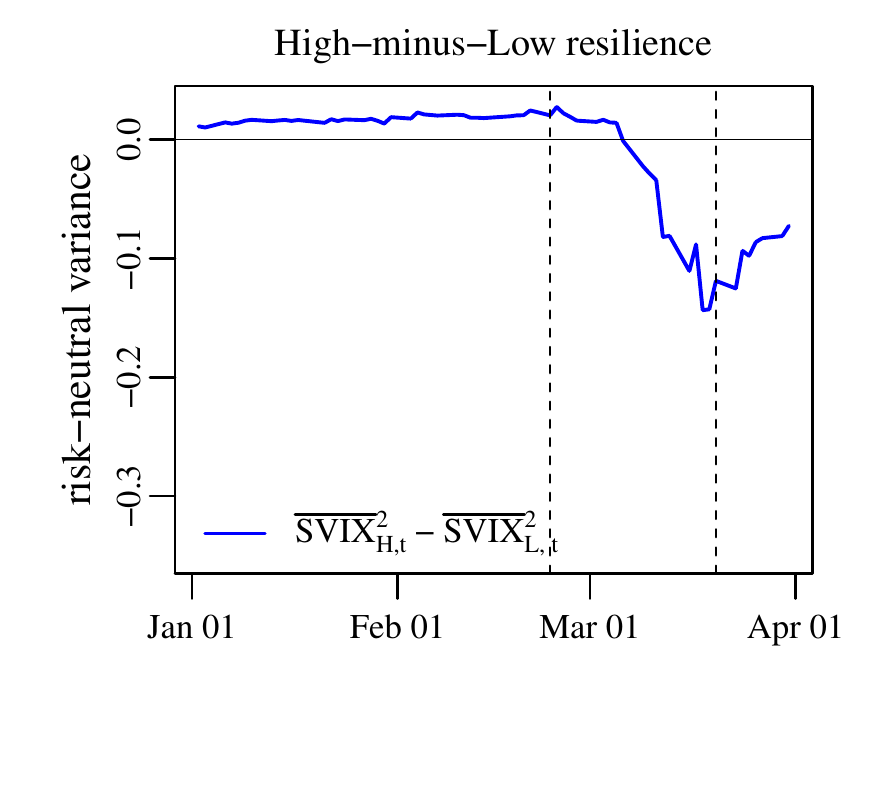}}

\vspace{-11mm}
	{Panel C. 365-day horizon}\\ \vspace{-1.5mm}
	{\includegraphics[scale = 0.775]{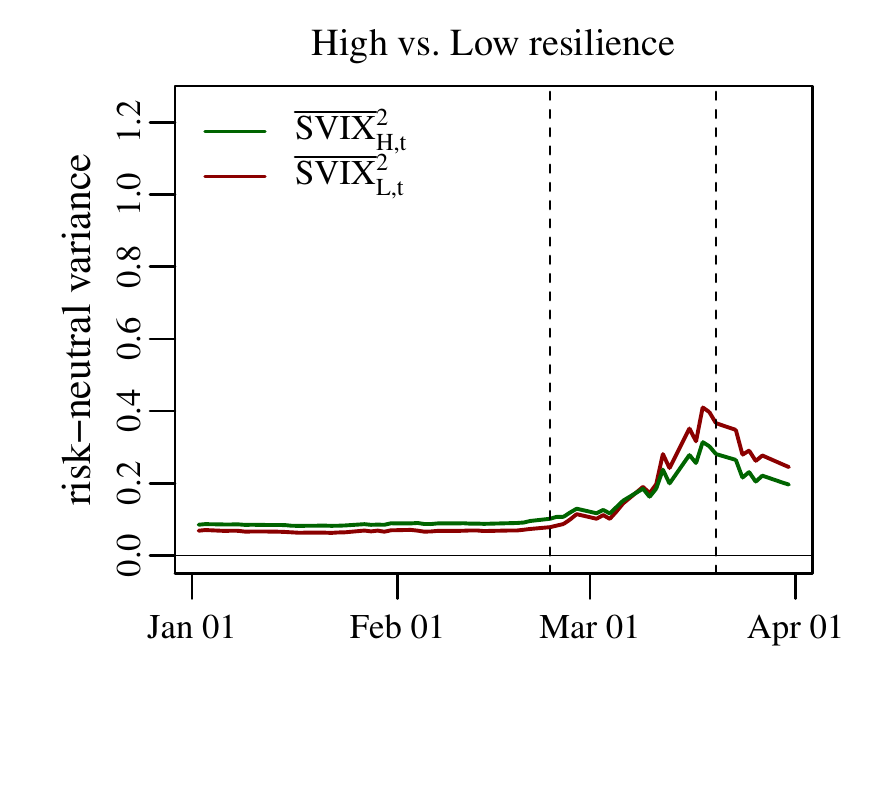}}
	{\includegraphics[scale = 0.775]{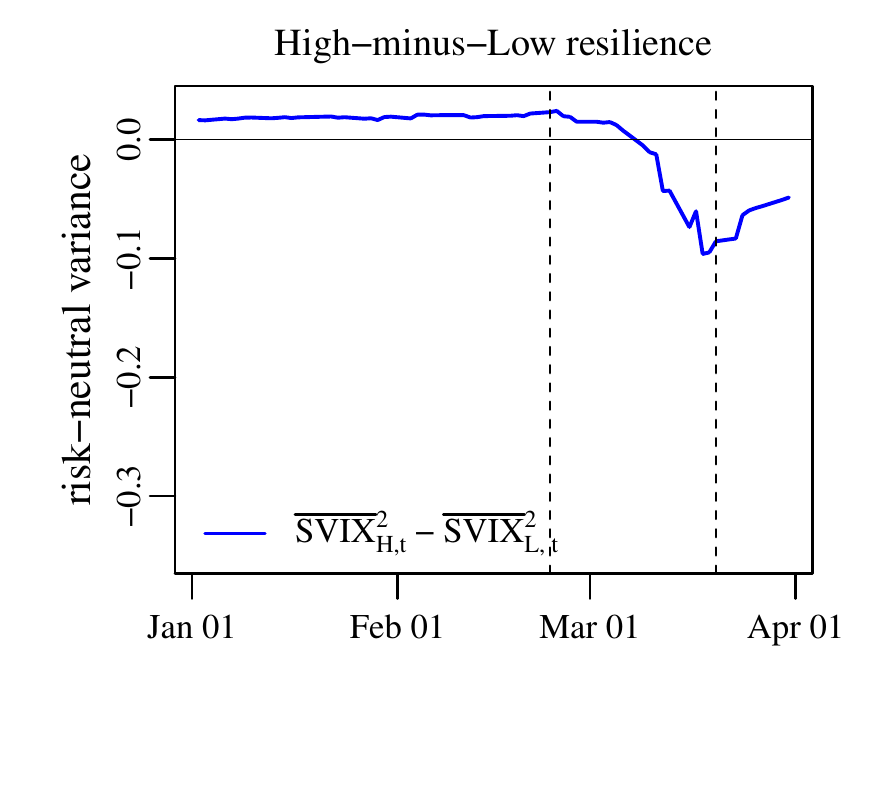}}

\vspace{-11mm}
	{Panel D. 730-day horizon}\\ \vspace{-1.5mm}
	{\includegraphics[scale = 0.775]{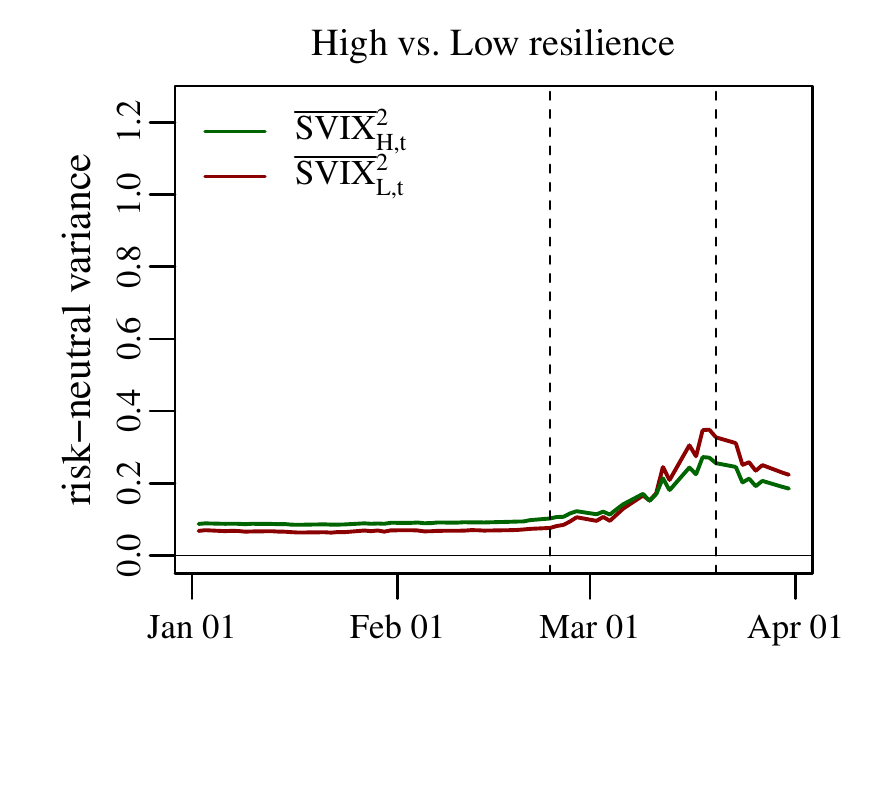}}
	{\includegraphics[scale = 0.775]{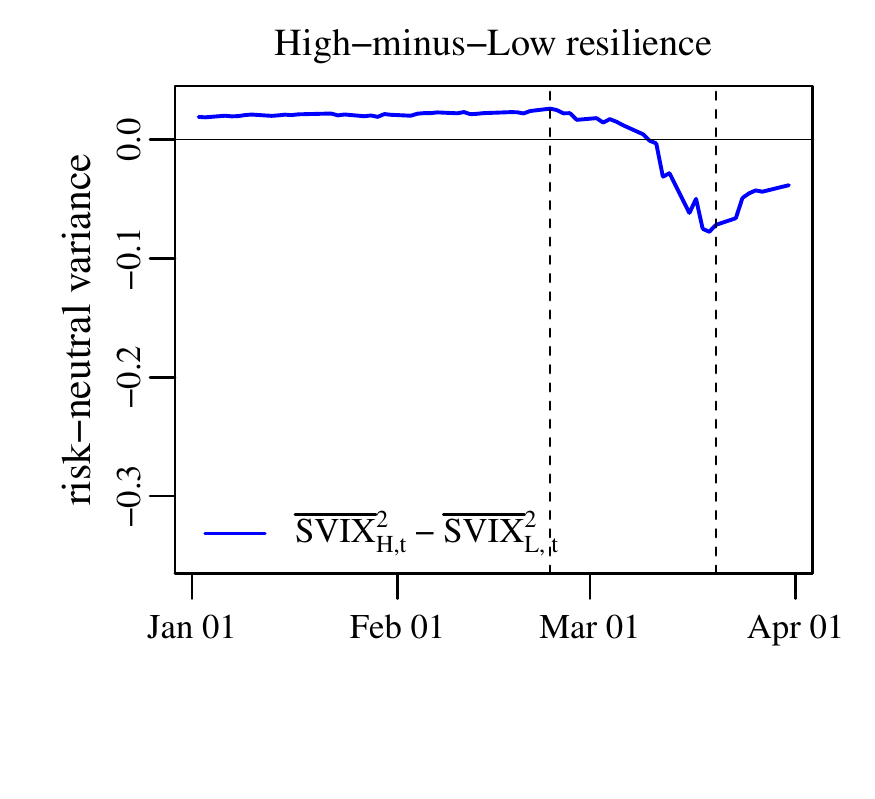}}

\end{adjustwidth}
\end{figure}

\clearpage
\newpage
\newpage
\clearpage
\begin{figure}\caption{Expected returns of selected S\&P 500 firms with high and low resilience (same y-axis)}\label{fig6a}\footnotesize

\vspace{-4mm}

\begin{adjustwidth}{-1.75cm}{-1.75cm}

\begin{spacing}{1.0}
This figure plots stock options-implied expected returns for selected S\&P 500 firms during the first quarter of 2020. The high resilience stocks we consider are Apple (AAPL), Google (GOOG), and Microsoft (MSFT), the low resilience stocks are Marriott (MAR), United Airlines (UAL) and Royal Caribbean (RCL). We compute the expected return on a stock in from the risk-neutral variance of the market and the stock's excess risk-neutral variance relative to that of the average stock following the approach of \citet{martin/wagner:19jf}.  Panels A to D present results for forecast horizons of 30, 91, 365 and 730 days, respectively. The dashed vertical lines mark February 24 and March 20. 
\end{spacing}

\end{adjustwidth}

\begin{adjustwidth}{-2cm}{-2cm}\centering

\vspace{-4mm}

	{Panel A. 30-day horizon}\\ \vspace{-1.5mm}
	{\includegraphics[scale = 0.775]{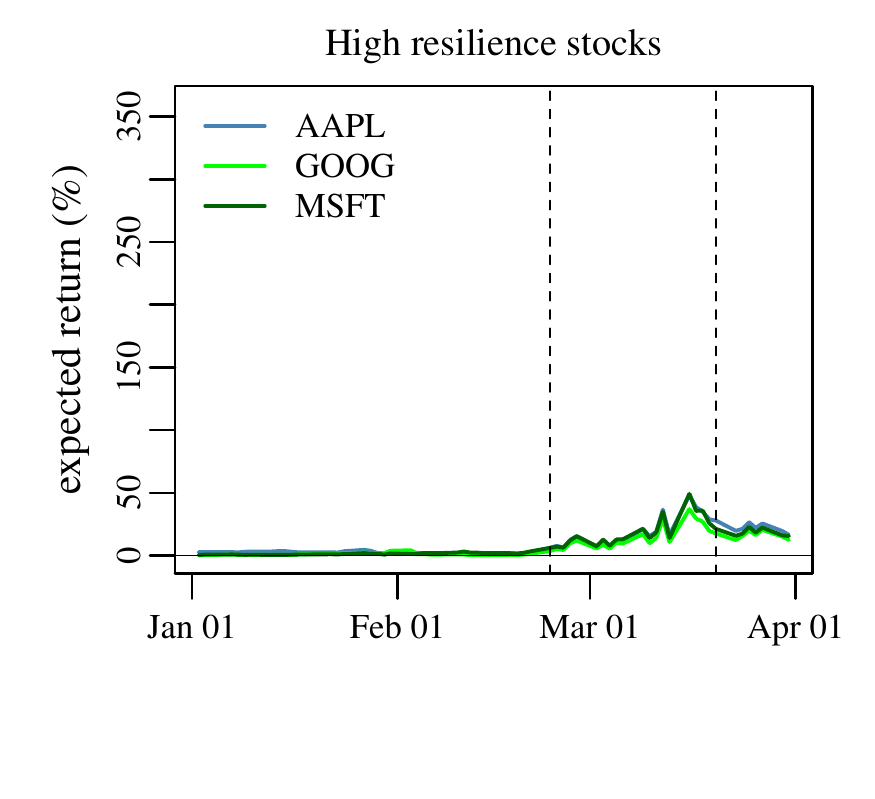}}
	{\includegraphics[scale = 0.775]{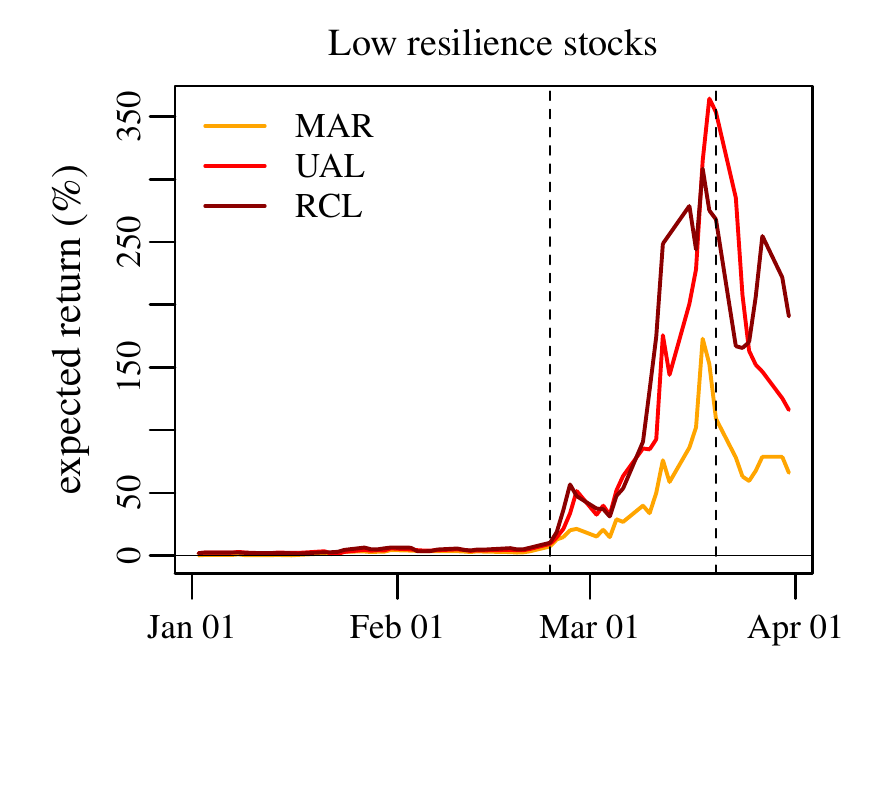}}

\vspace{-11mm}

	{Panel B. 91-day horizon}\\ \vspace{-1.5mm}
	{\includegraphics[scale = 0.775]{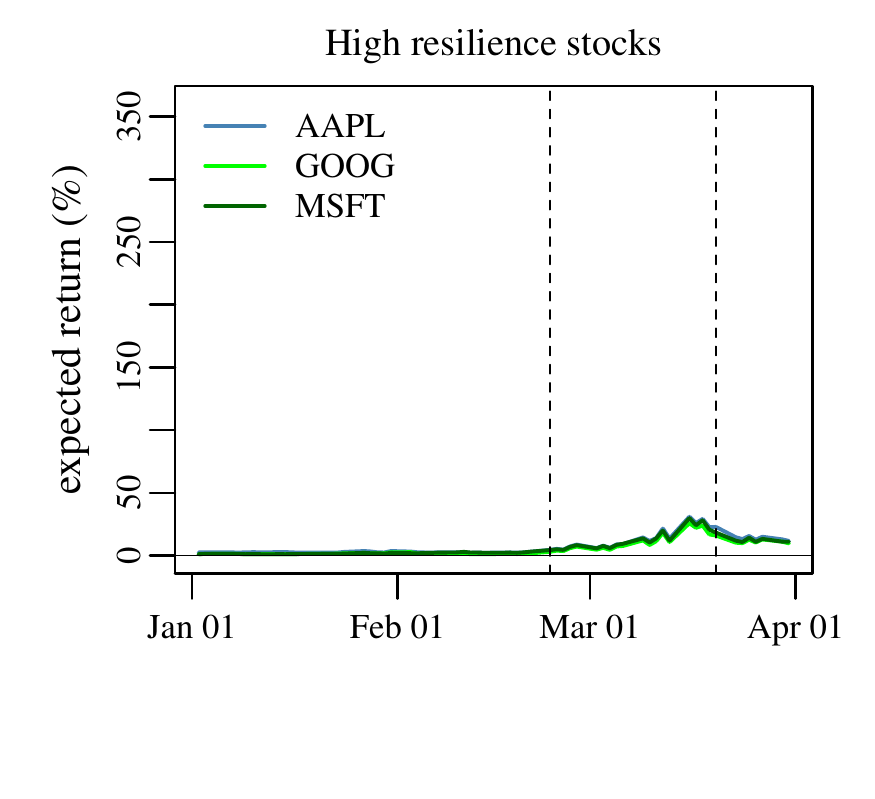}}
	{\includegraphics[scale = 0.775]{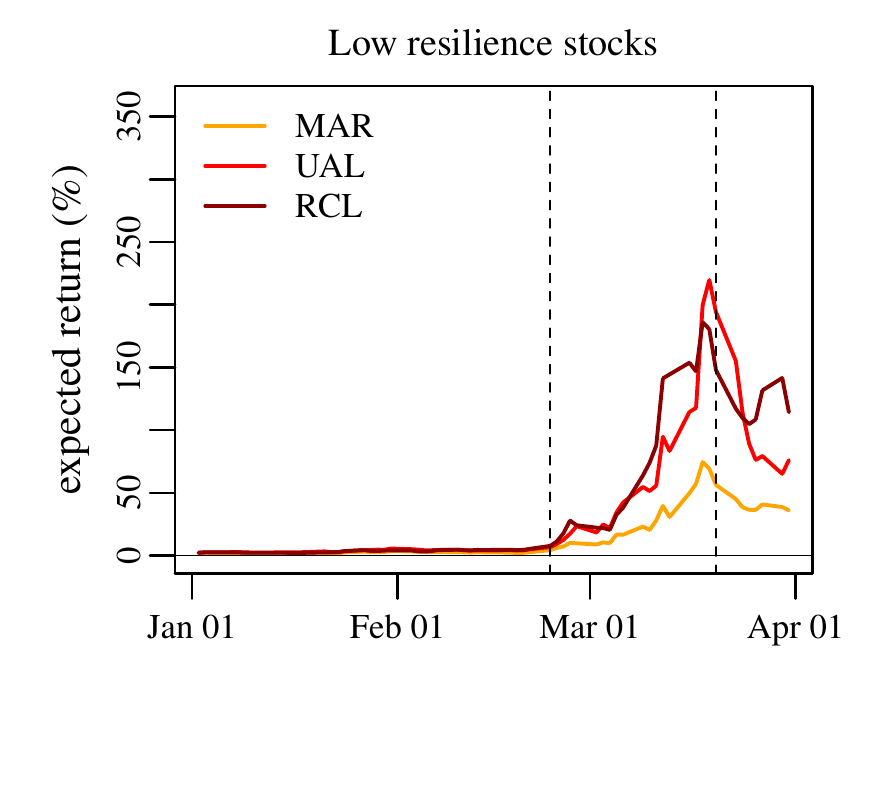}}

\vspace{-11mm}
	{Panel C. 365-day horizon}\\ \vspace{-1.5mm}
	{\includegraphics[scale = 0.775]{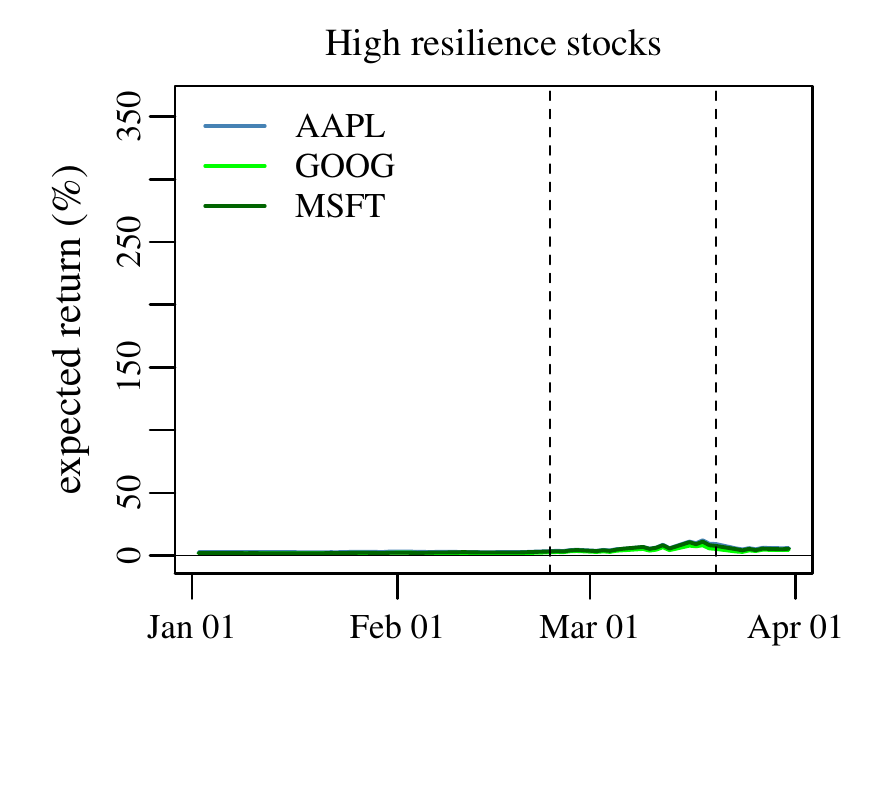}}
	{\includegraphics[scale = 0.775]{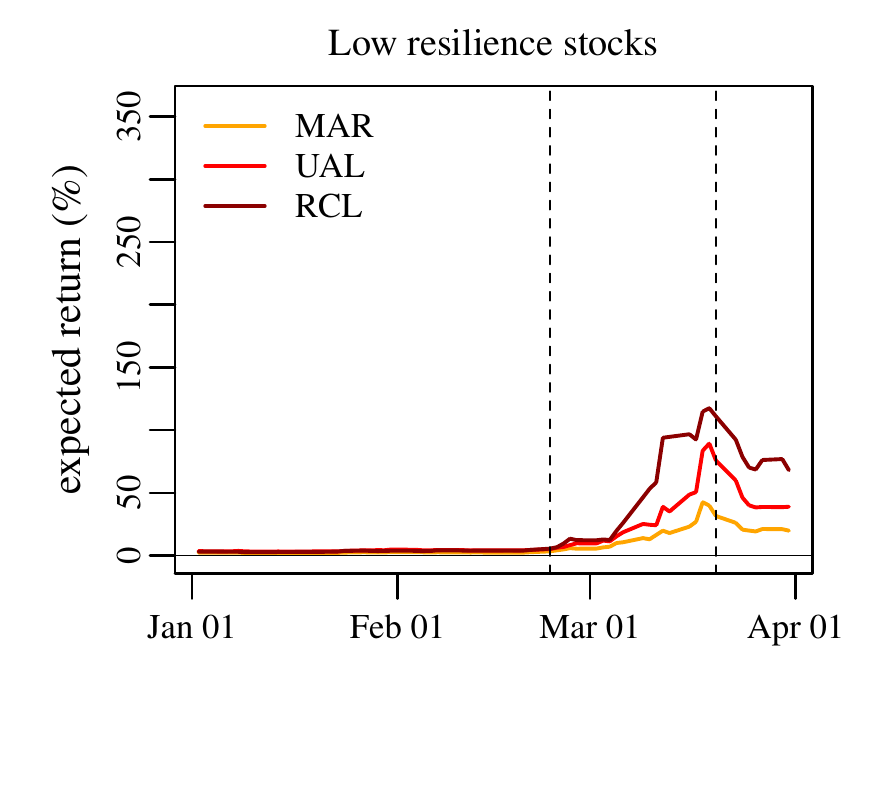}}

\vspace{-11mm}
	{Panel D. 730-day horizon}\\ \vspace{-1.5mm}
	{\includegraphics[scale = 0.775]{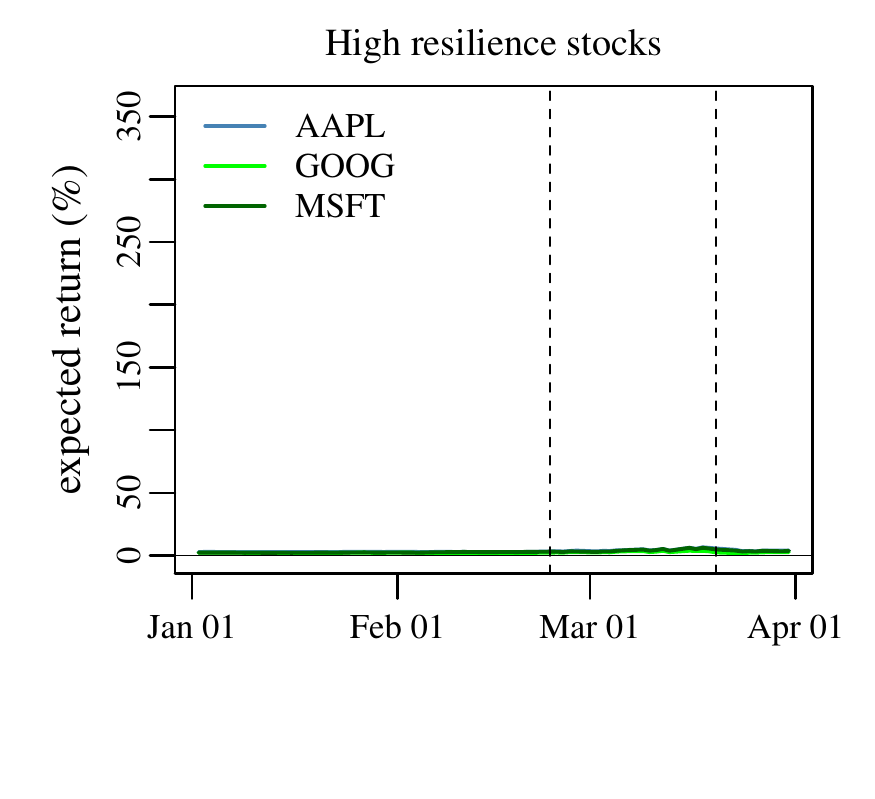}}
	{\includegraphics[scale = 0.775]{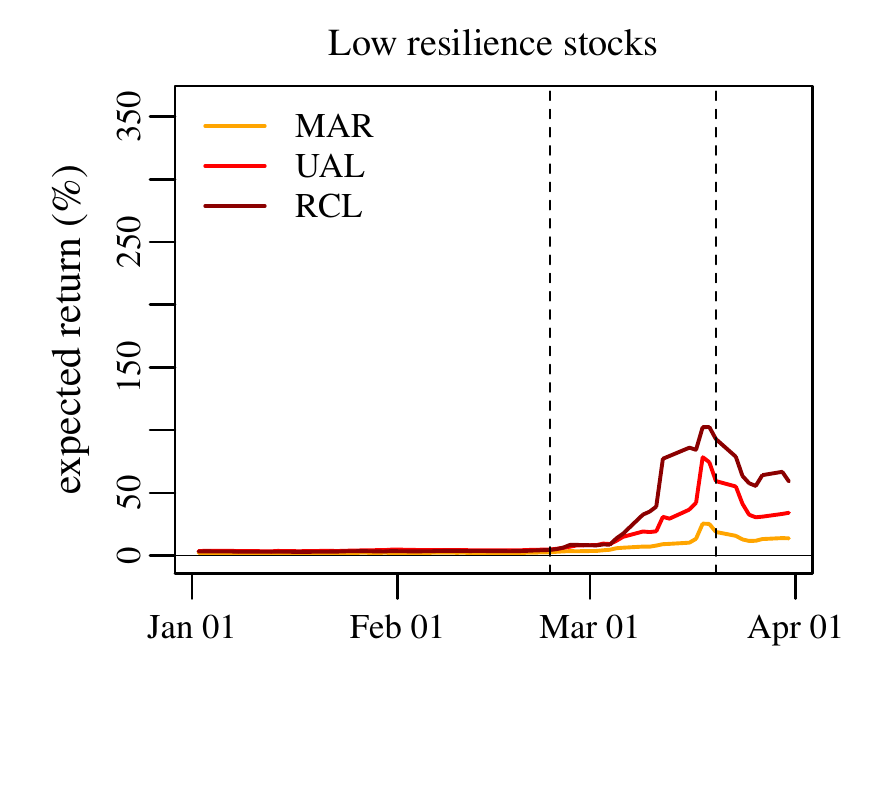}}

\end{adjustwidth}
\end{figure}

\clearpage
\newpage
\newpage
\clearpage
\begin{figure}\caption{Expected returns of selected S\&P 500 firms with high and low resilience}\label{fig6b}\footnotesize

\vspace{-4mm}

\begin{adjustwidth}{-1.75cm}{-1.75cm}

\begin{spacing}{1.0}
This figure plots stock options-implied expected returns for selected S\&P 500 firms during the first quarter of 2020. The high resilience stocks we consider are Apple (AAPL), Google (GOOG), and Microsoft (MSFT), the low resilience stocks are Marriott (MAR), United Airlines (UAL) and Royal Caribbean (RCL). We compute the expected return on a stock in from the risk-neutral variance of the market and the stock's excess risk-neutral variance relative to that of the average stock following the approach of \citet{martin/wagner:19jf}.  Panels A to D present results for forecast horizons of 30, 91, 365 and 730 days, respectively. The dashed vertical lines mark February 24 and March 20. 
\end{spacing}

\end{adjustwidth}

\begin{adjustwidth}{-2cm}{-2cm}\centering

\vspace{-4mm}

	{Panel A. 30-day horizon}\\ \vspace{-1.5mm}
	{\includegraphics[scale = 0.775]{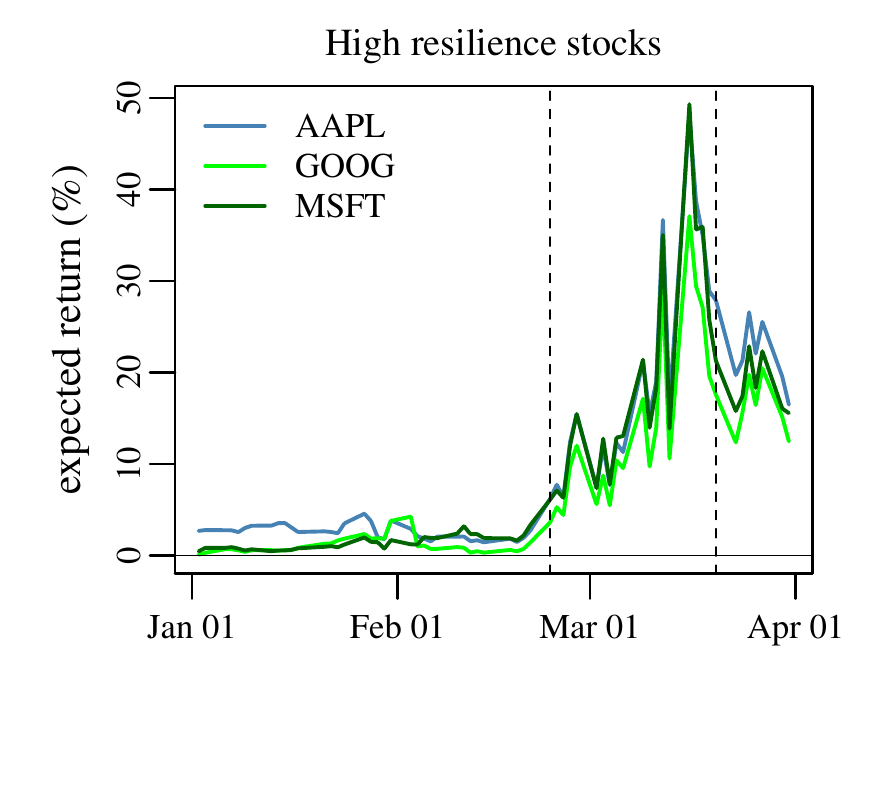}}
	{\includegraphics[scale = 0.775]{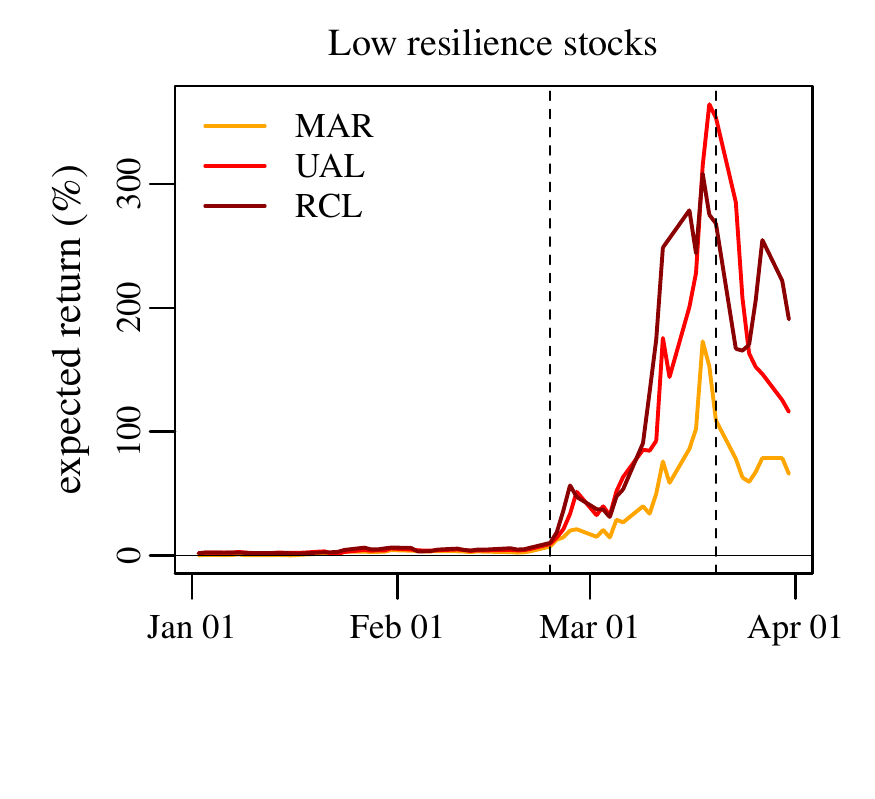}}

\vspace{-11mm}

	{Panel B. 91-day horizon}\\ \vspace{-1.5mm}
	{\includegraphics[scale = 0.775]{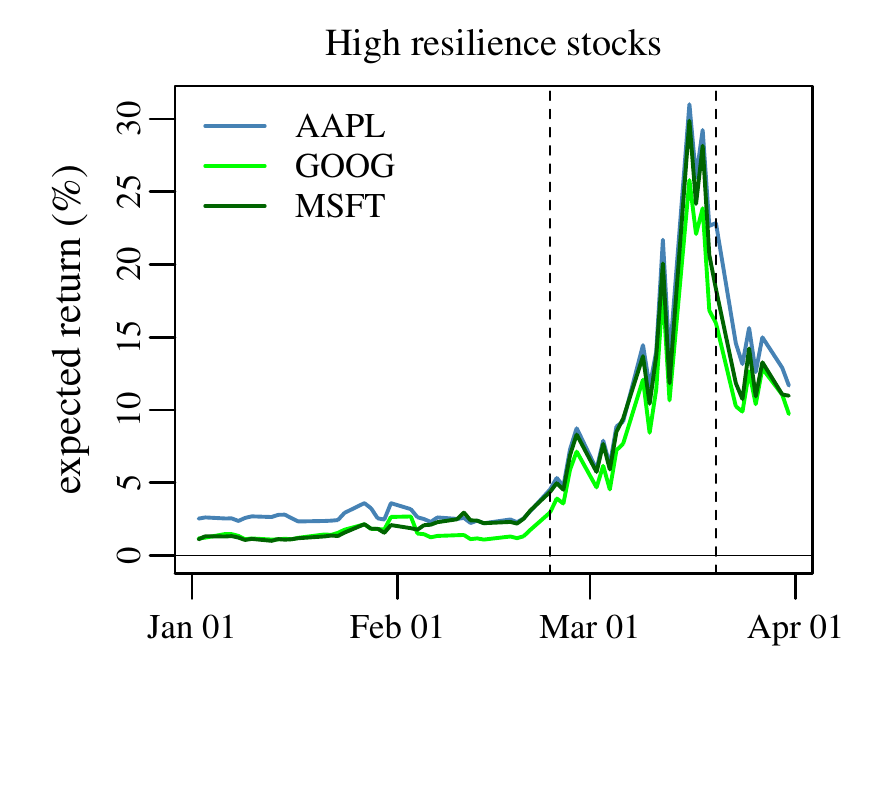}}
	{\includegraphics[scale = 0.775]{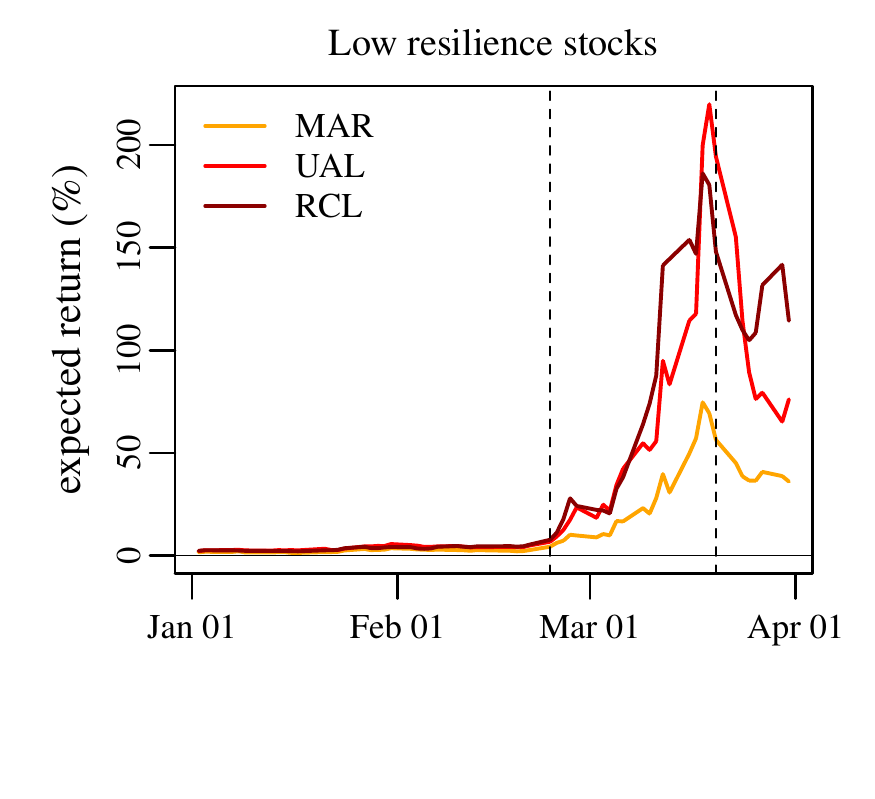}}

\vspace{-11mm}
	{Panel C. 365-day horizon}\\ \vspace{-1.5mm}
	{\includegraphics[scale = 0.775]{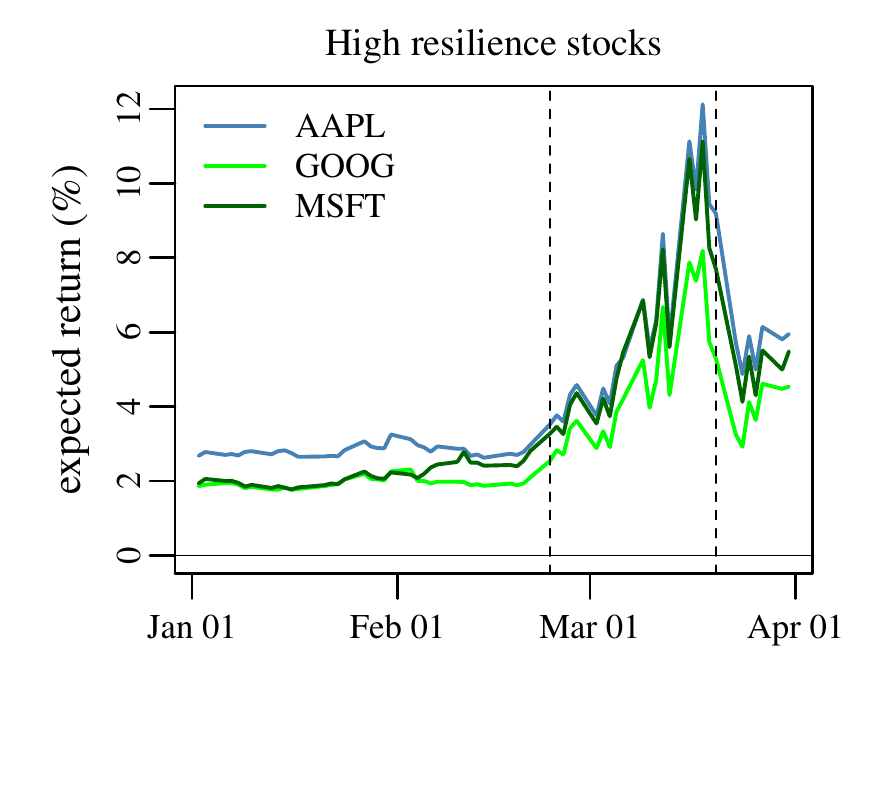}}
	{\includegraphics[scale = 0.775]{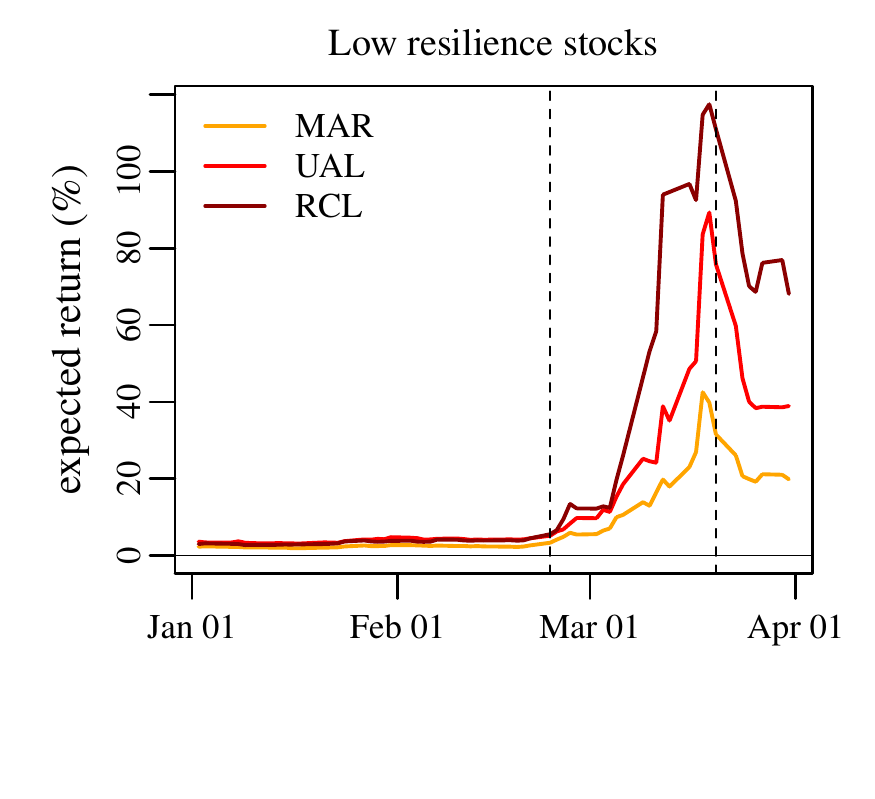}}

\vspace{-11mm}
	{Panel D. 730-day horizon}\\ \vspace{-1.5mm}
	{\includegraphics[scale = 0.775]{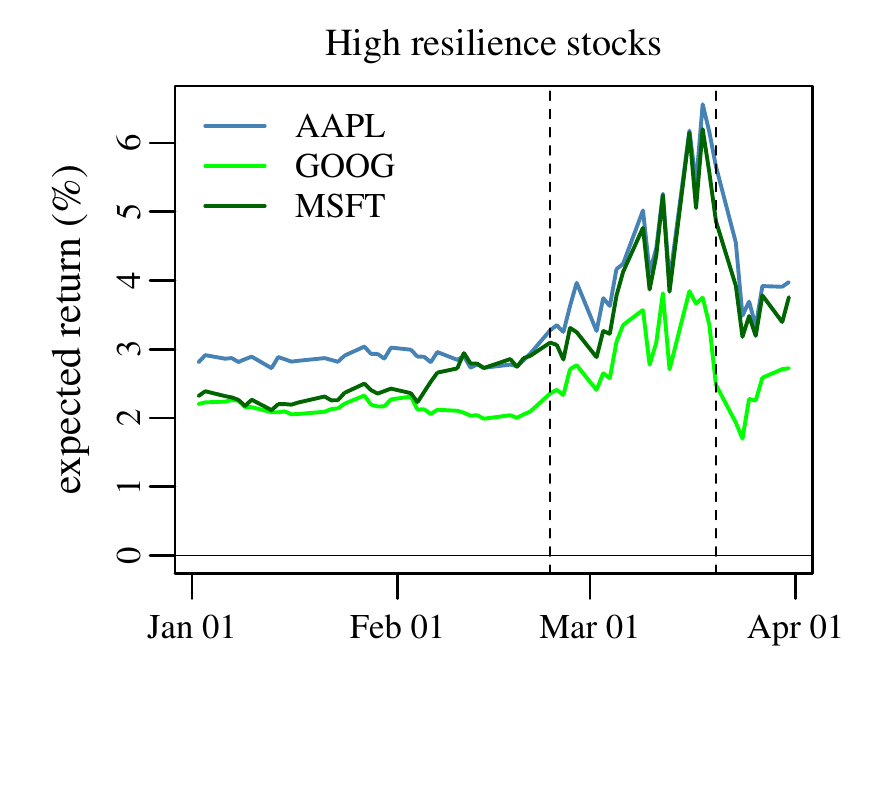}}
	{\includegraphics[scale = 0.775]{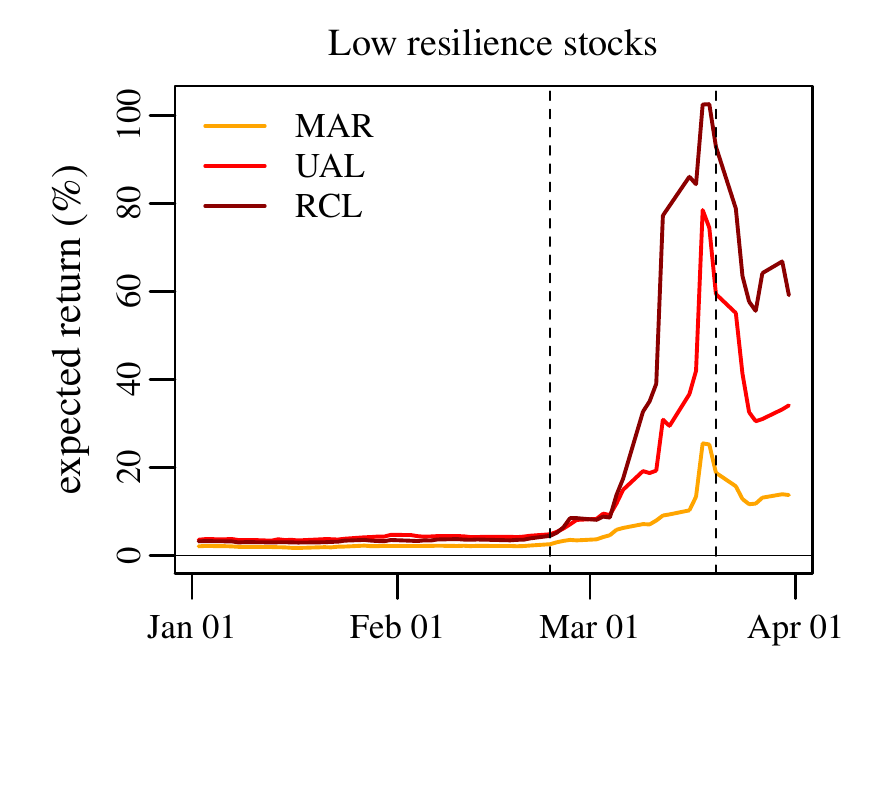}}

\end{adjustwidth}
\end{figure}

\clearpage
\newpage
\begin{table}[tbp]
    \caption{ Excess and risk-adjusted returns of high and low resilience stocks} \label{tab2} \footnotesize
    \bigskip
\begin{adjustwidth}{-1.75cm}{-1.75cm}
\begin{spacing}{1.0}
This table reports averages of daily returns for value-weighted portfolios of low and high resilience stocks from February 24 to March 20, 2020, i.e. from the day after Italy introduced its lockdown to the last trading day before the Fed announced its intervention. On any given day, we assign a firm to the `High' (`Low') portfolio, in Panel A, if its `affected\_share' \citep[as defined by][]{koren/peto:20covid} is below (above) the median value; in Panel B, if its `teleworkable\_manual\_wage' \citep[as defined by][]{dingel/neiman:20nber} is above (below) the median value; in Panel C, if its `dur\_workplace' \citep[as defined by][]{hensvik/lebarbanchon/rathelot:20cepr} is below (above) the median value.   and to the `Low' portfolio if it is above. In addition to rae excess returns (ret), we report CAPM-adjusted returns, i.e. controlling for exposure to market risk, returns adjusted for the Fama-French three factor model exposures (ff3, i.e. market, size, value) and five factor model exposures (ff5, i.e. market, size, value, investments, profitability), and the Fama-French models augmented by the momentum factor (ff4 and ff6). Values in square brackets are $t$-statistics based on standard errors following \citet{newey/west:87ecta}, where we choose the optimal truncation lag as suggested by \citet{andrews:91ecta}. $^{*}$, $^{**}$ and $^{***}$ denote statistical significance at the 10\%, 5\%, and 1\% level, respectively.
\end{spacing}
\end{adjustwidth}

  \begin{center}  \footnotesize
  \begin{spacing}{1.0}

	Panel A. Resilience based on `affected\_share' as in \citet{koren/peto:20covid} \\ \vspace{1mm}
	\begin{tabular}{L{5cm}d{2.2}d{2.2}d{2.2}d{2.2}d{2.2}d{2.2}} \toprule   &  \multicolumn{1}{C{1.4cm}}{ ret }  &   \multicolumn{1}{C{1.4cm}}{ capm }  &   \multicolumn{1}{C{1.4cm}}{ ff3 }  &   \multicolumn{1}{C{1.4cm}}{ ff4 }  &   \multicolumn{1}{C{1.4cm}}{ ff5 }  &   \multicolumn{1}{C{1.4cm}}{ ff6 }  \\ 
\midrule High resilience  &  -1.48^{***}  &  0.47^{***}  &  0.23^{***}  &  0.16^{**}  &  0.23^{***}  &  0.17^{**}  \\ 
  &  \multicolumn{ 1}{c}{[-3.02]}  &  \multicolumn{ 1}{c}{[6.65]}  &  \multicolumn{ 1}{c}{[2.69]}  &  \multicolumn{ 1}{c}{[2.06]}  &  \multicolumn{ 1}{c}{[3.04]}  &  \multicolumn{ 1}{c}{[2.29]}  \\ 
Low resilience  &  -1.88^{***}  &  -0.52^{***}  &  -0.52^{***}  &  -0.42^{***}  &  -0.53^{***}  &  -0.44^{***}  \\ 
  &  \multicolumn{ 1}{c}{[-3.84]}  &  \multicolumn{ 1}{c}{[-3.00]}  &  \multicolumn{ 1}{c}{[-2.89]}  &  \multicolumn{ 1}{c}{[-2.65]}  &  \multicolumn{ 1}{c}{[-2.90]}  &  \multicolumn{ 1}{c}{[-2.70]}  \\ 
\midrule High-minus-Low  &  0.40^{***}  &  0.99^{***}  &  0.75^{***}  &  0.59^{***}  &  0.77^{***}  &  0.61^{***}  \\ 
  &  \multicolumn{ 1}{c}{[3.22]}  &  \multicolumn{ 1}{c}{[4.26]}  &  \multicolumn{ 1}{c}{[2.94]}  &  \multicolumn{ 1}{c}{[2.65]}  &  \multicolumn{ 1}{c}{[3.12]}  &  \multicolumn{ 1}{c}{[2.85]}  \\  \bottomrule \end{tabular}

	\bigskip
	
	Panel B. Resilience based on `teleworkable\_manual\_wage' as in \citet{dingel/neiman:20nber} \\ \vspace{1mm}
	\begin{tabular}{L{5cm}d{2.2}d{2.2}d{2.2}d{2.2}d{2.2}d{2.2}} \toprule   &  \multicolumn{1}{C{1.4cm}}{ ret }  &   \multicolumn{1}{C{1.4cm}}{ capm }  &   \multicolumn{1}{C{1.4cm}}{ ff3 }  &   \multicolumn{1}{C{1.4cm}}{ ff4 }  &   \multicolumn{1}{C{1.4cm}}{ ff5 }  &   \multicolumn{1}{C{1.4cm}}{ ff6 }  \\ 
\midrule High resilience  &  -1.58^{***}  &  0.35^{***}  &  0.08  &  0.04  &  0.09  &  0.07  \\ 
  &  \multicolumn{ 1}{c}{[-3.65]}  &  \multicolumn{ 1}{c}{[3.36]}  &  \multicolumn{ 1}{c}{[0.59]}  &  \multicolumn{ 1}{c}{[0.32]}  &  \multicolumn{ 1}{c}{[1.24]}  &  \multicolumn{ 1}{c}{[1.06]}  \\ 
Low resilience  &  -1.77^{***}  &  -0.39^{**}  &  -0.40^{**}  &  -0.33^{*}  &  -0.41^{**}  &  -0.38^{**}  \\ 
  &  \multicolumn{ 1}{c}{[-3.43]}  &  \multicolumn{ 1}{c}{[-2.03]}  &  \multicolumn{ 1}{c}{[-2.11]}  &  \multicolumn{ 1}{c}{[-1.86]}  &  \multicolumn{ 1}{c}{[-2.36]}  &  \multicolumn{ 1}{c}{[-2.27]}  \\ 
\midrule High-minus-Low  &  0.19  &  0.74^{***}  &  0.48^{*}  &  0.37  &  0.49^{**}  &  0.45^{**}  \\ 
  &  \multicolumn{ 1}{c}{[1.16]}  &  \multicolumn{ 1}{c}{[2.69]}  &  \multicolumn{ 1}{c}{[1.66]}  &  \multicolumn{ 1}{c}{[1.28]}  &  \multicolumn{ 1}{c}{[2.22]}  &  \multicolumn{ 1}{c}{[2.09]}  \\  \bottomrule \end{tabular}

	\bigskip
		
	Panel C. Resilience based on  `dur\_workplace' as in \citet{hensvik/lebarbanchon/rathelot:20cepr} \\ \vspace{1mm}
	\begin{tabular}{L{5cm}d{2.2}d{2.2}d{2.2}d{2.2}d{2.2}d{2.2}} \toprule   &  \multicolumn{1}{C{1.4cm}}{ ret }  &   \multicolumn{1}{C{1.4cm}}{ capm }  &   \multicolumn{1}{C{1.4cm}}{ ff3 }  &   \multicolumn{1}{C{1.4cm}}{ ff4 }  &   \multicolumn{1}{C{1.4cm}}{ ff5 }  &   \multicolumn{1}{C{1.4cm}}{ ff6 }  \\ 
\midrule High resilience  &  -1.42^{***}  &  0.46^{***}  &  0.10  &  0.08  &  0.11  &  0.11  \\ 
  &  \multicolumn{ 1}{c}{[-3.34]}  &  \multicolumn{ 1}{c}{[4.64]}  &  \multicolumn{ 1}{c}{[0.77]}  &  \multicolumn{ 1}{c}{[0.63]}  &  \multicolumn{ 1}{c}{[1.11]}  &  \multicolumn{ 1}{c}{[1.12]}  \\ 
Low resilience  &  -1.89^{***}  &  -0.35^{**}  &  -0.31^{**}  &  -0.31^{**}  &  -0.31^{*}  &  -0.35^{**}  \\ 
  &  \multicolumn{ 1}{c}{[-3.73]}  &  \multicolumn{ 1}{c}{[-2.53]}  &  \multicolumn{ 1}{c}{[-2.10]}  &  \multicolumn{ 1}{c}{[-2.09]}  &  \multicolumn{ 1}{c}{[-2.01]}  &  \multicolumn{ 1}{c}{[-2.11]}  \\ 
\midrule High-minus-Low  &  0.46^{***}  &  0.81^{***}  &  0.41  &  0.39  &  0.42^{*}  &  0.46^{**}  \\ 
  &  \multicolumn{ 1}{c}{[3.40]}  &  \multicolumn{ 1}{c}{[3.90]}  &  \multicolumn{ 1}{c}{[1.67]}  &  \multicolumn{ 1}{c}{[1.61]}  &  \multicolumn{ 1}{c}{[1.92]}  &  \multicolumn{ 1}{c}{[1.99]}  \\  \bottomrule \end{tabular}

\end{spacing}      
  \end{center}
\end{table}

\clearpage
\newpage
\begin{table}[tbp]
   
    \caption{Summary statistics for industry portfolios}\label{tab3} \footnotesize

\begin{adjustwidth}{-1.75cm}{-1.75cm}
\begin{spacing}{1.0}
This table presents summary statistics for the returns of value-weighted industry portfolios from February 24 to March 20, 2020, i.e. from the day after Italy introduced its lockdown to the last trading day before the Fed announced its intervention. We define resilience as 100 (\%) minus the `affected\_share' defined by \citet{koren/peto:20covid} and present results for the 25 industries with the highest number of firms (in total 2,974). We report the industries' 3-digit NAICS code, their description, the number of firms in the respective industries and their cumulative return. Specifically, we present CAPM-adjusted returns, i.e. controlling for exposure to market risk as well as results controlling for the Fama-French three factor model exposures (ff3, i.e. market, size, value) and five factor model exposures (ff5, i.e. market, size, value, investments, profitability).

\end{spacing}
\end{adjustwidth}

\bigskip
  \begin{center}  \footnotesize
  \begin{spacing}{1.0}

\begin{adjustwidth}{-2cm}{-2cm}\centering

\begin{tabular}{cL{8cm}ccd{3.2}d{3.2}d{3.2}}
\toprule
\multicolumn{1}{c}{NAICS}	&	\multicolumn{1}{c}{description}	&	\multicolumn{1}{c}{firms}	&	\multicolumn{1}{c}{resilience}	&	\multicolumn{1}{c}{capm}	&	\multicolumn{1}{c}{ff3}	&	\multicolumn{1}{c}{ff5}	\\
\midrule
211	&	Oil and gas extraction	&	88	&	70	&	-26.14	&	-7.87	&	-6.44	\\
212	&	Mining, except oil and gas	&	87	&	29	&	-29.04	&	-34.09	&	-33.88	\\
213	&	Support activities for mining	&	37	&	46	&	-30.18	&	-8.02	&	-6.36	\\
221	&	Utilities	&	94	&	54	&	-24.79	&	-28.64	&	-29.01	\\
311	&	Food manufacturing	&	53	&	77	&	-3.80	&	-6.13	&	-6.73	\\
325	&	Chemicals	&	639	&	79	&	7.56	&	2.30	&	2.24	\\
332	&	Fabricated metal products	&	54	&	79	&	-7.50	&	-3.72	&	-3.44	\\
333	&	Machinery	&	118	&	80	&	3.37	&	12.06	&	12.45	\\
334	&	Computer and electronic products	&	327	&	87	&	19.52	&	15.50	&	15.59	\\
335	&	Electrical equipment and appliances	&	44	&	83	&	0.01	&	11.05	&	11.37	\\
336	&	Transportation equipment	&	97	&	81	&	-17.28	&	-14.74	&	-14.40	\\
339	&	Miscellaneous durable goods manufacturing	&	89	&	84	&	-1.32	&	-9.29	&	-9.26	\\
423	&	Wholesale trade: Durable goods	&	58	&	68	&	-3.44	&	2.79	&	3.19	\\
424	&	Wholesale trade: Nondurable goods	&	49	&	71	&	-14.89	&	-13.12	&	-12.80	\\
483	&	Water transportation	&	52	&	26	&	-31.38	&	-24.68	&	-23.87	\\
511	&	Publishing industries, except Internet	&	92	&	84	&	20.02	&	7.47	&	7.48	\\
515	&	Broadcasting, except Internet	&	71	&	65	&	-11.69	&	-9.90	&	-10.10	\\
518	&	Data processing, hosting and related services	&	71	&	81	&	12.38	&	2.19	&	2.73	\\
519	&	Other information services	&	161	&	76	&	11.39	&	-1.19	&	-0.71	\\
523	&	Securities, commodity contracts, investments, and funds and trusts	&	136	&	71	&	-2.62	&	5.68	&	5.70	\\
524	&	Insurance carriers and related activities	&	127	&	72	&	-12.24	&	-11.42	&	-11.57	\\
531	&	Real estate	&	222	&	48	&	-26.41	&	-32.80	&	-32.93	\\
541	&	Professional and technical services	&	104	&	77	&	-3.58	&	-9.49	&	-9.52	\\
561	&	Administrative and support services	&	57	&	65	&	-3.13	&	-8.49	&	-8.56	\\
722	&	Food services and drinking places	&	47	&	47	&	-19.93	&	-24.79	&	-25.38	\\
\bottomrule
\end{tabular}

\end{adjustwidth}
	
\end{spacing}      
  \end{center}
\end{table}

\clearpage
\newpage

\addtolength{\voffset}{+1.75cm}
\addtolength{\footskip}{-1.75cm}

\phantomsection
\addcontentsline{toc}{section}{Internet Appendix} 

\renewcommand{\thefootnote}{\fnsymbol{footnote}}

\begin{appendices}
\begin{center}
{\large Appendix for \vspace{0.25cm}\\
{\Large ``Disaster Resilience and Stock  Returns''} \\\vspace{0.25cm} Marco Pagano \hspace{1cm} Christian Wagner \hspace{1cm} Josef Zechner \vspace{0.25cm}\\
}
\end{center}

\renewcommand{\thepage}{Appendix -- \arabic{page}}
\setcounter{page}{1}

\setcounter{section}{0}
\setcounter{subsection}{0}

\renewcommand{\thesection}{A.\arabic{section}}
\renewcommand{\thesubsection}{\thesection.\arabic{subsection}}

\setcounter{equation}{0}
\renewcommand{\theequation}{\thesection.\arabic{equation}}

\renewcommand\thetable{A.\arabic{table}} \setcounter{table}{0}
\renewcommand\thefigure{A.\arabic{figure}} \setcounter{figure}{0}

\bigskip\bigskip\bigskip
  \noindent  This  Appendix provides additional results referred to in the paper.

\clearpage
\addtolength{\voffset}{-1.75cm}
\addtolength{\footskip}{1.75cm}


\clearpage
\newpage

\begin{figure}\caption{Risk-adjusted returns of stocks with high and low resilience to social distancing (DN) \label{figA1}}\footnotesize

\begin{adjustwidth}{-1.75cm}{-1.75cm}
\begin{spacing}{1.0}
This figure plots the cumulative risk-adjusted returns of portfolios sorted by firms' resilience to disaster risk during the first quarter of 2020. On any given day, we assign a firm to the `High' portfolio if its `teleworkable\_manual\_wage' \citep[as defined by][]{dingel/neiman:20nber} is above the median value and to the `Low' portfolio if it is below. In Panel A, we present CAPM-adjusted returns, i.e. controlling for exposure to market risk. Panels B and C present results controlling for the Fama-French three factor model exposures (i.e. market, size, value) and five factor model exposures (i.e. market, size, value, investments, profitability), respectively. We plot the cumulative value-weighted portfolio returns for the `High' portfolio (in green) and the Low portfolio (in red) as well as the High-Low differential return (in blue). The dashed vertical lines mark February 24, the day after Italy introduced its lockdown, and March 20, the last trading day before the Fed announced its intervention.
\end{spacing}
\end{adjustwidth}

\begin{adjustwidth}{-2cm}{-2cm}\centering
	
\vspace{-2mm}	{Panel A. CAPM-adjusted returns}\\ 
	{\includegraphics[scale = 0.85]{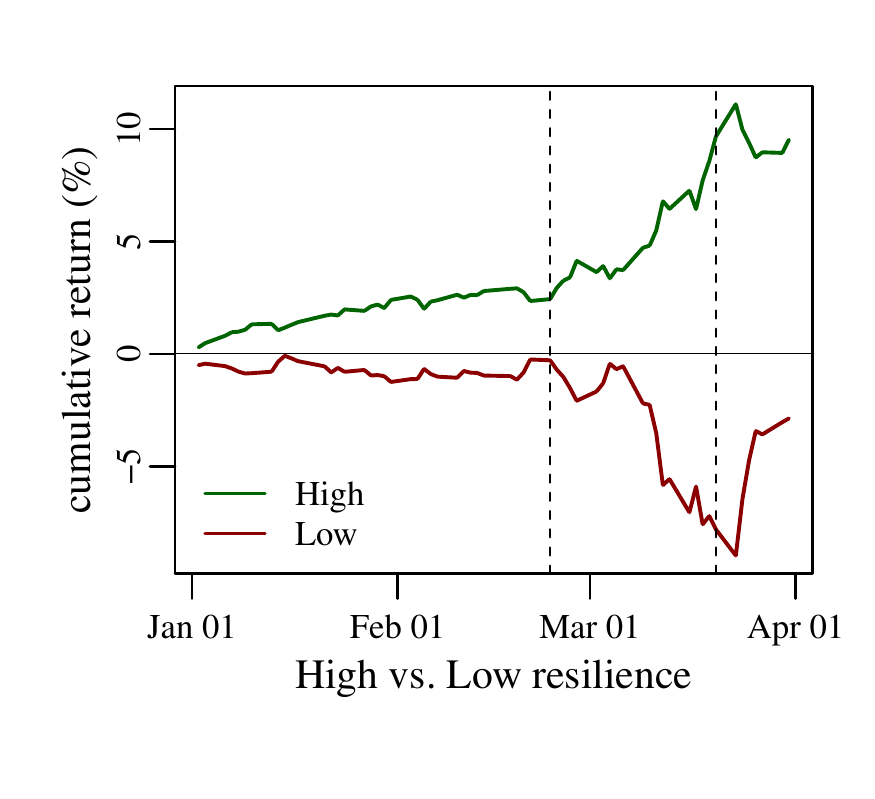}}
	{\includegraphics[scale = 0.85]{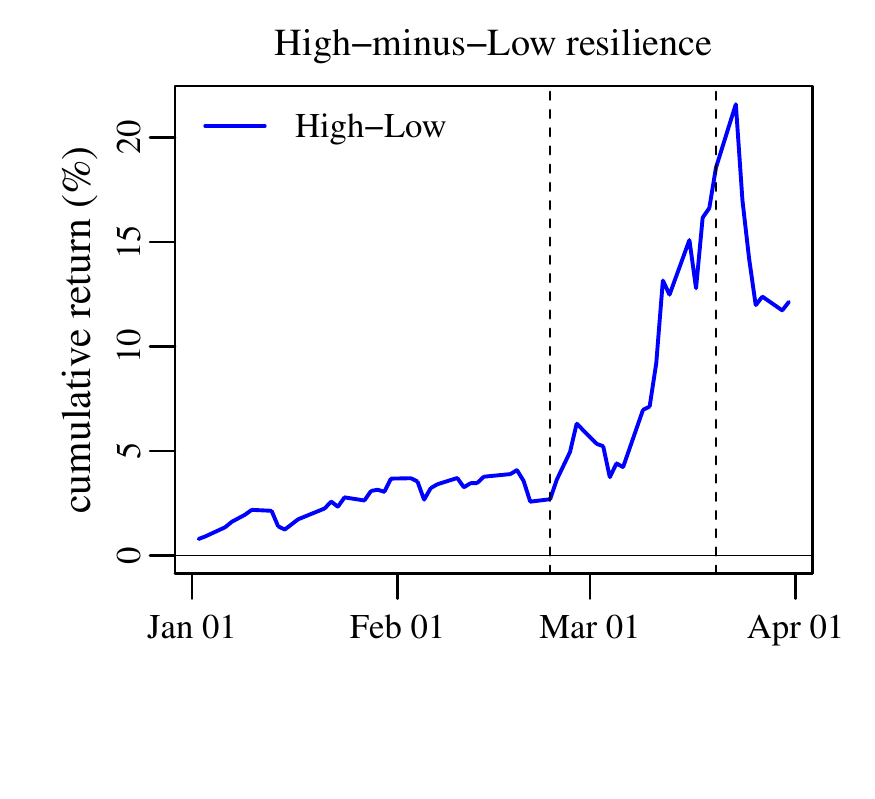}}
	
\vspace{-6mm}
	{Panel B. FF3-adjusted returns}\\ 
	{\includegraphics[scale = 0.85]{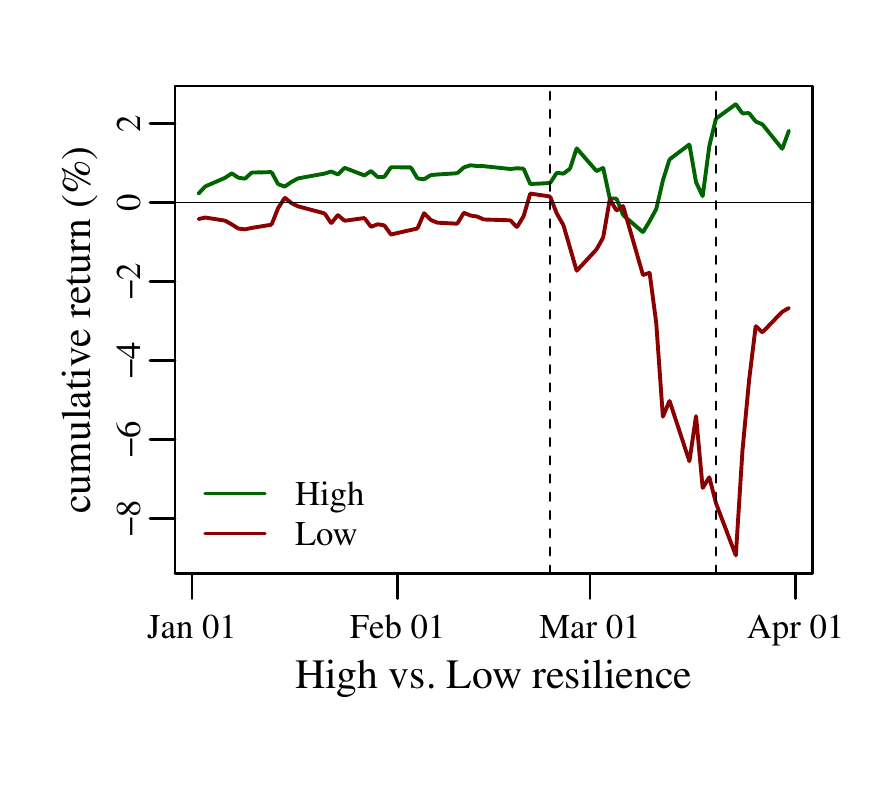}}
	{\includegraphics[scale = 0.85]{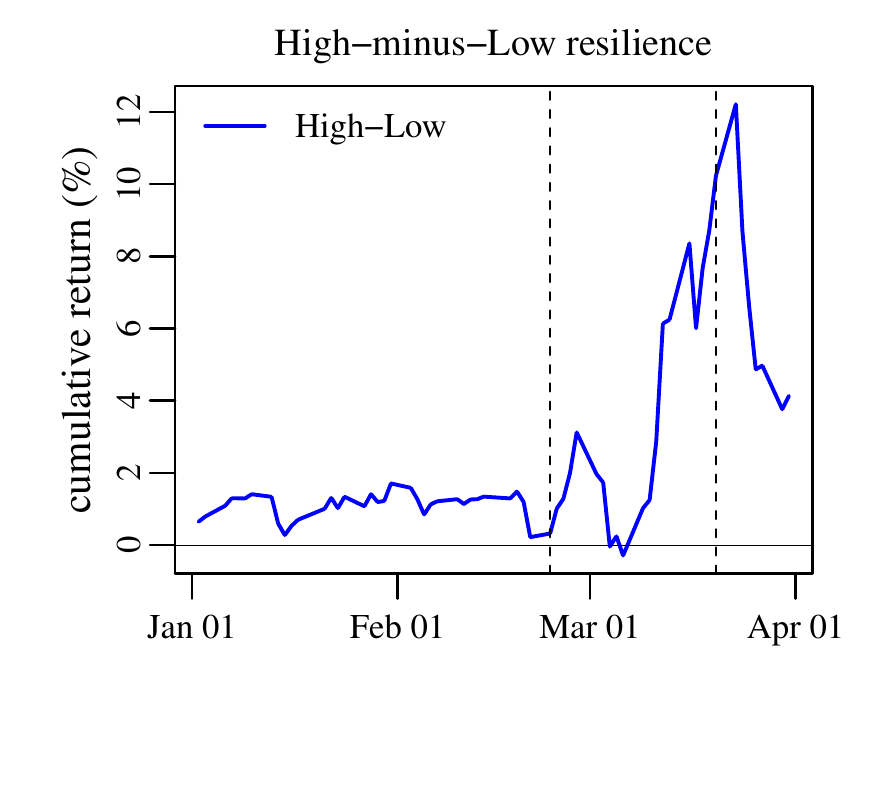}}
	
\vspace{-6mm}
	{Panel C. FF5-adjusted returns}\\ 
	{\includegraphics[scale = 0.85]{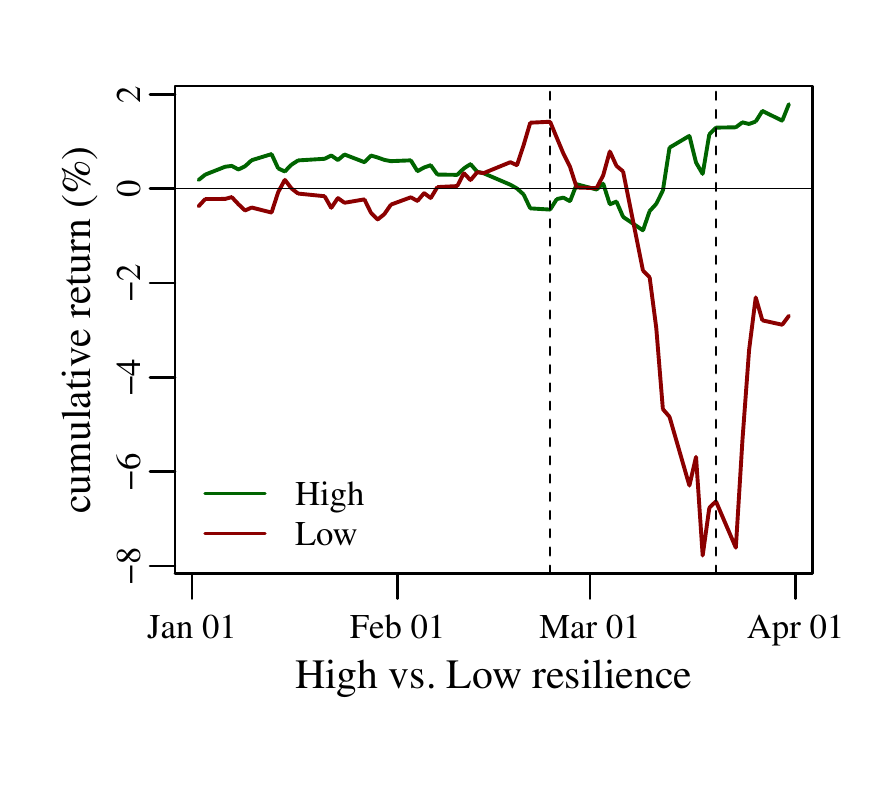}}
	{\includegraphics[scale = 0.85]{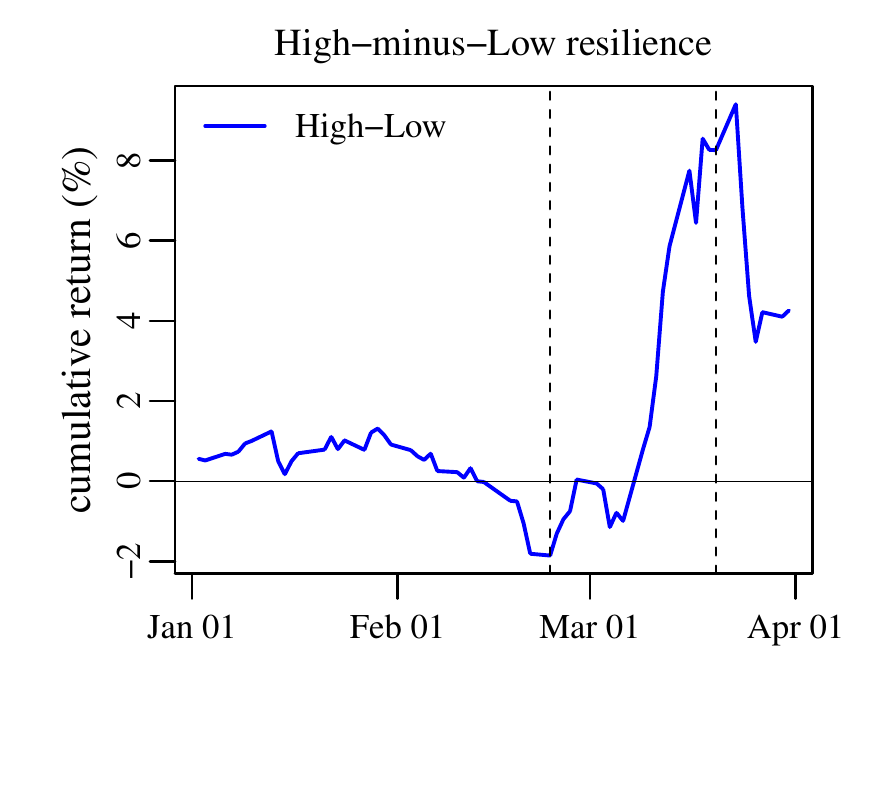}}
\end{adjustwidth}
\end{figure}

\clearpage
\newpage
\begin{figure}\caption{Risk-adjusted returns of stocks with high and low resilience to social distancing (HLRN) \label{figA2}}\footnotesize

\begin{adjustwidth}{-1.75cm}{-1.75cm}
\begin{spacing}{1.0}
This figure plots the cumulative risk-adjusted returns of portfolios sorted by firms' resilience to disaster risk during the first quarter of 2020. On any given day, we assign a firm to the `High' portfolio if its `dur\_workplace' \citep[as defined by][]{hensvik/lebarbanchon/rathelot:20cepr} is below the median value and to the `Low' portfolio if it is above. In Panel A, we present CAPM-adjusted returns, i.e. controlling for exposure to market risk. Panels B and C present results controlling for the Fama-French three factor model exposures (i.e. market, size, value) and five factor model exposures (i.e. market, size, value, investments, profitability), respectively. We plot the cumulative value-weighted portfolio returns for the `High' portfolio (in green) and the Low portfolio (in red) as well as the High-Low differential return (in blue). The dashed vertical lines mark February 24, the day after Italy introduced its lockdown, and March 20, the last trading day before the Fed announced its intervention.
\end{spacing}
\end{adjustwidth}

\begin{adjustwidth}{-2cm}{-2cm}\centering
	
\vspace{-2mm}
	{Panel A. CAPM-adjusted returns}\\ 
	{\includegraphics[scale = 0.85]{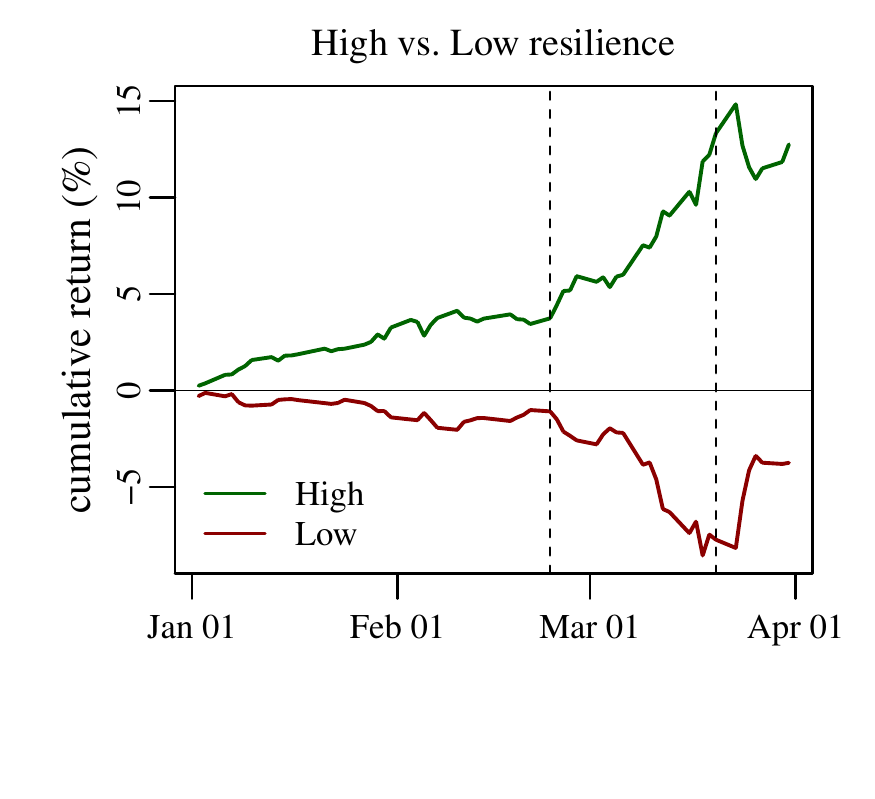}}
	{\includegraphics[scale = 0.85]{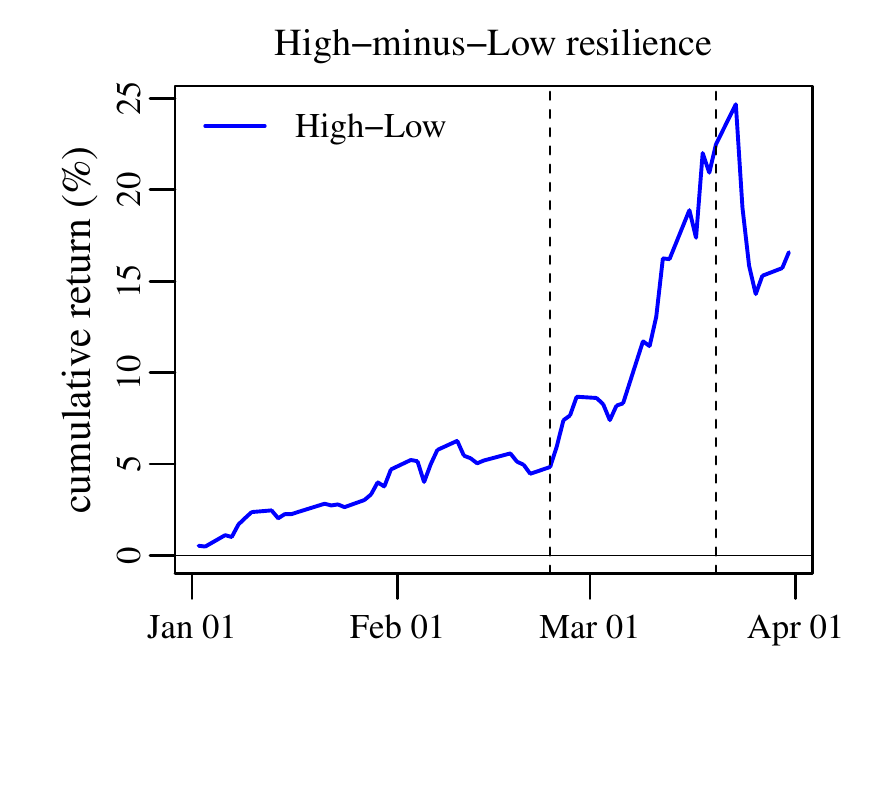}}
	
\vspace{-6mm}
	{Panel B. FF3-adjusted returns}\\ 
	{\includegraphics[scale = 0.85]{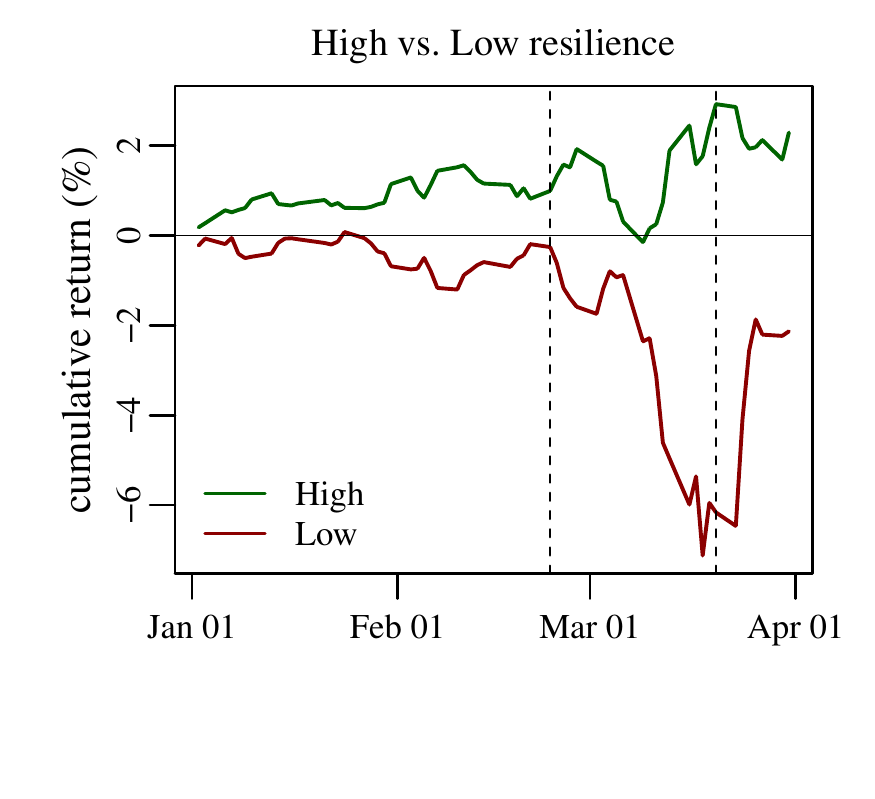}}
	{\includegraphics[scale = 0.85]{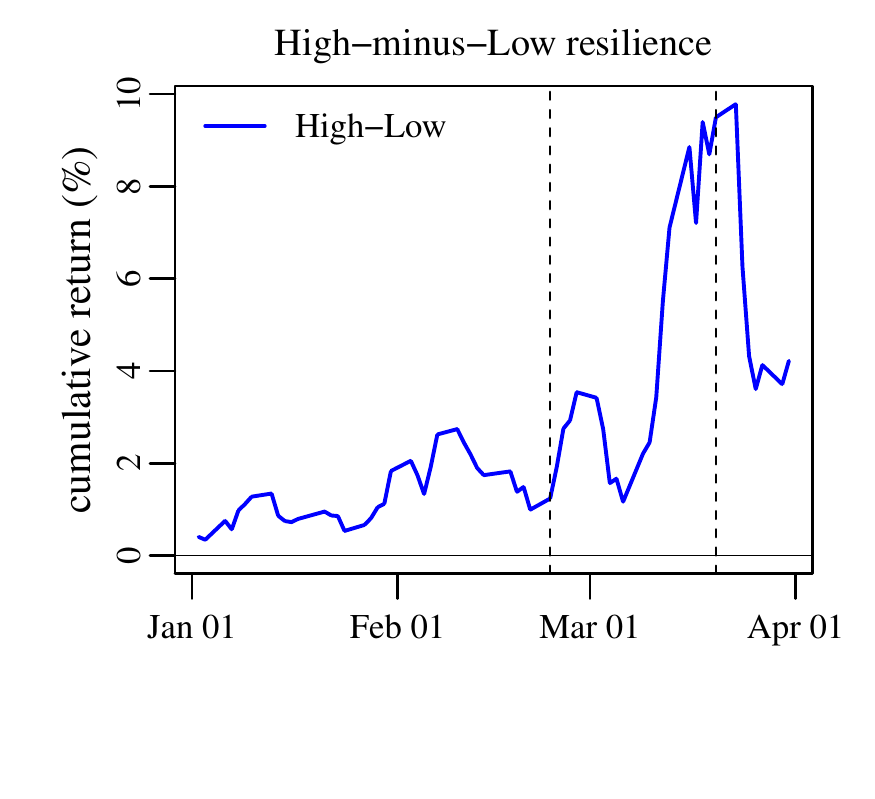}}
	
\vspace{-6mm}
	{Panel C. FF5-adjusted returns}\\ 
	{\includegraphics[scale = 0.85]{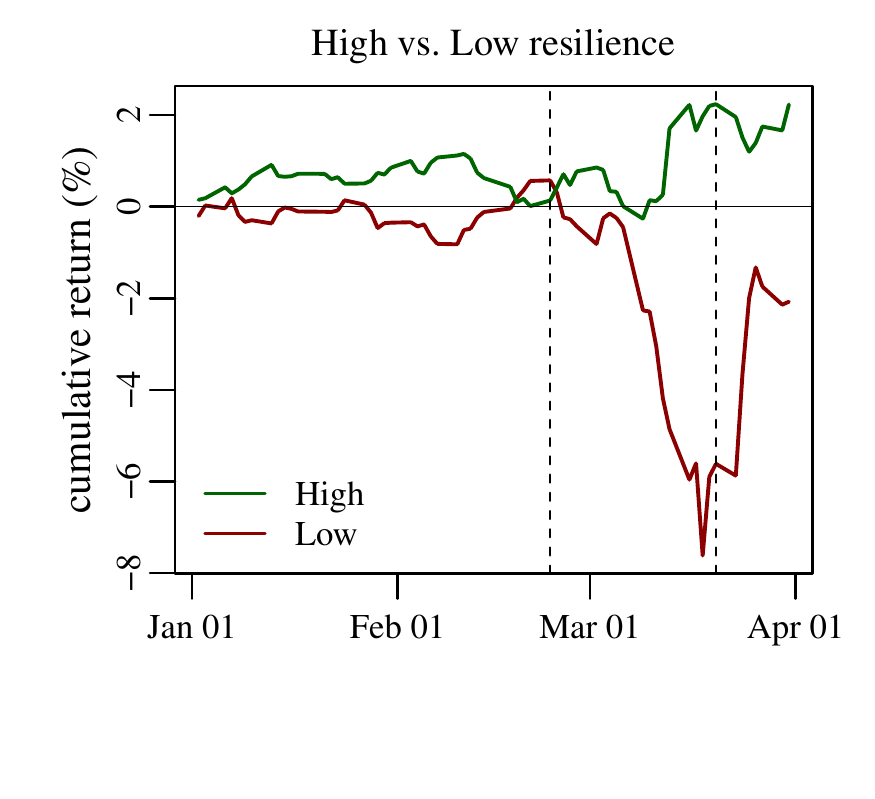}}
	{\includegraphics[scale = 0.85]{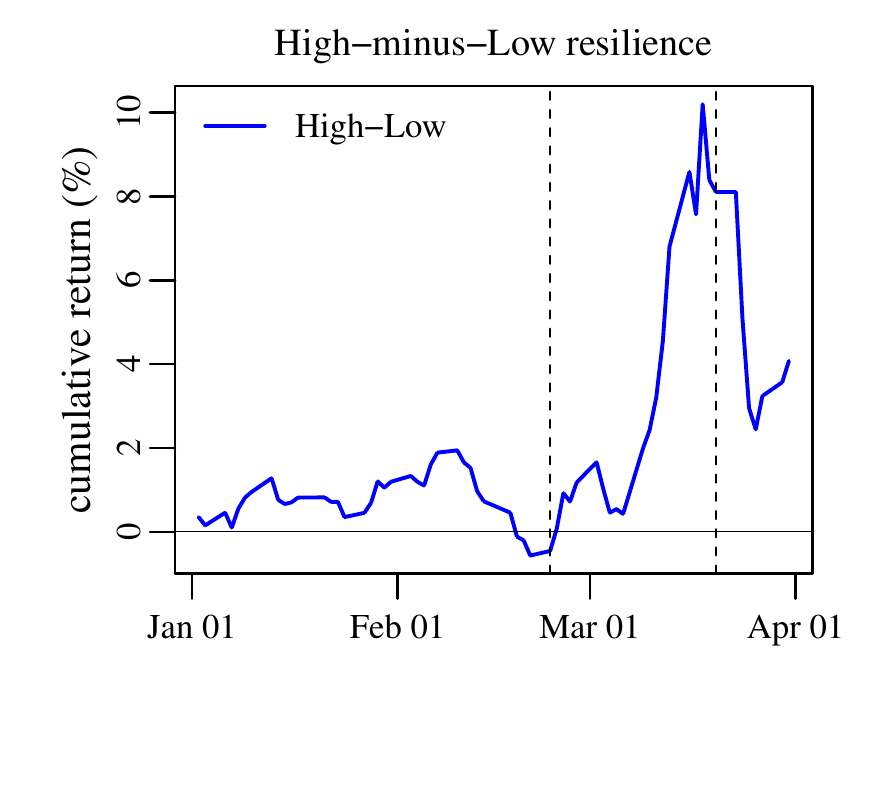}}
\end{adjustwidth}
\end{figure}

\clearpage
\newpage
\begin{figure}\caption{Resilience to social distancing and industry portfolio returns (DN) \label{figA3}}\footnotesize
\begin{adjustwidth}{-1.75cm}{-1.75cm}
\begin{spacing}{1.0}
This figure plots the cumulative risk-adjusted returns of value-weighted industry portfolios against the industries' resilience to disaster risk. The sample period is from February 24 to March 20, 2020, i.e. from the day after Italy introduced its lockdown to the last trading day before the Fed announced its intervention. We define resilience as 100 (\%) minus the `teleworkable\_manual\_wage' defined by \citet{dingel/neiman:20nber} and present results for the 25 industries with the highest number of firms (in total 2,974). In Panel A, we present CAPM-adjusted returns, i.e. controlling for exposure to market risk. Panels B and C present results controlling for the Fama-French three factor model exposures (i.e. market, size, value) and five factor model exposures (i.e. market, size, value, investments, profitability), respectively. The plot labels indicate the industries' 3-digit NAICS codes. The plot legends report results for cross-sectional regressions with $t$-statistics based on \citet{white:80ecta} standard errors in square brackets.
\end{spacing}
\end{adjustwidth}
\bigskip
\begin{adjustwidth}{-2cm}{-2cm}\centering

	{Panel A. CAPM-ajdusted returns}\\ \vspace{-5mm}
	{\includegraphics[scale = 0.75]{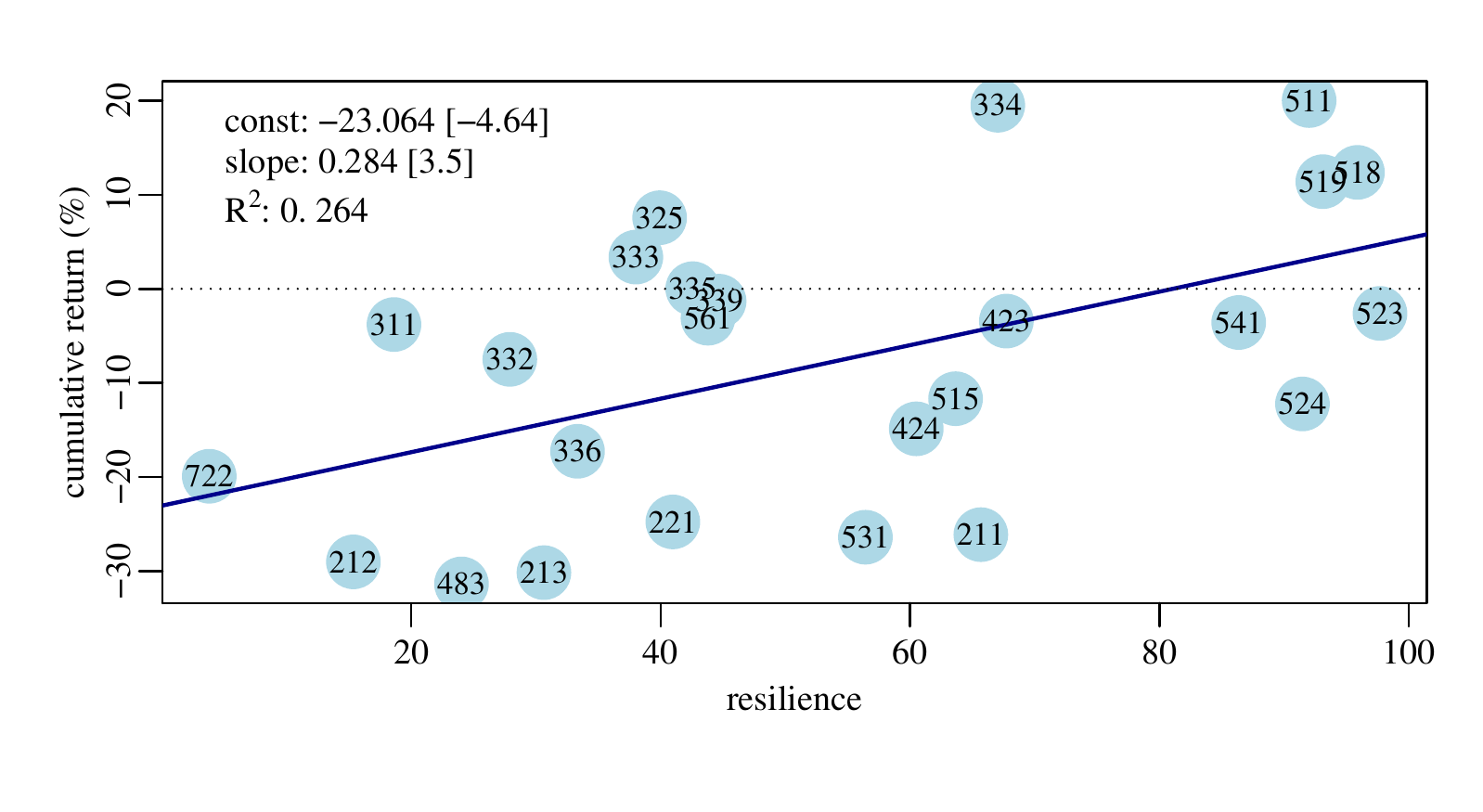}}

	{Panel B. FF3-ajdusted returns}\\ \vspace{-5mm}
	{\includegraphics[scale = 0.75]{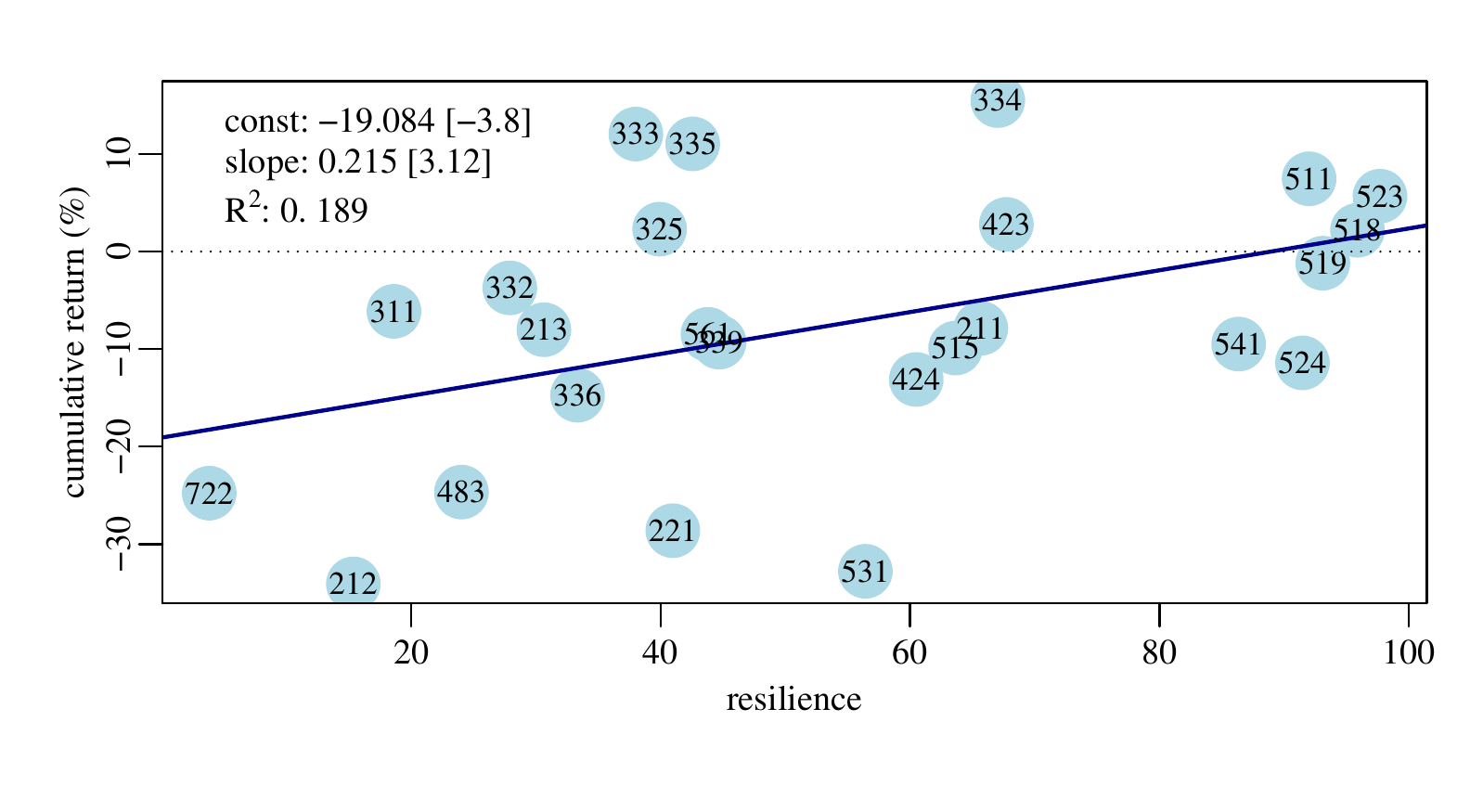}}

	{Panel C. FF5-ajdusted returns}\\ \vspace{-5mm}
	{\includegraphics[scale = 0.75]{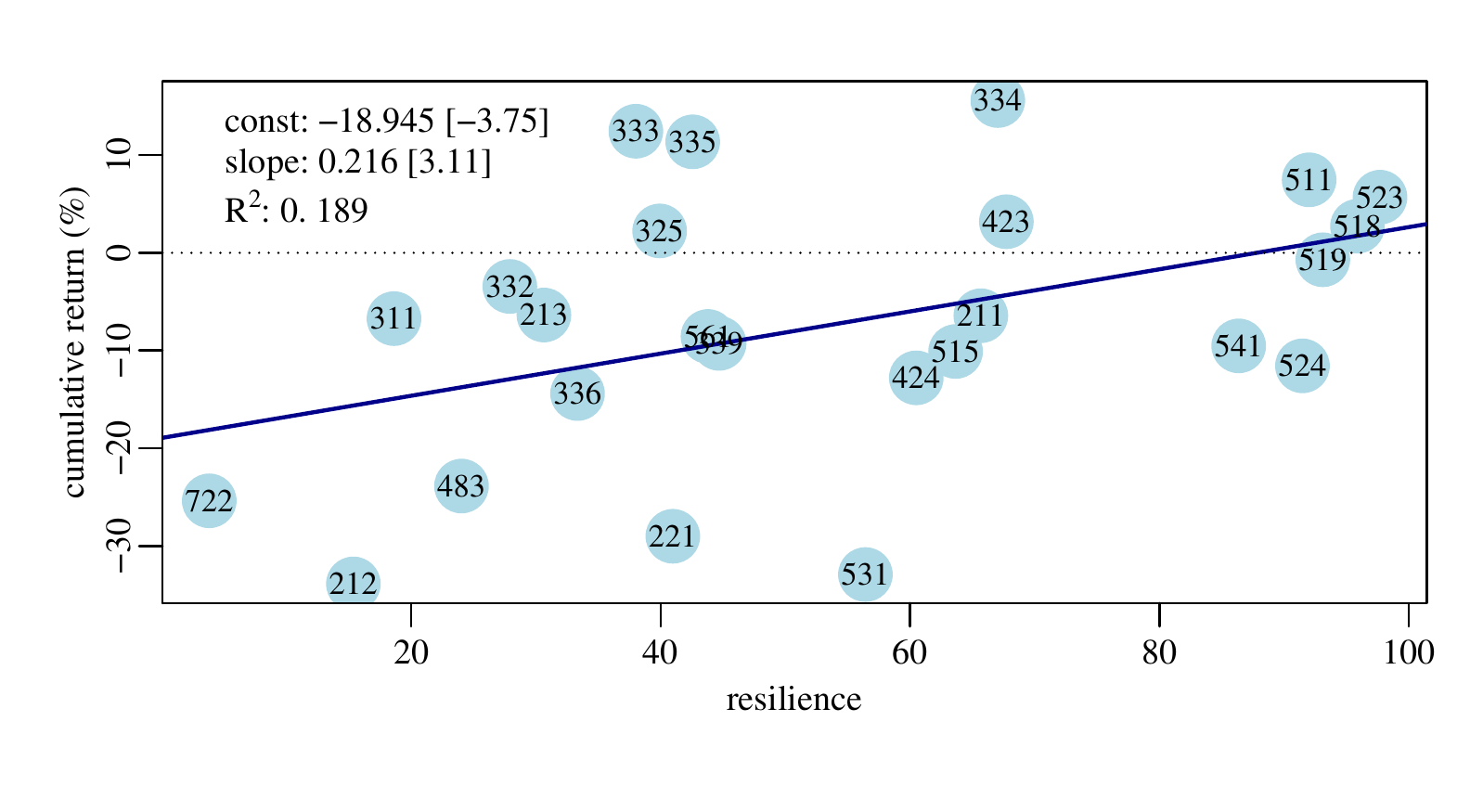}}

\end{adjustwidth}
\end{figure}


\clearpage
\newpage

\addtolength{\voffset}{-0.2cm}
\addtolength{\footskip}{0.75cm}

\begin{figure}\caption{Stock options-implied risk-neutral variances of S\&P 500 firms}\label{figA4}\footnotesize

\vspace{-3mm}

\begin{adjustwidth}{-1.75cm}{-1.75cm}
This figure plots stock options-implied risk-neutral variance indices for S\&P 500 firms with high and low resilience to social distancing during the first quarter of 2020. On any given day, we assign a firm to the high resilience index, $\overline{\SVS}^2_{H,t}$, if its `affected\_share' \citep[as defined by][]{koren/peto:20covid} is below the median value and to the low resilience index, $\overline{\SVS}^2_{L,t}$, if it is above. The indices are computed as the value-weighted sums of individual firms' risk-neutral variances, $\SVS^2_{i,t}$. The difference $\overline{\SVS}^2_{L,t} - \overline{\SVS}^2_{L,t}$ measures the expected return of low resilience in excess of high resilience stocks. Panels A to D present results using options maturities of 30, 91, 365 and 730 days, respectively. The dashed vertical lines mark February 24 and March 20. 

\begin{spacing}{1.0}
\end{spacing}

\end{adjustwidth}

\begin{adjustwidth}{-2cm}{-2cm}\centering

\vspace{-6mm}

	{Panel A. 30-day horizon}\\ \vspace{-1.5mm}
	{\includegraphics[scale = 0.775]{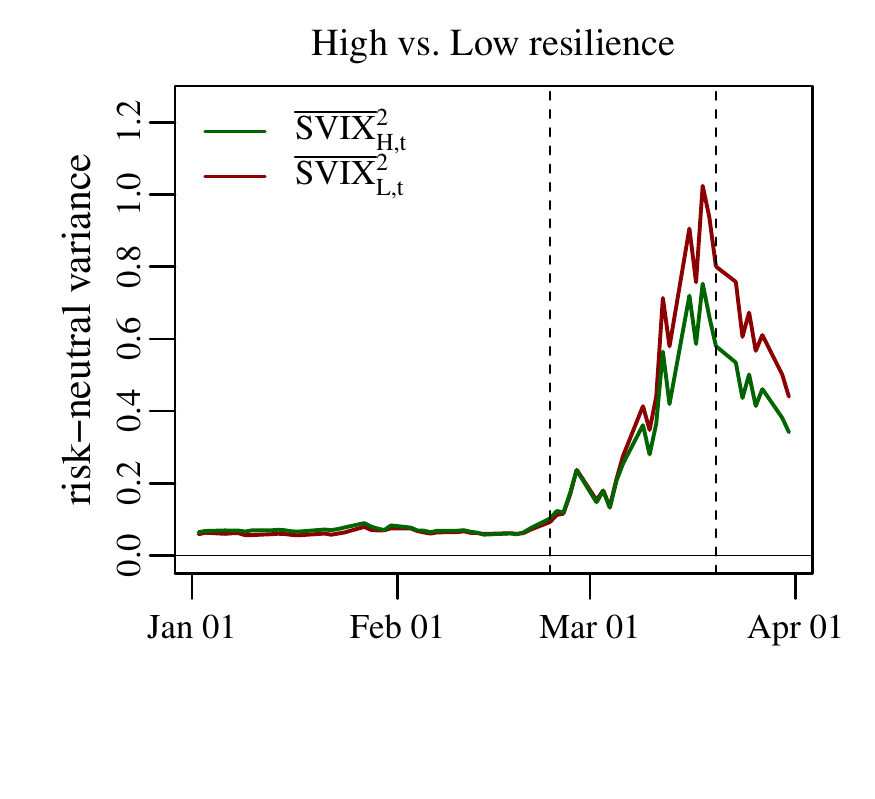}}
	{\includegraphics[scale = 0.775]{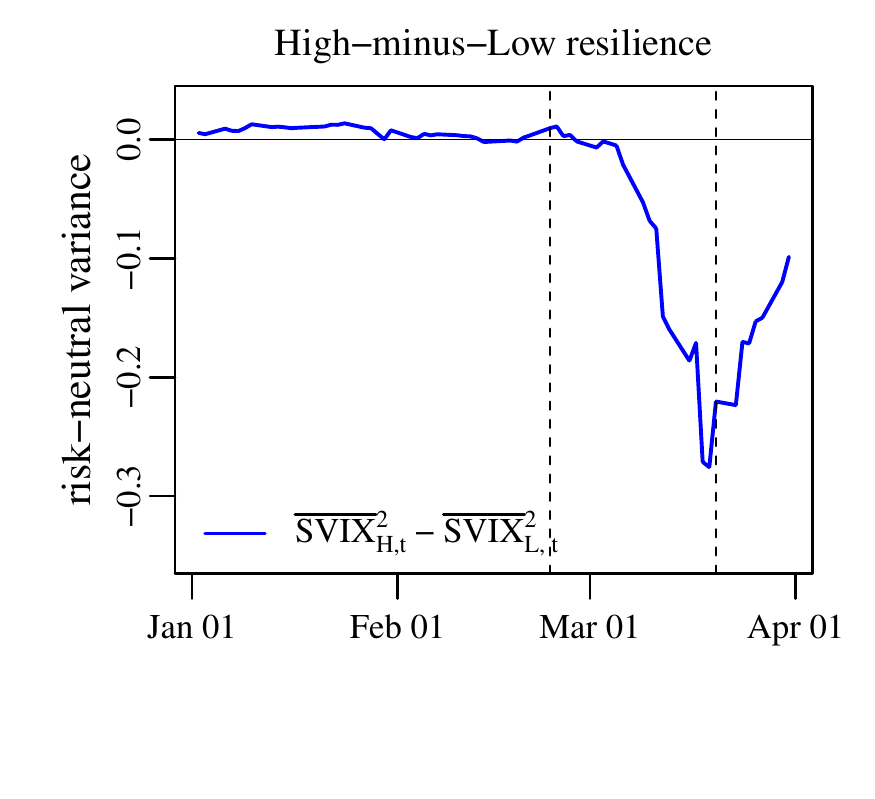}}

\vspace{-11mm}

	{Panel B. 91-day horizon}\\ \vspace{-1.5mm}
	{\includegraphics[scale = 0.775]{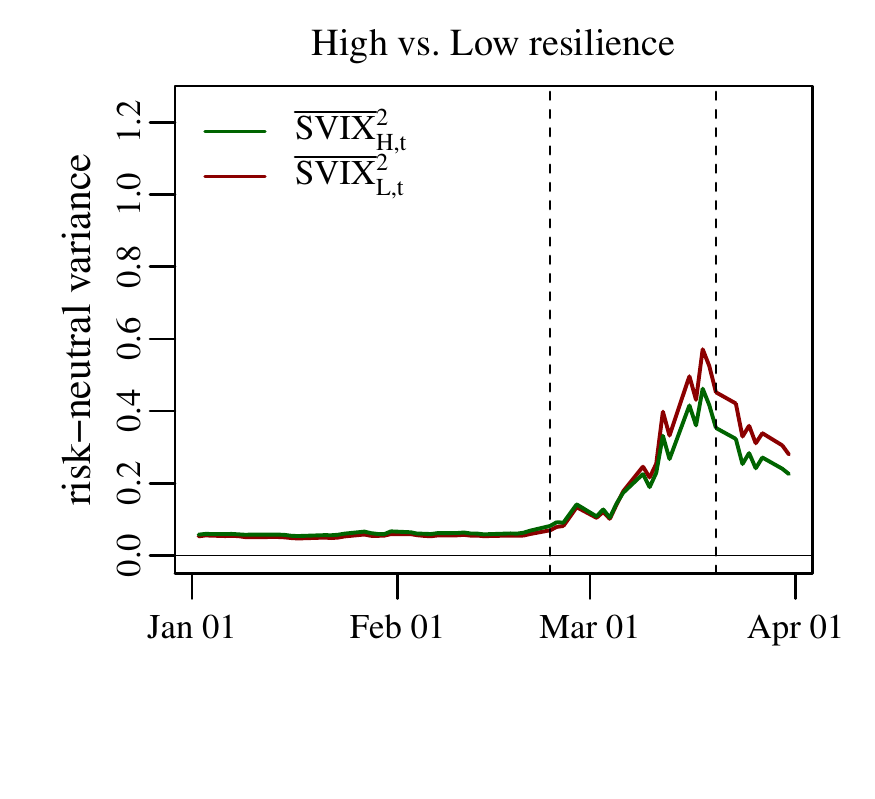}}
	{\includegraphics[scale = 0.775]{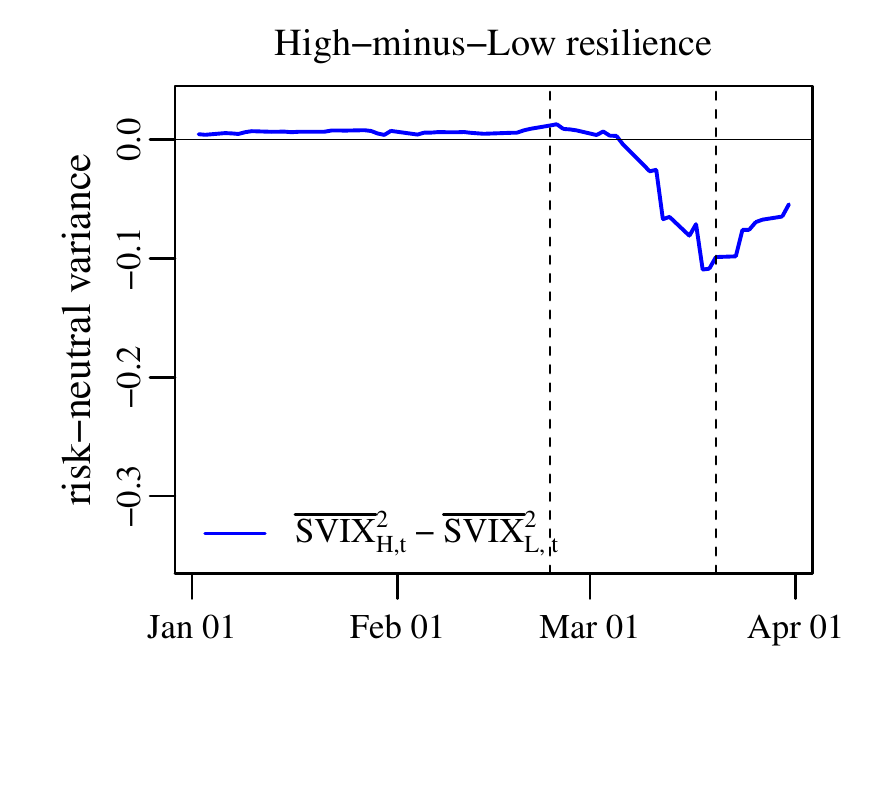}}

\vspace{-11mm}
	{Panel C. 365-day horizon}\\ \vspace{-1.5mm}
	{\includegraphics[scale = 0.775]{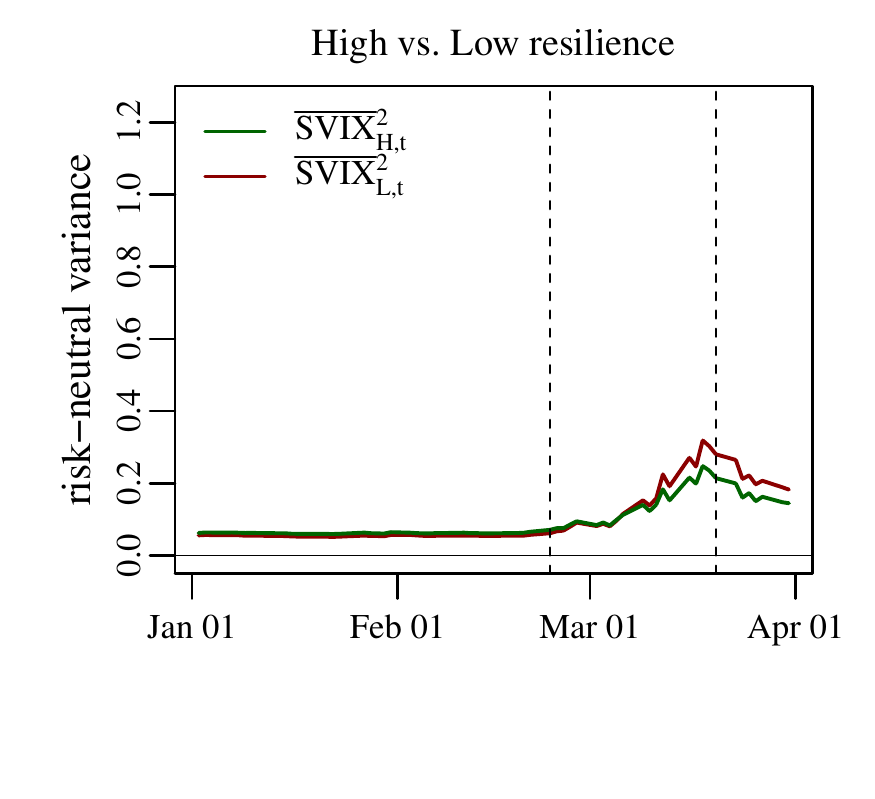}}
	{\includegraphics[scale = 0.775]{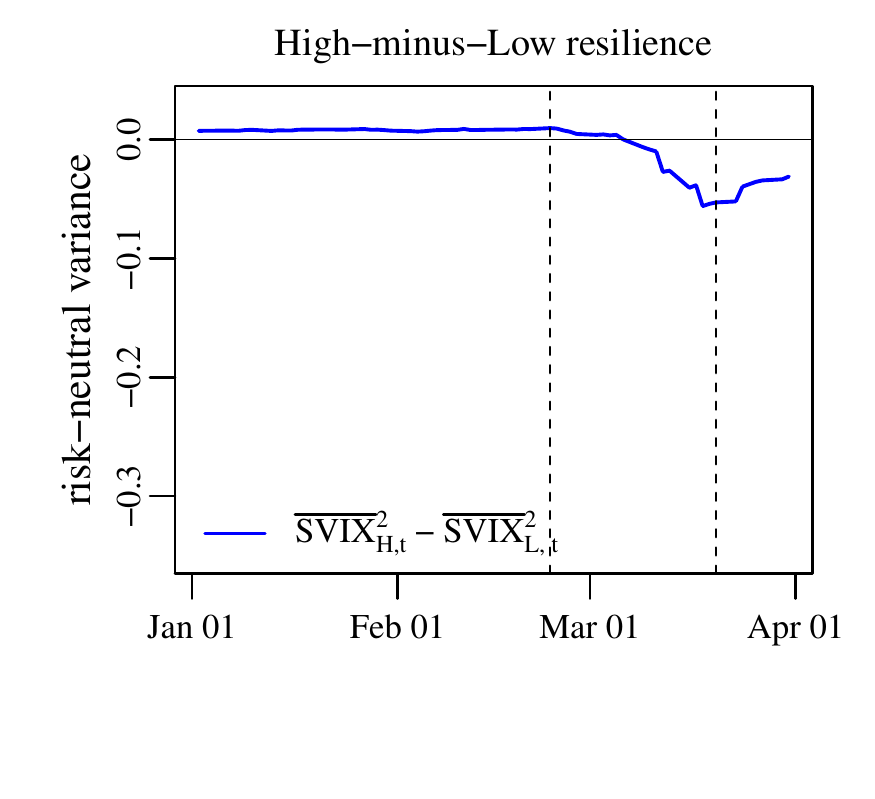}}

\vspace{-11mm}
	{Panel D. 730-day horizon}\\ \vspace{-1.5mm}
	{\includegraphics[scale = 0.775]{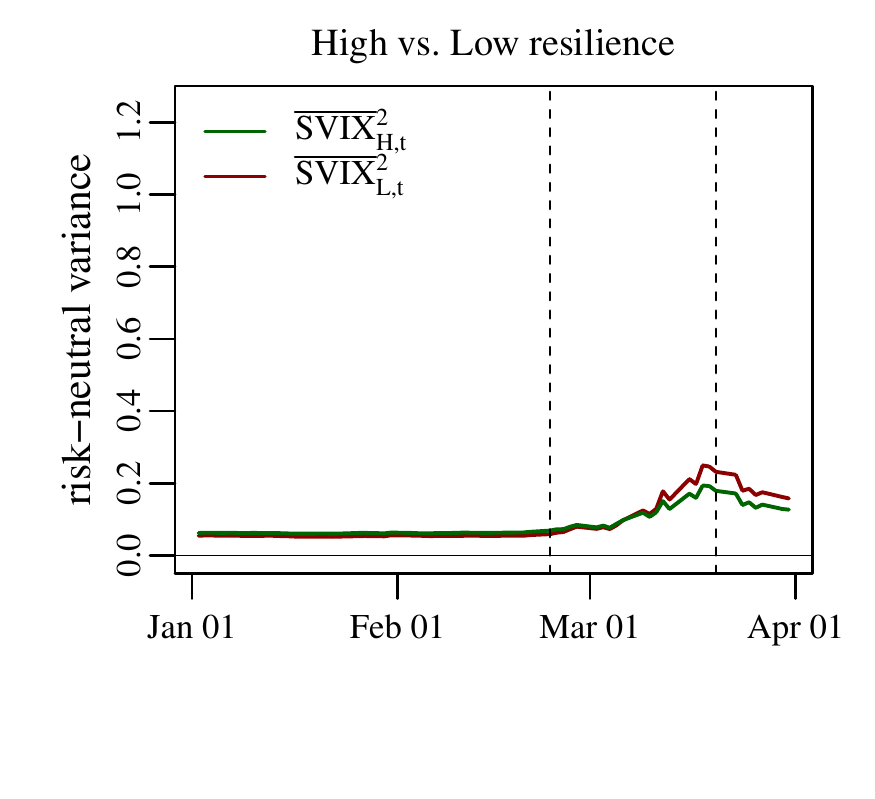}}
	{\includegraphics[scale = 0.775]{figures/figure_HLorLH---replaceTRFDNAwith1---01dec2016_03may2020---balanced_US_KP_affected_share_sp500firms_SVIXi730days_yscalefixed_timeseries_2020q1.pdf}}

\end{adjustwidth}
\end{figure}


\clearpage
\newpage

\addtolength{\voffset}{+0.2cm}
\addtolength{\footskip}{-0.75cm}

\begin{table}
\caption{Measures of teleworkability, working at home and at the workplace, and business face-to-face interactions}\label{tab:empirical.measures} \label{tabA1}
\footnotesize

\begin{adjustwidth}{-1.75cm}{-1.75cm}
\begin{spacing}{1.0}
This Table provides an overview of the empirical measures on which we base our analysis of stocks' disaster resilience. Panel A lists the teleworkability measures provided by  \citet{dingel/neiman:20nber} for 24 industries at the NAICS 2-digit level and for 88 industries at the NAICS 3-digit level. Panel B lists the work at home and work at the workplace measures provided by \citet{hensvik/lebarbanchon/rathelot:20cepr} for 310 industries at the NAICS 4-digit level. Panel C lists the communication-intensity and physical proximity measures suggested by \citet{koren/peto:20covid} for 84 industries at the NAICS 3-digit level.

\end{spacing}
\end{adjustwidth}

\bigskip

\begin{center}

Panel A. \citet{dingel/neiman:20nber}: \\
\begin{tabular}{L{5cm}L{10cm}}
\toprule
`teleworkable\_emp' & fraction of jobs that can be done from home estimated from  O*Net data\\
`teleworkable\_wage' & fraction of wages to jobs that can be done from home estimated from O*Net data \\
\midrule
`teleworkable\_manual\_emp' & fraction of jobs that can be done from home based on manual classification by the authors\\
`teleworkable\_manual\_wage' & fraction of wages to jobs that can be done from home based on manual classification by the authors\\
\bottomrule
\end{tabular}

\bigskip

Panel B. \citet{hensvik/lebarbanchon/rathelot:20cepr} \\
\begin{tabular}{L{5cm}L{10cm}}
\toprule
`home' & fraction of respondents that work at home\\
`workplace' & fraction of respondents that work at workplace\\
\midrule
`dur\_home' & hours worked at home per day\\
`dur\_workplace' & hours worked at workplace per day\\
\midrule
`share\_home' & hours worked at home divided by hours worked at home and at workplace \\
\bottomrule
\end{tabular}

\bigskip

Panel C. \citet{koren/peto:20covid} \\
\begin{tabular}{L{5cm}L{10cm}}
\toprule
`teamwork\_share' & percentage of workers in teamwork-intensive occupations, i.e. internal communication \\
`customer\_share' & percentage of workers in customer-facing occupations, i.e. external communication \\
`communication\_share' & percentage of workers in teamwork-intensive and/or  customer-facing occupations\\
\midrule
`presence\_share' & percentage of workers whose jobs require physical presence in close proximity to others \\
\midrule
`affected\_share' & percentage of workers in occupations that are communication-intensive and/or require physical presence in close proximity to others\\
\bottomrule
\end{tabular}

\end{center}

\end{table}

\clearpage
\newpage
\begin{table}[tbp]
    \caption{Other measures of \citet{koren/peto:20covid}  } \label{tabA2}
\footnotesize
    \bigskip
\begin{adjustwidth}{-1.75cm}{-1.75cm}
\begin{spacing}{1.0}
This table presents results analogous to Panel A in Table \ref{tab2} but using the other measures constructed by \citet{koren/peto:20covid} instead of their `affected\_share' variable. 

\end{spacing}
\end{adjustwidth}    
\bigskip
  \begin{center}  \footnotesize
  \begin{spacing}{1.0}

	Panel A. Portfolios sorted by `teamwork\_share'  \\ \vspace{1mm}
	\begin{tabular}{L{5cm}d{2.2}d{2.2}d{2.2}d{2.2}d{2.2}d{2.2}} \toprule   &  \multicolumn{1}{C{1.4cm}}{ ret }  &   \multicolumn{1}{C{1.4cm}}{ capm }  &   \multicolumn{1}{C{1.4cm}}{ ff3 }  &   \multicolumn{1}{C{1.4cm}}{ ff4 }  &   \multicolumn{1}{C{1.4cm}}{ ff5 }  &   \multicolumn{1}{C{1.4cm}}{ ff6 }  \\ 
\midrule High resilience  &  -1.67^{***}  &  0.30^{***}  &  0.09  &  0.05  &  0.10  &  0.07  \\ 
  &  \multicolumn{ 1}{c}{[-3.79]}  &  \multicolumn{ 1}{c}{[2.72]}  &  \multicolumn{ 1}{c}{[0.75]}  &  \multicolumn{ 1}{c}{[0.44]}  &  \multicolumn{ 1}{c}{[1.48]}  &  \multicolumn{ 1}{c}{[1.58]}  \\ 
Low resilience  &  -1.86^{***}  &  -0.51^{**}  &  -0.59^{**}  &  -0.57^{**}  &  -0.59^{**}  &  -0.58^{**}  \\ 
  &  \multicolumn{ 1}{c}{[-3.38]}  &  \multicolumn{ 1}{c}{[-2.00]}  &  \multicolumn{ 1}{c}{[-2.42]}  &  \multicolumn{ 1}{c}{[-2.35]}  &  \multicolumn{ 1}{c}{[-2.37]}  &  \multicolumn{ 1}{c}{[-2.33]}  \\ 
\midrule High-minus-Low  &  0.19  &  0.81^{***}  &  0.68^{**}  &  0.61^{**}  &  0.69^{**}  &  0.66^{**}  \\ 
  &  \multicolumn{ 1}{c}{[1.01]}  &  \multicolumn{ 1}{c}{[2.58]}  &  \multicolumn{ 1}{c}{[2.21]}  &  \multicolumn{ 1}{c}{[2.06]}  &  \multicolumn{ 1}{c}{[2.35]}  &  \multicolumn{ 1}{c}{[2.27]}  \\  \bottomrule \end{tabular}

	\bigskip
	
	Panel B. Portfolios sorted by `customer\_share' \\ \vspace{1mm}
	\begin{tabular}{L{5cm}d{2.2}d{2.2}d{2.2}d{2.2}d{2.2}d{2.2}} \toprule   &  \multicolumn{1}{C{1.4cm}}{ ret }  &   \multicolumn{1}{C{1.4cm}}{ capm }  &   \multicolumn{1}{C{1.4cm}}{ ff3 }  &   \multicolumn{1}{C{1.4cm}}{ ff4 }  &   \multicolumn{1}{C{1.4cm}}{ ff5 }  &   \multicolumn{1}{C{1.4cm}}{ ff6 }  \\ 
\midrule High resilience  &  -1.60^{***}  &  0.21^{***}  &  0.16^{***}  &  0.02  &  0.16^{***}  &  -0.01  \\ 
  &  \multicolumn{ 1}{c}{[-3.33]}  &  \multicolumn{ 1}{c}{[4.51]}  &  \multicolumn{ 1}{c}{[4.10]}  &  \multicolumn{ 1}{c}{[0.26]}  &  \multicolumn{ 1}{c}{[3.24]}  &  \multicolumn{ 1}{c}{[-0.14]}  \\ 
Low resilience  &  -1.68^{***}  &  -0.05  &  -0.29^{***}  &  -0.20^{***}  &  -0.30^{***}  &  -0.18^{***}  \\ 
  &  \multicolumn{ 1}{c}{[-3.36]}  &  \multicolumn{ 1}{c}{[-0.74]}  &  \multicolumn{ 1}{c}{[-3.49]}  &  \multicolumn{ 1}{c}{[-2.87]}  &  \multicolumn{ 1}{c}{[-4.14]}  &  \multicolumn{ 1}{c}{[-3.94]}  \\ 
\midrule High-minus-Low  &  0.08  &  0.26^{***}  &  0.45^{***}  &  0.22^{***}  &  0.46^{***}  &  0.17^{***}  \\ 
  &  \multicolumn{ 1}{c}{[0.96]}  &  \multicolumn{ 1}{c}{[15.17]}  &  \multicolumn{ 1}{c}{[6.51]}  &  \multicolumn{ 1}{c}{[2.69]}  &  \multicolumn{ 1}{c}{[9.94]}  &  \multicolumn{ 1}{c}{[5.07]}  \\  \bottomrule \end{tabular}

	\bigskip
		
	Panel C. Portfolios sorted by `communication\_share' \\ \vspace{1mm}
	\begin{tabular}{L{5cm}d{2.2}d{2.2}d{2.2}d{2.2}d{2.2}d{2.2}} \toprule   &  \multicolumn{1}{C{1.4cm}}{ ret }  &   \multicolumn{1}{C{1.4cm}}{ capm }  &   \multicolumn{1}{C{1.4cm}}{ ff3 }  &   \multicolumn{1}{C{1.4cm}}{ ff4 }  &   \multicolumn{1}{C{1.4cm}}{ ff5 }  &   \multicolumn{1}{C{1.4cm}}{ ff6 }  \\ 
\midrule High resilience  &  -1.59^{***}  &  0.23^{***}  &  0.17^{***}  &  0.05  &  0.17^{***}  &  0.01  \\ 
  &  \multicolumn{ 1}{c}{[-3.39]}  &  \multicolumn{ 1}{c}{[5.46]}  &  \multicolumn{ 1}{c}{[15.28]}  &  \multicolumn{ 1}{c}{[0.89]}  &  \multicolumn{ 1}{c}{[3.61]}  &  \multicolumn{ 1}{c}{[0.19]}  \\ 
Low resilience  &  -1.77^{***}  &  -0.20^{**}  &  -0.40^{***}  &  -0.30^{***}  &  -0.40^{***}  &  -0.29^{***}  \\ 
  &  \multicolumn{ 1}{c}{[-3.63]}  &  \multicolumn{ 1}{c}{[-2.06]}  &  \multicolumn{ 1}{c}{[-3.74]}  &  \multicolumn{ 1}{c}{[-3.67]}  &  \multicolumn{ 1}{c}{[-3.85]}  &  \multicolumn{ 1}{c}{[-3.96]}  \\ 
\midrule High-minus-Low  &  0.18^{**}  &  0.43^{***}  &  0.56^{***}  &  0.35^{***}  &  0.57^{***}  &  0.30^{***}  \\ 
  &  \multicolumn{ 1}{c}{[2.02]}  &  \multicolumn{ 1}{c}{[5.62]}  &  \multicolumn{ 1}{c}{[5.61]}  &  \multicolumn{ 1}{c}{[6.72]}  &  \multicolumn{ 1}{c}{[6.14]}  &  \multicolumn{ 1}{c}{[12.44]}  \\  \bottomrule \end{tabular}

	\bigskip
		
	Panel D. Portfolios sorted by `presence\_share' \\ \vspace{1mm}
	\begin{tabular}{L{5cm}d{2.2}d{2.2}d{2.2}d{2.2}d{2.2}d{2.2}} \toprule   &  \multicolumn{1}{C{1.4cm}}{ ret }  &   \multicolumn{1}{C{1.4cm}}{ capm }  &   \multicolumn{1}{C{1.4cm}}{ ff3 }  &   \multicolumn{1}{C{1.4cm}}{ ff4 }  &   \multicolumn{1}{C{1.4cm}}{ ff5 }  &   \multicolumn{1}{C{1.4cm}}{ ff6 }  \\ 
\midrule High resilience  &  -1.57^{***}  &  0.41^{***}  &  0.16  &  0.11  &  0.16^{**}  &  0.14^{***}  \\ 
  &  \multicolumn{ 1}{c}{[-3.63]}  &  \multicolumn{ 1}{c}{[3.09]}  &  \multicolumn{ 1}{c}{[1.07]}  &  \multicolumn{ 1}{c}{[0.85]}  &  \multicolumn{ 1}{c}{[2.49]}  &  \multicolumn{ 1}{c}{[2.80]}  \\ 
Low resilience  &  -2.06^{***}  &  -0.79^{***}  &  -0.74^{***}  &  -0.68^{***}  &  -0.75^{***}  &  -0.73^{***}  \\ 
  &  \multicolumn{ 1}{c}{[-3.78]}  &  \multicolumn{ 1}{c}{[-3.31]}  &  \multicolumn{ 1}{c}{[-3.05]}  &  \multicolumn{ 1}{c}{[-2.98]}  &  \multicolumn{ 1}{c}{[-3.16]}  &  \multicolumn{ 1}{c}{[-3.14]}  \\ 
\midrule High-minus-Low  &  0.49^{***}  &  1.20^{***}  &  0.90^{***}  &  0.79^{**}  &  0.92^{***}  &  0.87^{***}  \\ 
  &  \multicolumn{ 1}{c}{[2.90]}  &  \multicolumn{ 1}{c}{[3.57]}  &  \multicolumn{ 1}{c}{[2.68]}  &  \multicolumn{ 1}{c}{[2.50]}  &  \multicolumn{ 1}{c}{[3.27]}  &  \multicolumn{ 1}{c}{[3.23]}  \\  \bottomrule \end{tabular}

\end{spacing}      
  \end{center}
\end{table}

\clearpage
\newpage
\begin{table}[tbp]
    \caption{Other measures of \citet{dingel/neiman:20nber}  } \label{tabA3} \footnotesize

\begin{adjustwidth}{-1.75cm}{-1.75cm}
\begin{spacing}{1.0}
This table presents results analogous to Panel B in Table \ref{tab2} but using the other measures constructed by \citet{dingel/neiman:20nber} instead of their `teleworkable\_manual\_wage' variable. 

\end{spacing}
\end{adjustwidth}  

\bigskip
  \begin{center}  \footnotesize
  \begin{spacing}{1.0}
  	
	Panel A. Portfolios sorted by ``teleworkable\_wage'' \\ \vspace{1mm}
	\begin{tabular}{L{5cm}d{2.2}d{2.2}d{2.2}d{2.2}d{2.2}d{2.2}} \toprule   &  \multicolumn{1}{C{1.4cm}}{ ret }  &   \multicolumn{1}{C{1.4cm}}{ capm }  &   \multicolumn{1}{C{1.4cm}}{ ff3 }  &   \multicolumn{1}{C{1.4cm}}{ ff4 }  &   \multicolumn{1}{C{1.4cm}}{ ff5 }  &   \multicolumn{1}{C{1.4cm}}{ ff6 }  \\ 
\midrule High resilience  &  -1.65^{***}  &  0.29^{***}  &  0.05  &  -0.01  &  0.05  &  0.02  \\ 
  &  \multicolumn{ 1}{c}{[-3.67]}  &  \multicolumn{ 1}{c}{[2.80]}  &  \multicolumn{ 1}{c}{[0.40]}  &  \multicolumn{ 1}{c}{[-0.12]}  &  \multicolumn{ 1}{c}{[1.46]}  &  \multicolumn{ 1}{c}{[0.65]}  \\ 
Low resilience  &  -1.66^{***}  &  -0.31  &  -0.34^{*}  &  -0.26  &  -0.35^{**}  &  -0.31^{*}  \\ 
  &  \multicolumn{ 1}{c}{[-3.24]}  &  \multicolumn{ 1}{c}{[-1.59]}  &  \multicolumn{ 1}{c}{[-1.77]}  &  \multicolumn{ 1}{c}{[-1.39]}  &  \multicolumn{ 1}{c}{[-2.08]}  &  \multicolumn{ 1}{c}{[-1.89]}  \\ 
\midrule High-minus-Low  &  0.01  &  0.60^{**}  &  0.39  &  0.24  &  0.41^{**}  &  0.33^{*}  \\ 
  &  \multicolumn{ 1}{c}{[0.06]}  &  \multicolumn{ 1}{c}{[2.20]}  &  \multicolumn{ 1}{c}{[1.36]}  &  \multicolumn{ 1}{c}{[0.81]}  &  \multicolumn{ 1}{c}{[1.99]}  &  \multicolumn{ 1}{c}{[1.67]}  \\  \bottomrule \end{tabular}
	
	\bigskip

	Panel B. Portfolios sorted by ``teleworkable\_manual\_emp''\\ \vspace{1mm}
	\begin{tabular}{L{5cm}d{2.2}d{2.2}d{2.2}d{2.2}d{2.2}d{2.2}} \toprule   &  \multicolumn{1}{C{1.4cm}}{ ret }  &   \multicolumn{1}{C{1.4cm}}{ capm }  &   \multicolumn{1}{C{1.4cm}}{ ff3 }  &   \multicolumn{1}{C{1.4cm}}{ ff4 }  &   \multicolumn{1}{C{1.4cm}}{ ff5 }  &   \multicolumn{1}{C{1.4cm}}{ ff6 }  \\ 
\midrule High resilience  &  -1.58^{***}  &  0.25^{***}  &  -0.02  &  -0.03  &  -0.02  &  -0.01  \\ 
  &  \multicolumn{ 1}{c}{[-3.34]}  &  \multicolumn{ 1}{c}{[3.74]}  &  \multicolumn{ 1}{c}{[-0.23]}  &  \multicolumn{ 1}{c}{[-0.32]}  &  \multicolumn{ 1}{c}{[-0.37]}  &  \multicolumn{ 1}{c}{[-0.14]}  \\ 
Low resilience  &  -1.77^{***}  &  -0.26^{*}  &  -0.24^{*}  &  -0.23  &  -0.24  &  -0.28^{*}  \\ 
  &  \multicolumn{ 1}{c}{[-3.60]}  &  \multicolumn{ 1}{c}{[-1.80]}  &  \multicolumn{ 1}{c}{[-1.74]}  &  \multicolumn{ 1}{c}{[-1.70]}  &  \multicolumn{ 1}{c}{[-1.60]}  &  \multicolumn{ 1}{c}{[-1.78]}  \\ 
\midrule High-minus-Low  &  0.18  &  0.51^{***}  &  0.22  &  0.20  &  0.22  &  0.27  \\ 
  &  \multicolumn{ 1}{c}{[1.56]}  &  \multicolumn{ 1}{c}{[2.93]}  &  \multicolumn{ 1}{c}{[1.02]}  &  \multicolumn{ 1}{c}{[0.96]}  &  \multicolumn{ 1}{c}{[1.21]}  &  \multicolumn{ 1}{c}{[1.41]}  \\  \bottomrule \end{tabular}
	
	\bigskip

	Panel C. Portfolios sorted by ``teleworkable\_emp'' \\ \vspace{1mm}
	\begin{tabular}{L{5cm}d{2.2}d{2.2}d{2.2}d{2.2}d{2.2}d{2.2}} \toprule   &  \multicolumn{1}{C{1.4cm}}{ ret }  &   \multicolumn{1}{C{1.4cm}}{ capm }  &   \multicolumn{1}{C{1.4cm}}{ ff3 }  &   \multicolumn{1}{C{1.4cm}}{ ff4 }  &   \multicolumn{1}{C{1.4cm}}{ ff5 }  &   \multicolumn{1}{C{1.4cm}}{ ff6 }  \\ 
\midrule High resilience  &  -1.64^{***}  &  0.20^{***}  &  -0.04  &  -0.07  &  -0.04  &  -0.05  \\ 
  &  \multicolumn{ 1}{c}{[-3.25]}  &  \multicolumn{ 1}{c}{[3.14]}  &  \multicolumn{ 1}{c}{[-0.51]}  &  \multicolumn{ 1}{c}{[-0.81]}  &  \multicolumn{ 1}{c}{[-0.94]}  &  \multicolumn{ 1}{c}{[-1.00]}  \\ 
Low resilience  &  -1.68^{***}  &  -0.20  &  -0.21^{*}  &  -0.17  &  -0.21  &  -0.22^{*}  \\ 
  &  \multicolumn{ 1}{c}{[-3.53]}  &  \multicolumn{ 1}{c}{[-1.44]}  &  \multicolumn{ 1}{c}{[-1.66]}  &  \multicolumn{ 1}{c}{[-1.41]}  &  \multicolumn{ 1}{c}{[-1.65]}  &  \multicolumn{ 1}{c}{[-1.67]}  \\ 
\midrule High-minus-Low  &  0.04  &  0.40^{***}  &  0.16  &  0.10  &  0.17  &  0.17  \\ 
  &  \multicolumn{ 1}{c}{[0.54]}  &  \multicolumn{ 1}{c}{[2.56]}  &  \multicolumn{ 1}{c}{[0.85]}  &  \multicolumn{ 1}{c}{[0.50]}  &  \multicolumn{ 1}{c}{[1.24]}  &  \multicolumn{ 1}{c}{[1.21]}  \\  \bottomrule \end{tabular}
	
\end{spacing}      
  \end{center}
\end{table}

\clearpage
\newpage
\begin{table}[tbp]
    \caption{Other measures of \citet{hensvik/lebarbanchon/rathelot:20cepr}  } \label{tabA4} 

\footnotesize

\begin{adjustwidth}{-1.75cm}{-1.75cm}
\begin{spacing}{1.0}
This table presents results analogous to Panel C in Table \ref{tab2} but using the other measures constructed by \citet{hensvik/lebarbanchon/rathelot:20cepr} instead of their `dur\_workplace' variable. 

\end{spacing}
\end{adjustwidth}  

\bigskip
  \begin{center}  \footnotesize
  \begin{spacing}{1.0}

	Panel A. Portfolios sorted by `workplace' \\ \vspace{1mm}
	\begin{tabular}{L{5cm}d{2.2}d{2.2}d{2.2}d{2.2}d{2.2}d{2.2}} \toprule   &  \multicolumn{1}{C{1.4cm}}{ ret }  &   \multicolumn{1}{C{1.4cm}}{ capm }  &   \multicolumn{1}{C{1.4cm}}{ ff3 }  &   \multicolumn{1}{C{1.4cm}}{ ff4 }  &   \multicolumn{1}{C{1.4cm}}{ ff5 }  &   \multicolumn{1}{C{1.4cm}}{ ff6 }  \\ 
\midrule High resilience  &  -1.47^{***}  &  0.44^{***}  &  0.13  &  0.06  &  0.13  &  0.08  \\ 
  &  \multicolumn{ 1}{c}{[-3.45]}  &  \multicolumn{ 1}{c}{[5.16]}  &  \multicolumn{ 1}{c}{[0.92]}  &  \multicolumn{ 1}{c}{[0.47]}  &  \multicolumn{ 1}{c}{[1.40]}  &  \multicolumn{ 1}{c}{[0.86]}  \\ 
Low resilience  &  -1.84^{***}  &  -0.35^{**}  &  -0.34^{**}  &  -0.28^{**}  &  -0.35^{**}  &  -0.30^{**}  \\ 
  &  \multicolumn{ 1}{c}{[-3.68]}  &  \multicolumn{ 1}{c}{[-2.40]}  &  \multicolumn{ 1}{c}{[-2.22]}  &  \multicolumn{ 1}{c}{[-1.91]}  &  \multicolumn{ 1}{c}{[-2.22]}  &  \multicolumn{ 1}{c}{[-1.99]}  \\ 
\midrule High-minus-Low  &  0.37^{***}  &  0.80^{***}  &  0.47^{*}  &  0.33  &  0.49^{**}  &  0.38^{*}  \\ 
  &  \multicolumn{ 1}{c}{[3.17]}  &  \multicolumn{ 1}{c}{[3.77]}  &  \multicolumn{ 1}{c}{[1.87]}  &  \multicolumn{ 1}{c}{[1.28]}  &  \multicolumn{ 1}{c}{[2.11]}  &  \multicolumn{ 1}{c}{[1.75]}  \\  \bottomrule \end{tabular}
	
	\bigskip

	Panel B. Portfolios sorted by `dur\_home' \\ \vspace{1mm}
	\begin{tabular}{L{5cm}d{2.2}d{2.2}d{2.2}d{2.2}d{2.2}d{2.2}} \toprule   &  \multicolumn{1}{C{1.4cm}}{ ret }  &   \multicolumn{1}{C{1.4cm}}{ capm }  &   \multicolumn{1}{C{1.4cm}}{ ff3 }  &   \multicolumn{1}{C{1.4cm}}{ ff4 }  &   \multicolumn{1}{C{1.4cm}}{ ff5 }  &   \multicolumn{1}{C{1.4cm}}{ ff6 }  \\ 
\midrule High resilience  &  -1.63^{***}  &  0.19^{***}  &  -0.10  &  -0.05  &  -0.10  &  -0.03  \\ 
  &  \multicolumn{ 1}{c}{[-3.46]}  &  \multicolumn{ 1}{c}{[2.53]}  &  \multicolumn{ 1}{c}{[-0.93]}  &  \multicolumn{ 1}{c}{[-0.43]}  &  \multicolumn{ 1}{c}{[-1.54]}  &  \multicolumn{ 1}{c}{[-0.40]}  \\ 
Low resilience  &  -1.82^{***}  &  -0.19  &  -0.16  &  -0.23^{*}  &  -0.17  &  -0.25  \\ 
  &  \multicolumn{ 1}{c}{[-3.66]}  &  \multicolumn{ 1}{c}{[-1.58]}  &  \multicolumn{ 1}{c}{[-1.32]}  &  \multicolumn{ 1}{c}{[-1.76]}  &  \multicolumn{ 1}{c}{[-1.12]}  &  \multicolumn{ 1}{c}{[-1.56]}  \\ 
\midrule High-minus-Low  &  0.19^{*}  &  0.38^{***}  &  0.06  &  0.18  &  0.07  &  0.22  \\ 
  &  \multicolumn{ 1}{c}{[1.88]}  &  \multicolumn{ 1}{c}{[2.71]}  &  \multicolumn{ 1}{c}{[0.33]}  &  \multicolumn{ 1}{c}{[0.89]}  &  \multicolumn{ 1}{c}{[0.40]}  &  \multicolumn{ 1}{c}{[1.16]}  \\  \bottomrule \end{tabular}
	
	\bigskip

	Panel C. Portfolios sorted by `home' \\ \vspace{1mm}
	\begin{tabular}{L{5cm}d{2.2}d{2.2}d{2.2}d{2.2}d{2.2}d{2.2}} \toprule   &  \multicolumn{1}{C{1.4cm}}{ ret }  &   \multicolumn{1}{C{1.4cm}}{ capm }  &   \multicolumn{1}{C{1.4cm}}{ ff3 }  &   \multicolumn{1}{C{1.4cm}}{ ff4 }  &   \multicolumn{1}{C{1.4cm}}{ ff5 }  &   \multicolumn{1}{C{1.4cm}}{ ff6 }  \\ 
\midrule High resilience  &  -1.58^{***}  &  0.27^{***}  &  -0.01  &  0.01  &  -0.01  &  0.02  \\ 
  &  \multicolumn{ 1}{c}{[-3.49]}  &  \multicolumn{ 1}{c}{[5.89]}  &  \multicolumn{ 1}{c}{[-0.11]}  &  \multicolumn{ 1}{c}{[0.13]}  &  \multicolumn{ 1}{c}{[-0.20]}  &  \multicolumn{ 1}{c}{[0.37]}  \\ 
Low resilience  &  -1.94^{***}  &  -0.39^{**}  &  -0.32^{*}  &  -0.37^{*}  &  -0.32  &  -0.38^{*}  \\ 
  &  \multicolumn{ 1}{c}{[-3.77]}  &  \multicolumn{ 1}{c}{[-2.12]}  &  \multicolumn{ 1}{c}{[-1.80]}  &  \multicolumn{ 1}{c}{[-1.98]}  &  \multicolumn{ 1}{c}{[-1.64]}  &  \multicolumn{ 1}{c}{[-1.84]}  \\ 
\midrule High-minus-Low  &  0.36^{**}  &  0.66^{***}  &  0.31  &  0.38  &  0.31  &  0.41  \\ 
  &  \multicolumn{ 1}{c}{[2.01]}  &  \multicolumn{ 1}{c}{[3.31]}  &  \multicolumn{ 1}{c}{[1.23]}  &  \multicolumn{ 1}{c}{[1.52]}  &  \multicolumn{ 1}{c}{[1.22]}  &  \multicolumn{ 1}{c}{[1.56]}  \\  \bottomrule \end{tabular}
	
	\bigskip

	Panel D. Portfolios sorted by `share\_home' \\ \vspace{1mm}
	\begin{tabular}{L{5cm}d{2.2}d{2.2}d{2.2}d{2.2}d{2.2}d{2.2}} \toprule   &  \multicolumn{1}{C{1.4cm}}{ ret }  &   \multicolumn{1}{C{1.4cm}}{ capm }  &   \multicolumn{1}{C{1.4cm}}{ ff3 }  &   \multicolumn{1}{C{1.4cm}}{ ff4 }  &   \multicolumn{1}{C{1.4cm}}{ ff5 }  &   \multicolumn{1}{C{1.4cm}}{ ff6 }  \\ 
\midrule High resilience  &  -1.62^{***}  &  0.27^{***}  &  -0.01  &  -0.03  &  -0.01  &  -0.01  \\ 
  &  \multicolumn{ 1}{c}{[-3.73]}  &  \multicolumn{ 1}{c}{[3.69]}  &  \multicolumn{ 1}{c}{[-0.08]}  &  \multicolumn{ 1}{c}{[-0.28]}  &  \multicolumn{ 1}{c}{[-0.14]}  &  \multicolumn{ 1}{c}{[-0.08]}  \\ 
Low resilience  &  -1.84^{***}  &  -0.30^{*}  &  -0.28^{*}  &  -0.26  &  -0.28^{*}  &  -0.28^{*}  \\ 
  &  \multicolumn{ 1}{c}{[-3.64]}  &  \multicolumn{ 1}{c}{[-1.88]}  &  \multicolumn{ 1}{c}{[-1.79]}  &  \multicolumn{ 1}{c}{[-1.67]}  &  \multicolumn{ 1}{c}{[-1.66]}  &  \multicolumn{ 1}{c}{[-1.63]}  \\ 
\midrule High-minus-Low  &  0.22  &  0.56^{***}  &  0.27  &  0.23  &  0.27  &  0.28  \\ 
  &  \multicolumn{ 1}{c}{[1.63]}  &  \multicolumn{ 1}{c}{[2.85]}  &  \multicolumn{ 1}{c}{[1.09]}  &  \multicolumn{ 1}{c}{[0.90]}  &  \multicolumn{ 1}{c}{[1.23]}  &  \multicolumn{ 1}{c}{[1.19]}  \\  \bottomrule \end{tabular}

\end{spacing}      
  \end{center}
\end{table}

\clearpage

\end{appendices}
\end{document}